\documentclass[useAMS,usenatbib]{mn2e}
\usepackage{rotating}
\usepackage{journals}
\usepackage{graphicx, subfigure}
\usepackage{xspace}
\usepackage{amssymb}

\LARGE \normalsize \title[Accreting binaries in the GBS]{Identification of twenty-three  accreting binaries in the Galactic Bulge Survey}

\author[Torres et al.]  {M.A.P.~Torres$^{1}$\thanks{email : M.Torres@sron.nl}, P.G.~Jonker$^{1,2,3}$, 
C.T. Britt$^{4,5}$,  C. B. Johnson$^{4}$, R.I.~Hynes$^{4}$,   S. Greiss$^{6}$,   \newauthor D.~Steeghs$^{6}$,   T. J. Maccarone$^{5}$,  
F.~$\ddot{\rmn{O}}$zel$^{7}$,  C. Bassa$^{8}$, G. Nelemans$^{3,9}$\\
$^1$SRON, Netherlands Institute for Space Research, Sorbonnelaan 2, 3584~CA, Utrecht, The Netherlands\\
$^2$Harvard--Smithsonian Center for Astrophysics, 60 Garden Street, Cambridge, MA~02138, U.S.A.\\
$^3$Department of Astrophysics/ IMAPP, Radboud University Nijmegen, Heyendaalseweg 135,6525 AJ, Nijmegen, The Netherlands \\
$^4$Department of Physics and Astronomy, Louisiana State University, Baton Rouge, LA, 70803-4001, USA \\
$^5$ Department of Physics, Texas Tech University, Box 41051, Lubbock, TX 79409-1051, USA  \\
$^6$Department of Physics, University of Warwick, Coventry CV4 7AL, UK \\
$^7$University of Arizona, Department of Astronomy, 933 N. Cherry Ave., Tucson, AZ 85721, U.S.A \\
$^8$Jodrell Bank Centre for Astrophysics, The University of Manchester, Manchester M13 9PL \\
$^9$Institute for Astronomy, KU Leuven, Celestijnenlaan 200D, B-3001 Leuven, Belgium
}

\begin{document}

\maketitle

\begin{abstract} \noindent We are undertaking a survey to characterize
  the X-ray sources found with the {\it Chandra} X-ray observatory in
  a strip of fields at $ -3^\circ < l < 3^\circ$, $b=+1.5^\circ$ and
  $-3^\circ < l < 3^\circ$, $b=-1.5^\circ$. This so-called Galactic
  Bulge Survey (GBS) targets X-ray emitting binaries in the bulge with
  the primary purpose of finding quiescent X-ray binaries. The aims of
  this survey are to quantify dynamically the mass of compact objects
  in these X-ray binaries in order to constrain the neutron star
  equation of state and to test black hole formation models. In
  addition, using the survey number counts of various sources we aim
  to test models for binary formation and evolution.  Here, we present
  the identification of optical counterparts to twenty-three GBS X-ray
  sources.  We report their accurate coordinates and medium resolution
  optical spectra acquired at the Very Large Telescope and Magellan.
  All sources are classified as accreting binaries according to their
  emission line characteristics. To distinguish  accreting binaries
  from chromospherically  active objects we develop and explain
  criteria based on H$\alpha$ and He{\sc i} $\lambda\lambda 5786,6678$
  emission line properties available in the literature.  The
  spectroscopic properties and photometric variability of all the
  objects are discussed and a classification of the source is given
  where possible. Among the twenty-three systems, at least nine of them
  show an accretion-dominated optical spectrum (CX28, CX63, CX70,
  CX128, CX142, CX207, CX522, CX794 CX1011) and another six show
  photospheric lines from a late-type donor star in addition to
  accretion disc emission (CX44, CX93, CX137, CX154, CX377 and CX1004)
  indicating that they are probably accreting binaries in quiescence
  or in a low accretion rate state. Two sources are confirmed to be
  eclipsing: CX207 and CX794.  CX207 shows a  broad asymmetric
  H$\alpha$ profile blue-shifted by $>300$ km s$^{-1}$. Such line
  profile characteristics are consistent with a magnetic (Polar)
  cataclysmic variable. CX794 is an eclipsing nova-like cataclysmic
  variable  in the period gap.  Time-resolved photometry and the large
  broadening of the H$\alpha$ emission lines in CX446  (2100 km
  s$^{-1}$ Full-Width at Half Maximum; FWHM) suggest that this is also
  an eclipsing or high-inclination accreting binary.   Finally, the
  low-accretion rate source CX1004 shows a double-peaked H$\alpha$
  profile with a FWHM of 2100 km s$^{-1}$.  This supports a high
    inclination or even eclipsing system. Whether the compact object
    is a white dwarf in an eclipsing cataclysmic variable  or a black
    hole primary in a high-inclination low-mass X-ray binary remains
    to be established.

\end{abstract}


\section{Introduction}

The Galactic Bulge Survey (GBS) is a   multi-wavelength survey
designed to search for both X-ray sources and their optical
counterparts in the direction towards and in the Milky Way bulge. The GBS
covers a pair of $6^\circ \times 1^\circ$ areas centered $1.5^\circ$
above and below the Galactic Plane. These regions are chosen to avoid the
high optical extinction and crowding found in the mid-plane.   The
full survey area has been observed in X-rays with the  {\it Chandra}
X-ray observatory to  a   0.5 - 10 keV flux upper limit of  $(1-3) \times
10{^{-14}}$ erg cm${^{-2}}$ s$^{-1}$.  The GBS area has been also
imaged in optical bands with  the CTIO 4-m Blanco telescope to typical
upper limits  of $r^\prime$, $i^\prime \lesssim 23$.

A total of 1640 unique X-ray sources have been found (Jonker et
al.~2011; Jonker et al.~in prep.), providing a large sample to fulfill
the two main goals of this project. In what follows we briefly
describe these goals, while the specific details concerning the GBS
can be found in Jonker et al. (2011). The first GBS goal involves
measuring masses of neutron stars and black holes in eclipsing X-ray
binaries in order to constrain the neutron star equation of state and
black hole formation models. This requires us to identify new X-ray
binaries with optical counterparts suitable for dynamical studies.
These studies will allow one to measure the masses of compact objects,
with unprecedented accuracy in the case of eclipsing systems. To
facilitate achieving this objective, the limiting sensitivity for the
X-ray imaging was selected to maximize the number of detected
quiescent low-mass X-ray binaries (qLMXBs) with respect to new
Cataclysmic Variables (CVs). The expectation is to detect in X-rays
more than 200 qLMXBs with 120 of them having detectable counterparts
in the GBS optical images.  In comparison, the current number of known
LMXBs in the Galaxy exceeds 170 (see e.g.~\"Ozel et al. 2010). More
than half of them are persistently X-ray bright, making them
  usually unsuitable for  dynamical studies due to the lack of
spectral features from the donor star in their optical spectra.  In
contrast, the photosphere from the donor star is expected to
contribute significantly to the optical spectra of qLMXBs.  Nowadays
the census of LMXBs grows each year by a few with the discovery of
transient LMXBs during unpredictable X-ray outbursts.  Once in
quiescence, optical counterparts are detected for sources not affected
by severe reddening and/or crowding issues.  The GBS instead
represents a systematic effort to identify a large sample of new
qLMXBs with optical counterparts suitable for further follow-up. 
  A sample consisting of systems that have not undergone a recent
  outburst will allow us to establish if there exist selection effects that
  arise from the current sample of  persistent objects and transients that have suffered  outbursts in
  the past few decades.

The second GBS objective is to investigate binary formation and
evolution scenarios, including the study of kicks during the supernova
event that produces the compact primary in an LMXB. The method
employed to achieve this goal involves comparing the observed number
of sources per source class with the number of sources predicted by
binary population synthesis calculations.  The calculations presented
in Jonker et al. (2011) predicted a total of 1648 X-ray sources in the
GBS area. This is in very good agreement with the 1640 sources
detected in X-rays.  To allow detailed comparisons, multiwavelength and
variability studies are in progress to identify and characterize
counterparts to the GBS sources. See e.g. Udalski et al. (2012) and Hynes et al. (2012a)
for a photometric search and study of GBS sources with bright or moderately bright optical counterparts, Greiss et al. (2013) for the identification  of counterparts at infrared  bands, Britt et al. (2013) for the optical spectroscopic identificacion of five accreting binaries,  Maccarone et al. (2012b) 
for the characterization of GBS sources with bright radio counterparts, Ratti et al. (2013a) for the first dynamical study of a GBS source and Hynes et al. (2013) for the identification of a symbiotic X-ray binary.  The GBS and related data are also being used to study  recently discovered X-ray transients  or previously known (but poorly studied)  sources showing  renewed activity (e.g.~Greiss et al. 2011a,b; Hynes et al. 2012a, Maccarone et al. 2012a, Rojas et al. 2012)

In order to identify the optical counterparts we are acquiring spectra
for the candidate counterparts found within the error circle for the
precise {\it Chandra} X-ray position.  Large aperture telescopes are
being used to target qLMXBs which are expected to be faint as they
suffer from reddenening and are located at kiloparsec distances.  For
instance the unreddened apparent magnitude and colours of a (donor)
star at a distance of 8 kpc are
${(r',r'-i',r'-K)_0}={(17.4,0.1,0.9)_0}$ for F0 dwarfs and
$(23.6,1.3,3.4)_0$ for M2 dwarfs.  Taking into account reddening the
predicted apparent magnitudes for active LMXBs in the GBS area is $18
< i' < 21$, while qLMXBs are expected to have $i' >20 $ (Jonker et
al. 2011).

In this paper we focus on twenty-three GBS objects that we have
selected as new accreting binaries on the basis of our results
obtained from spectroscopy with the VLT and Magellan telescopes.  We
begin by describing in Section 2 the spectroscopic and photometric
observations.  In Section 3 we explain how the data were processed. In
Section 4 we present and discuss the results for the twenty-three sources.
Accurate coordinates, finding charts, optical light curves from the
Blanco 4m telescope and spectra of medium resolution ($\sim5, 10$ \AA)
and relatively large wavelength coverage are provided for each object.
Finally, in Section 5 we consider how to overcome the selection bias
toward emission line objects that single epoch spectroscopic surveys
will suffer from. In this paper we will follow the designation to the
sources introduced in Jonker et al. (2001): CX $N$ where $N$ is the
source number in the count rate sorted GBS catalogue.

\section{Observations and data reduction}

\subsection{{\it VIMOS (VIsible Multi-Object Spectrograph}}

VIMOS (Le F\`evre et al., 2003) is a 4-channel imager and multi-object
spectrograph mounted on the Nasmyth focus of the 8.2-m ESO Unit 3 Very
Large Telescope at Paranal, Chile.   Each channel (a.k.a.~quadrant)
covers a field of view of about $7 \times 8 $ arcmin$^2$. The four
quadrants are separated by approximately 2 arcmin wide gaps.  The
current VIMOS detectors are  $2048 \times 4096$ pixels EEV CCDs that
provide a 0.205 arcsec/pixel scale.   Prior to  August 2011 VIMOS was
equipped with EEV CCDs of same format as the new detector, but
with both lower sensitivity and strong fringing in the red. The GBS
sources CX73 and CX772 were observed using the old CCDs.

During the observations VIMOS was operated in multi-object
spectroscopy (MOS) mode. In this mode laser-cut slit masks are
inserted at the entrance focal plane permitting spectroscopy of
optical counterpart candidates to several GBS sources in a single
exposure.  The slit masks are prepared using the VIMOS Mask
Preparation Software (VMMPS) tool (see Bottini et al. 2005).  VMMPS
uses pre-images of the fields as reference frames to define the slit and secure objects
well-centered at the slit positions after telescope acquisition of the
field. All slits were designed with a  1.0 arcsec width. Their length and location 
were  determined per GBS source to include more than one 
optical counterpart candidates if  possible, to secure 
areas for sky subtraction during the data reduction and/or to avoid strong saturation of bright sources. Thus 
the identified counterparts were not necessarily centered on their slits.   

We selected the medium resolution (MR) grism that yields a 2.5
\AA/pixel dispersion and a wavelength coverage of $\sim 4800-10000$
\AA. The exact coverage depends on the location of the small slit with
respect to the CCD boundaries as some part of the dispersed spectrum
may fall off the chip. The GG475 order-sorting filter was used to
avoid overlap between the first and second grating orders when
observing bright sources. The use of 1.0 arcsec width slits provided a
spectral resolution of about 10 \AA~FWHM corresponding to 460 km
s$^{-1}$ at H$\alpha$ and 350 km s$^{-1}$ at 8600 \AA.

The spectroscopic observations were obtained in service mode under
  programs 085.D-0441(A)  and 087.D-0596(A). One hour observing blocks
(OBs) were executed per pointing, consisting of acquisition imaging,
two spectroscopic integrations of 875 s, three flat-field exposures
and a helium-argon lamp exposure for wavelength calibration.  The
standard data reduction steps were  performed with the ESO-VIMOS
pipeline (Izzo et al. 2004).  These include  bias subtraction,
bad-pixel correction, wavelength calibration and removal of the
instrumental response. The typical root mean square (RMS) for the
  wavelength calibration fit was $<0.1$  pixels.  The pipeline
averages the two spectra and extracts automatically the objects found
in each slit. 

The pipeline data products were examined to identify any problems with
the extraction of the spectra.  This examination showed that the
output spectra for isolated objects were satisfactory in terms  of
extraction quality. This was not the case for partially resolved
sources where the choice of the aperture size for the object, sky
subtraction regions and tracing become critical steps during the
extraction process.  Thus we extracted interactively with the {\sc
  iraf kpnoslit} package all reduced 2-D frames that contain both
stellar and sky spectra (pipeline files with product code SSEM). For
partially resolved sources in crowded fields,  the object and sky
apertures were sized to maximize the number of pixels contributing to
the object and sky, respectively, and to minimize the contribution
from any field star/s.   Checks for the stability of the wavelength
calibration were made for each object looking at the strongest
atmospheric emission lines present in the spectra. We made use of the
[O{\sc i}] $\lambda 6300.3$ line and/or the OH emission at
$\lambda\lambda 6863.96, 9872.14 $ \AA~ (Osterbrock et al. 1996,1997).
In this way we estimate that the median amplitude of the radial
velocity error is $\lesssim 20$ km s$^{-1}$. Zero point corrections
were applied only to the spectra of the sources CX70 and CX377 which
showed radial velocity offsets in the [O{\sc i}] and OH emission lines 
wavelengths well above the distribution of the internal radial velocity error.

In a few cases the observing conditions were not good during the OB
execution and the OB was repeated later under the weather conditions
requested for the program.  Spectra   of an optical counterpart
  extracted from different OBs were combined if they were acquired
close in time and if there were no significant changes in the spectra
between observations.  A log of the VIMOS observations is presented in
Table 1 in which identification numbers for the GBS sources,
pre-imaging OBs and spectroscopic OBs are given.  The observing date
and the quadrant where the object was detected are also provided.

\subsection{{\it IMACS (Inamori-Magellan Areal Camera and Spectrograph)}}

The GBS sources CX44 and CX1011 were observed with IMACS  (Bigelow \&
Dressler 2003) which is mounted on the 6.5-m Baade-Magellan telescope
at Las Campanas Observatory.  Both objects were observed on the night
of 15 May 2011 UT with  IMACS operated in short-camera mode. In this
mode the spectra are dispersed along the long axis of two of the eight
$2112 \times 4160$ pixels E2V CCDs with 0.20 arcsec/pixel scale.  We
used a 300 line mm$^{-1}$ grism centered at 6700  \AA~ yielding a
dispersion of  1.15 \AA~ pixel$^{-1}$ (CCD No. 3) and 1.31 \AA~
pixel$^{-1}$ (CCD No. 8) in the spectral intervals $3450-6660$ and
$6750-10410$  \AA, respectively.  Together with a 1.0 arcsec wide slit, 
the instrumental setup achieved  a spectral resolution of
$\sim 5$ \AA~ FWHM corresponding to 230 km s$^{-1}$ at H$\alpha$
and 175 km s$^{-1}$  at 8600 \AA.  A single spectrum was acquired per
source with an  integration time of 1200 and 900 s  for CX44 and
CX1011, respectively.  The flux standards Feige 56, Feige
57 and LTT 3864 were also observed in order to correct for the
instrumental response.

The IMACS frames were bias and flat-field corrected with standard {\sc
  iraf} routines. The spectra were extracted from each CCD frame with
the {\sc iraf kpnoslit} package. The pixel-to-wavelength calibration
was derived from cubic spline fits to HeNeAr arc lines. The RMS
deviation of the fit was $\lesssim 0.06$ \AA .  The sky [O{\sc i}]
$\lambda 6300.3$ line and the OH emission blend at 7316.3 \AA~show
that the accuracy of the wavelength calibration of the spectra from both
CCDs is $\lesssim 0.08$~\AA.

\subsection{NOAO Mosaic-II imager}

We acquired time-resolved photometry from 12 to 18 July 2010 with the
Mosaic-II instrument mounted on the Blanco 4-m telescope at the Cerro
Tololo Inter-American Observatory. Multiple 120\,s exposures of 45
overlapping fields were taken to cover a nine square degree area in
the Sloan $r'$-band containing the GBS X-ray sources reported in
Jonker et al. (2011). The order in which the fields were visited was
randomized to minimize aliasing caused by regular sampling. Typical
seeing conditions during the observations were around 1 arcsec.  The
last column of Table 1 provides the number of frames containing a
detection of the optical counterpart to a GBS source. No photometric
data are available for CX522 as the source is unresolved from a bright
field star in the Mosaic-II data.

The data were reduced via the NOAO Mosaic Pipeline  which searches and
corrects for instrumental artifacts in the image such as cross talk
between CCDs. The pipeline  applies bias and flat field corrections,
and adds a world coordinate system  for each image based on USNO-B1
stars in the field.   An estimate of the photometric  zero-point of
  the images is also made by comparing the instrumental magnitude of
  the field star to their USNO B1.0 apparent magnitudes.  A detailed
explanation of the data  reduction procedures can be found in chapter
2 of the NOAO Data Handbook (Shaw 2009).

\section{Data analysis}
\subsection{Spectroscopy}
Given the uncertain reddening towards the GBS sources and thereby
uncertain intrinsic colours, the classification of optical
counterparts in this paper was based on the identification of spectral
lines and not on the optical continuum measurements.  Thus the reduced
spectra were normalized by fitting cubic splines to the continuum
after masking emission lines, telluric bands and instrumental
artifacts.   Relevant spectral features were identified and studied.
These include late, early or white dwarf type photospheric lines,
emission lines and diffuse interstellar bands (DIBs).  The
normalized spectra of the accreting binaries  discussed in this paper
are presented in Fig.~1 to 3.  For the sake of  identification,  the
rest wavelengths for the following lines are marked in all figures:
H$\beta$,  He {\sc i} $\lambda 5876$, H$\alpha$ and He {\sc i}
$\lambda\lambda 6678, 7065$ (left panels). In the right panels the
markers correspond to  the Na {\sc i} doublet $\lambda\lambda 8183,
8195$, Ca {\sc ii} $\lambda 8498$, P16,  Ca {\sc ii} $\lambda 8542$,
P15,  P14,  Ca {\sc ii} $\lambda 8662$, and P13/12/11/10/9 and 8
(here, P$n$ stands for the Paschen $n - 3$ transition).   

Strong features were examined manually with the {\it splot} task in
{\sc iraf} obtaining their radial velocity (RV), FWHM and equivalent
width (EW).  The two first parameters were measured by fitting the
line profiles with a Gaussian function, whereas the EW was measured by
means of the {\it splot} keystroke $e$.  Table 2 reports the values
for the strongest emission lines except for lines which do not have a
well-defined profile due to noise and/or telluric contamination
problems. Their uncertainties were estimated by looking at the scatter
in the values when selecting different wavelength intervals to set the
local continuum level.  Finally note that all radial velocities provided in the work
are heliocentric ones. 
 
\subsection{Astrometry}

  Coordinates for the optical counterparts to the twenty-three GBS sources
  discussed in the present work are provided in Table 2.  Except for
  CX44, they were obtained from the VIMOS pre-images after performing
  an astrometric calibration that delivered a position accurate to  RMS $< 0.1$
  arcsec. The position for the optical counterpart to CX44 was derived
  from Mosaic-II  images (see Sec 2.3).  Finding charts 20 arcsec
  $\times$ 20 arcsec wide are shown  in the appendix. Observers
  requiring wider charts can find in Table 1 the  necessary details to
  retrieve from the ESO archive the pre-images for the target of interest.

\subsection{Photometry}

Photometry was performed using the image subtraction package ISIS
(Alard \& Lupton 1998,  Alard 2000). ISIS uses a reference image which
is convolved with a kernel in an effort to match a subsequent image of
the same field. This subsequent image is then subtracted from the
convolved reference image. Stars which do not vary in magnitude should
subtract cleanly, such that the subtracted image is clear of
non-variable objects. Therefore, any residual flux is due to an
inherent change in brightness of a source.    ISIS performs PSF photometry on
the subtracted images. The  model of the PSF used in each image is build  by convolving
the PSF model  in the reference image with the kernel solution.

In order to save computation time, $104'' \times 104''$ cutouts of the full Mosaic-II
  images around the position of each X-ray source were taken for processing.
The resulting ISIS images were examined by eye to identify variable
objects near or at the X-ray positions of the GBS sources.  Light curves
were then built for these variables and  Lomb-Scargle
periodograms were used to search for periodicities if enough data
points were available.  Ellipsoidal modulation is expected in close
binaries, in particular   in accreting binaries in which the donor
star dominates the light curve. We took into account that this type of
photometric variability has two maxima and minima in a single orbital
cycle and  checked periods twice as long as prominent peaks on a
periodogram. We also consider both aliases and harmonics, as higher
harmonics can sometimes appear at a higher power than the fundamental
frequency.

At present, we lack an absolute calibration for these $r'$- band
Mosaic-II data. All apparent magnitudes cited in this paper are a
pipeline calibration product (see Sec 2.3). They are to be considered
rough estimates and thereby used with caution until secondary
standards are established for the Mosaic-II fields. The pipeline
zero-point calibration carries an estimated uncertainty of 0.5 mag.

\section{Results and Discussion}

\subsection{Criteria for the spectroscopic identification of accreting binaries}

All counterparts were selected for showing (at least) H$\alpha$ in emission above the
continuum.  A key motivation of the present work is to report the
secure and likely accreting binaries found in this sample and describe the selection
criteria used for their identification. We will report on additional
candidate counterparts to X-ray sources that showed no H$\alpha$
emission line in their spectrum elsewhere (see e.g. Hynes et al. 2013). 

Typically, accreting binaries with hydrogen-rich mass donor stars are
unambiguously identified by the presence of broad Hydrogen and Helium
emission lines formed in the gas being accreted.  Emission from this
gas and/or from a white dwarf primary (in the case of CVs) produces a
characteristic excess in the optical continuum at blue wavelengths.
We have based the identification of accreting binaries on the
detection of strong (EW$> 18$ \AA, see below) H$\alpha$ in emission,
the Doppler-broadening of the Hydrogen line profiles or the detection
of He{\sc i} $\lambda 6678$ emission in the case of observing weak and narrow Hydrogen lines.  A
search for Carbon, Nitrogen and Oxygen emission lines was also
performed.  Emission from these heavy elements in a spectrum lacking
Hydrogen lines is observational evidence for a Carbon-Oxygen white dwarf donor in
an ultra-compact X-ray binary (see e.g. Werner et al. 2006, Nelemans
et al. 2004, 2006). None were found so far in our sample.

Chromospherically active binaries and stars can be distinguished from
qLMXBs and low-accretion rate CVs by  the presence of H$\alpha$
emission that does not exceed 15 - 18 \AA~ in EW. This limit is
observed in M-type stars (Mohanty  \& Basri 2003, Covey et al. 2008)
and post-common envelope binaries (Bleach et al. 2002),  but does not
account  for the possibility of finding larger EWs during  a flare
event.   In active stars line broadening of the H$\alpha$ profile can
be well above the rotational broadening measured in the photospheric
lines and reach  up to 9 \AA~(= 400 km s$^{-1}$) FWHM in extremely
active objects (see e.g. Walter \& Basri 1982, Torres et al. 2005).
The lack of Helium in emission above the continuum is also a common
characteristic of chromospherically active sources. Out of a flare, He {\sc i}  
emission above the continuum has been found  in a few very active objects
(see Bleach et al. 2002 and references therein).  From Bleach et al.
one can establish that when present in emission, the He {\sc i} $\lambda\lambda 5876,6678$ lines do not exceed 
3 \AA~in EW. The He {\sc i}  to H$\alpha$ line ratios are $< 0.2$ even during flares.

Pre-main sequence stars that host a circumstellar disc (Classic T
Tauri stars) may show permitted lines in emission and mimic the
optical and X-ray characteristics of reddened CVs and qLMXBs at a low
accretion rate   state.  For instance, they can reach X-ray luminosities of up
to 10$^{31}$  erg s$^{-1}$ (Telleschi et al. 2007), a value typically
observed in quiescent black hole LMXBs, and their optical  spectra
can exhibit broad Hydrogen lines that can appear asymmetric or
double-peaked and have EWs  of hundreds of \AA~ (e.g.  Reipurth et
al. 1996,  Hamann \& Persson 1992, Barrado et al. 2005). Helium lines
are often present in the optical spectrum. Fortunately, this young
class of objects can be identified on the basis of their association
with star formation regions.

\subsection{Application to the current sample}

Through the analysis of the VIMOS and IMACS spectroscopy we have
found optical counterparts to 54 GBS sources. Of the 52 optical counterparts with an H$\alpha$ emission line found
in the data, 29 of them are likely chromospherically active objects as
they show narrow (unresolved) H$\alpha$ emission with EW $< 3$ \AA~ 
and a continuum clearly marked by absorption features
from a late-type star or binary components\footnote{These sources will
  be presented elsewhere once their chromospheric active nature is
  fully established through further analysis and/or observation. The
  objects in question are CX25, 29, 47, 57, 74, 124, 126, 161, 251,
  294, 321, 325, 381, 393, 409, 450, 469, 502, 634, 642, 712,
  820, 852, 858, 881, 895, 1014, 1088 and 1169.}.  Using the
SIMBAD database we have not found known star forming regions, star
clusters, OB associations or H{\sc ii} regions within one arcmin
radius of any of the twenty-three remaining sources. Applying the aforementioned
EW limit of 18 \AA, fifteen of the 23 sources are classified as
accreting binaries. Six of the remaining eight objects (CX70,
CX73, CX93, CX137, CX377 and CX1011) have H$\alpha$ profiles with FWHM
ranging $15-30$ \AA.  When subtracting in quadrature the instrumental
width, they correspond to intrinsic broadenings of $11-28$ \AA~($\sim
500-1300$ km s$^{-1})$.  Such values are above the 9 \AA~maximum
broadening observed in chromospherically active sources. This
indicates that the line profiles in these sources are broadened due to
the motion of emitting material in an accretion flow or wind.
Finally, the GBS sources CX154 and CX781 qualify as likely accreting binaries
on the basis of the presence of He{\sc i} $\lambda 6678$ in emission with a He{\sc i}/H$\alpha$ 
ratio above the  value expected from lines powered by chromospheric activity. 
Thus, we conclude that all twenty-three sources are most likely accreting
binaries.

Of the 23 GBS sources, five (CX44, CX93, CX137, CX154, CX377 and CX1004)
show, besides the H$\alpha$ emission from the accretion disc 
clear evidence  of photospheric absorption lines from the
donor star.  On the other hand, the spectra of CX28, CX63, CX70, CX128,
CX142, CX207, CX522, CX784 and CX1011 lack absorption features from such a donor.   In
these nine systems emission from the accreted gas is the dominant
contribution to the total optical light.  The other  eight objects
require other or more observations to establish with certainty the accretion
state and the presence of the donor star in the optical spectrum.

\subsection{Further constraints on  binary parameters from emission line properties}

The measured emission line properties provide not only a method to
discern accreting binaries from other classes of objects, but also
encode a wealth of information on binary parameters such as the system
inclination, orbital period, compact object nature or mass accretion
rate (see e.g. Warner 1995).  In particular the centroid of the line
profiles originating in a disc with gas in Keplerian motion will be
shifted by the systemic and primary's radial velocities at the time of
the observation: $\gamma +K{_1} \sin \phi $, where $\phi$ is the
orbital phase, $K_1$ the radial velocity semi-amplitudes of the
primary and $\gamma$ is the systemic radial velocity.  Additional
velocities incurred by the gas motion in the accretion disc may be
present as well.  Furthermore, the line broadening scales with the
mass of the compact object, the system inclination and the orbital
period as $\left ( M_1 / P_{\rm orb}\right )^{1/3}\sin i$. In this
regard, the low intrinsic FWHM of the H$\alpha$ line in CX44 ($510$ km
s$^{-1}$) suggests a moderate to low inclination for this system and thereby a low
l$K_1$.  Thus the large radial velocity offset in the emission lines  ($|RV| \sim 200$ km s$^{-1}$ )
likely reflect a high $\gamma$ given that the RV component due to the motion of the primary 
is expected to be low.  On this basis, CX44 is a good
candidate for an LMXB that experienced a natal kick - CVs in general
have low $\gamma$ (van Paradijs et al. 1996, Ak et al. 2010).  The
line profiles in CX207 and CX1004 have also significant radial
velocity offsets and exhibit the highest line broadening in the sample
(FWHM $> 1600$ km s$^{-1}$) in favor of a high inclination for both
binaries (both sources are discussed below).  However, caution must be
taken with this interpretation of the line properties as departures
for the simplistic picture of a Keplerian disc are expected and can be
dominating the observed $RV and$ FWHM. For instance, complexities in the line
profile morphology can result from tidal effects on the disc (Foulkes
et al. 2004) and winds that will cause asymmetric profiles. Also,
emission originating in regions of the binary system other than the
disc can contribute or even dominate the line profile shape and
variability. For instance,  this is the case in  high-magnetic CVs
(known as Polars)  where the magnetic field of the white dwarf
prevents the formation of an accretion disc. Emission lines in these systems can
originate from the accretion stream  that is directed towards the white dwarf's magnetic poles and 
from the (irradiated) donor star.  As discussed below, a high-magnetic CV is the 
favoured scenario to explain the emission line properties in CX207.

\section{Individual Objects}

In what follows, we discuss the  spectroscopy and photometry of the
twenty-three accreting binaries. We describe remarkable features and
provide a characterization of the GBS source on their basis.  When
useful we will employ the absolute magnitudes   and colors of stars as
given in Allen's Astrophysical Quantities (Cox 2000) and apply  the
transformation equations between Johnson-Cousins and SDSS magnitude
derived by Jester et al. (2005) and Bilir, Karaali \& Tun{\c c}el
(2005). We will quote from Jonker et al. (2011) the X-ray harness
ratio or/and the number of counts detected  in the 0.3 - 8 keV band
during the discovery observations with {\it Chandra}.  The hardness
ratio is defined as HR= (H-S)/(H+S), where S and H are the soft 0.3 -
2.5 keV and hard 2.5- 8 keV band count rates.   We provide here counts
to 0.5 - 10 keV unabsorbed flux conversion factors  of $6.6 \times
10^{-15}$  and $1.4 \times 10^{-14}$ erg cm$^{-2}$ s$^{-1}$
count$^{-1}$. They were obtained using a power-law spectra with index
$\Gamma = 2$ modified by a Galactic absorption  column  density $N_H$
of $10^{21}$ and $10^{22}$ cm$^{-2}$, respectively.  The adoption of
this spectral shape  for the X-ray sources yields a rough but useful
conversion except for soft sources suffering significant reddening in
which case the flux will  be underestimated by a factor of $\sim 10$.
Additionally, we give expressions to compute the unabsorbed 0.5 - 10
keV X-ray luminosity as a function of the distance  to the source $d$
and the X-ray counts: $L{_x} (d,N{_H}=10^{21}) = 7.9 \times 10^{29}
\times counts \times (\frac{d}{1.0~kpc})^2$ and  $L{_x}
(d,N{_H}=10^{22}) = 1.6 \times 10^{30} \times counts \times
(\frac{d}{1.0~kpc})^2$ erg s$^{-1}$.  We will adopt $N{_H}=10^{21}$
cm${^{-2}}$ for objects with bright optical counterparts or lacking
strong DIBs and  $N{_H}=10^{22}$ cm${^{-2}}$ to faint optical
counterparts.   The  inferred luminosities are  below those observed
in very-faint persistent neutron star LMXBs (e.g. Armas Padilla et
al. 2013)  and within the broad range of  luminosities  observed in
CVs (Baskill et al. 2005, Byckling et al. 2010, Reis et al. 2013) and
qLMXBs (see e.g. Jonker et al. 2012).

\subsection*{CX28 = CXOGBS J173946.9-271809, a high accretion rate CV}

Results based in part on the VIMOS spectrum (Fig. 1) and the Mosaic-II
time-resolved photometry suggest that CX28 is a high accretion rate
CV. This classification is supported by the detection of
the high-excitation emission line He {\sc ii} $\lambda\lambda 4686,
5412$.  We refer the reader to Britt et al. (2013) for a full
  discussion.  The absorption lines detected at $\lambda\lambda 8349, 8536,
8659$ in the VIMOS spectrum are caused by contamination of a late type
field star NE from this GBS source. A finding chart for the optical
  counterpart to CX28 can be found in Britt et al. (2013) as well.

\subsection*{CX39 = CXOGBS J174140.0-271738, an outbursting CV}

The VIMOS spectrum of CX39 (Fig. 1) exhibits H$\alpha$ and He {\sc i}
$\lambda\lambda 6678, 7065$ emission with single-peaked asymmetric
line profiles of intrinsic FWHM ranging from 700 to 840 km
s$^{-1}$. Note here that the apparent narrow emission features near
$\lambda 5900$ are sky-line residuals. Photospheric lines from the
donor star are not obvious in the spectrum. The presence of DIBs at
$\lambda\lambda6284, 8620$  is uncertain. The soft X-ray hardness
ratio $HR = -0.35 \pm 0.25$ indicates low reddening and thereby a
nearby object.

The optical light curve (Fig. 4) shows, what appears to be a brief
outburst of 1.2 mag amplitude, rising on the first day of observations
and decaying to $r'=20.2$ over the next 3 days.  After that, the
counterpart flickers with an RMS scatter of 0.1 mag.  We examined the
spatial profile of both the optical counterpart spectrum and the
spectra of field stars contained in the slit. This together with
comparison of the VIMOS pre-images obtained on 31 Mar 2010 and 17 Aug
2010 UT shows that CX39 was not in outburst during the spectroscopic
observations.

Recurrent low-amplitude brightenings with fast return to minimum light
have been observed in the long-term light curves of dwarf novae (see
e.g. Shears et al. 2006) and with lower amplitude in nova-like
CVs (Rodriguez-Gil et al. 2012a).  In addition, a small number of weak
magnetic CVs (Intermediate Polars, IPs) are  known to undergo
relatively short duration and low amplitude outbursts (see
e.g. Ishioka et al. 2002 and references therein).   

The lack of stellar features in the quiescent optical spectrum
  together with this photometric behaviour suggests that CX39 is a
    dwarf nova or IP with a  disc-dominated optical spectrum.  The
    X-ray flux detected with {\it Chandra} (39 counts) implies  $L{_x}
    (d,N{_H}=10^{21}) = 3.1 \times 10^{31} \times
    (\frac{d}{1.0~kpc})^2$  erg s$^{-1}$.

\subsection*{CX44 = CXOGBS J175542.8-281809 = AX J1755.7-2818, a qLMXB 
or a low accretion rate CV}

The IMACS spectrum of CX44 (Fig. 1) shows H$\alpha$ and H$\beta$ in
emission. He {\sc i} $\lambda 5876$ is present while the location of
the He {\sc i} $\lambda 6678$ line fell in a chip gap.  The IMACS
spectrum also covers the high-excitation line He {\sc ii} $\lambda
4686$, which is not detected.  As reported in Sec 4.2, the emission
lines exhibit velocity offsets of about $-200$ km s$^{-1}$.  Moreover,
photospheric features from a K-type companion star are visible in the
spectrum, the strongest being the molecular bands of the
TiO$\gamma^{'}$ system situated longward of $\lambda 5840$, Ca {\sc i}
$\lambda 6162$ and the blend of Fe {\sc i} + Ca {\sc i} at rest
wavelength $\lambda 6495$.  Gaussian fits to the metal-line profiles
yield a -50 km s$^{-1}$ radial velocity. In the red, the spectrum is dominated by
numerous telluric features and shows evidence for absorption from the
Ca {\sc ii} $\lambda 8542$ line.  There is no obvious detection of
the Na {\sc i} $\lambda 8190$ doublet or Paschen lines (neither in
emission nor in absorption). The lack of strong photospheric lines
from the donor in this part of the spectrum suggests a significant
contribution to the I-band continuum by the accretion disc.
    
The optical light curve (Fig. 4) shows aperiodic variability with an RMS of 0.05
mag.~and with an observed maximum variability of $\Delta r'$ =
r$'{_{max}}$ - r$'{_{min}}$ = 0.44 mag.

CX44 is the unique GBS source within the 50''-radius error circle
  of AX J1755.7-2818.  This ASCA object was found at a 0.7 - 10 keV  flux
  level of $1.2 \times 10^{-12}$ erg cm$^{-2}$ s$^{-1}$ during a
  pointing observation on 2 Nov 1999 (Sakano et al. 2002).
  Thirty-two counts were observed from CX44  with {\it Chandra} on 4
  May 2008. Using an absorbed power law model with $\Gamma = 2$ and
  $N{_H}$ ranging $10^{21} - 10^{22}$ cm$^{-2}$ we derive an absorbed
  flux  of $(1.7-2.5) \times 10^{-13}$ erg cm$^{-2}$  s$^{-1}$ in the
  0.7 - 10 keV band. This flux  is  at least 5 times lower than the
  value found during  the ASCA observation. 

Given the lack of strong DIBs in the optical spectrum and the soft
X-ray  hardness ratio ($HR = -0.19 \pm 0.19$), we expect CX44 to be a
nearby (few kpc) binary with a non-evolved donor star.  From the
apparent optical magnitude ($r'=18.7$)  we obtain a rough estimation
of 2 kpc for its distance by  assuming a K5 V companion star (M$_{r'}
= 7.0)$ and by neglecting both the effects of interstellar reddening
and veiling from the accretion disc. We also derive   $L{_x}
(d,N{_H}=10^{21}) \sim 1 \times 10^{32} \times (\frac{d}{2.0~kpc})^2$.
This X-ray brightness of the source together with, the detection of
the donor star and the dominant flickering in the optical light curve
suggests that we may have found a nearby low accretion rate CV or a
quiescent neutron star LMXB where the accretion disc contributes in a
large fraction to the optical light.

\subsection*{CX45 = CXOGBS J173538.5-285251, a CV with an accretion-dominated optical spectrum}

The VIMOS spectrum of CX45 (Fig. 1) shows H$\alpha$ and He {\sc i}
$\lambda 6678$ emission. The profiles for both lines are asymmetric
and have an intrinsic $\approx 900$ km s$^{-1}$ (FWHM) broadening.
The narrow emission features near $\lambda 5900$ are sky-line
subtraction residuals. Longward of H$\alpha$ there are no obvious
emission lines. There is no clear detection of absorption features
from the donor star in the full spectral range covered by VIMOS.  The
Mosaic-II light curve (Fig. 4) shows flickering with an RMS scatter of
0.39 mag. The amplitude of the observed variability is large: $\Delta
r' = 1.5$ mag.
 
The strong flickering at optical wavelengths together with the
  apparent lack of photospheric lines from the donor star supports
  that CX45 is a CV in a state of high mass transfer
  rate. Alternatively, CX45 could be a low accretion rate CV with the
  optical spectrum dominated by emission from the accreting flow as
  for instance observed in short orbital period dwarf novae (see CX128
  for more discussion on this type of CVs).  The brightness of the
  X-ray source (32 counts in the {\it Chandra} detection, $HR= 0.66
  \pm 0.24$) yields $L{_x} (d,N{_H}=10^{21}) = 2.5 \times 10^{31}
  \times (\frac{d}{1.0~kpc})^2$).

\subsection*{CX63 = CXOGBS J173411.3-293117, a CV with an accretion-dominated optical spectrum}

The optical spectrum of CX63 (Fig. 1) is dominated by the accretion
disc, there are no signs of photospheric lines from the donor star.
H$\alpha$ and He {\sc i} $\lambda\lambda 5876, 6678, 7065$ are in
emission with intrinsic broadening between $350-500$ km s$^{-1}$
(FWHM). H$\beta$ is also present and appears to be unresolved.  In the
red, P8 to P13 emission lines are seen.  The intrinsic broadening for
the Paschen lines ranges from 630 to 1000 km s$^{-1}$ (FWHM).

The optical light curve of CX63 (see Fig. 4) displays events of
increased/decreased brightness of up to 0.5 mag from its 21.3 mean
magnitude.  The RMS scatter in the Mosaic-II photometry is 0.16 mag,
while $\Delta r' = 0.85$ mag.

CX63 is bright in X-rays with 26 counts detected with {\it Chandra}
($HR= 0.16 \pm 0.23$, $L{_x} (d,N{_H}=10^{21}) = 2.1 \times 10^{31} \times (\frac{d}{1.0~kpc})^2$). 
The characteristics of the optical spectrum together with  the flickering observed in the light curve support 
that CX63 is a high accretion rate CV or a low accretion rate CV with an accretion-dominated optical spectrum
The low FWHM values measured from the emission line profiles support a moderate to low inclination system.

\subsection*{CX64 = CXOGBS J173802.7-283126, a CV in a low state of accretion or a CV in the Bulge region}

The H$\alpha$ emission line (intrinsic FWHM = $500 \pm 10$ km
s$^{-1}$) is the only feature detected in the blue part of the VIMOS spectrum of CX64
(Fig. 1).  The line profile  appears to be redshifted by 200  km s$^{-1}$.
Longward of H$\alpha$, lines of the Paschen serie may be in emission.
 It is unclear whether DIBs  or photospheric lines from the donor star are
present due to the noise in the spectrum. The optical counterpart to
CX64 is faint with $r'=22.3$.  The Mosaic-II light curve allows
us to place an upper limit in the photometric variability of 0.14 mag
(RMS).

If H$\alpha$ is originated in the accretion disc, its
low intrinsic broadening implies a moderate to low system inclination and thereby $K_1$. 
Given the moderate $K_1$ and that  CVs are expected to have low systemic velocities (see CX207 below for more details),
the  large velocity offset observed in the line can be explained if CX64
  is  a bulge CV (accreting at high mass transfer rate).  On the other
  hand, both the radial velocity offset and low line broadening are
  also consistent with a more nearby CV in  a low state. High accretion
  rate CVs n a low state of mass transfer show narrow emission lines
  originating in the irradiated donor stars. The radial velocities of these lines reflect 
therefore the motion of the donor star. 
The non-detection of He{\sc i} $\lambda 6678$ in
  the optical spectrum of CX64 is in support of this possibility as
  magnetic  and nova-like CVs have shown weak He{\sc
    i} $\lambda 6678$ or absence of  the line during low-state periods 
  (e.g. Kafka et al. 2005, Rodriguez-Gil et al. 2012b ). Unfortunately we cannot confirm this scenario as we  lack the signal to search for absorption
  features from the stellar components. On the basis of the X-ray brightness (25 {\it Chandra} counts, $HR=
0.22 \pm 0.23$) we derive  $L{_x} (d,N{_H}=10^{21}) = 2.0 \times
10^{32} \times (\frac{d}{1.0~kpc})^2$ erg s$^{-1}$).

\subsection*{CX70 = CXOGBS J173535.1-295942, a CV with an accretion-dominated optical spectrum}

The spectrum of the optical counterpart to CX70 shows the H$\alpha$
and P10 to P15 Hydrogen lines in emission with intrinsic broadening of
$\sim 700-900$ km s$^{-1}$ (see Fig. 1 and line parameters in
Table 2).   The presence of He {\sc i} $\lambda 6678$ is 
uncertain as the line position fell on a region of bad rows in the
detector.  While artifact features make difficult to address the presence of  photospheric lines
and DIBs in the blue part of the spectrum, it is clear that stellar lines  are not present at wavelengths  redward of $\lambda 8000$. 
 
The Mosaic-II light curve (Fig. 4) shows significant variability with
an RMS of 0.07 mag and $\Delta r' = 0.32 $ mag.  A Lomb-Scargle
periodogram yields a peak at a period of 5.6 hr. However, we find a
False Alarm Probability of 7.8\% for this period by running Monte
Carlo simulations on randomly ordered light curves.

The non-detection of absorption features from the donor star and  the
photometric flickering suggest a high accretion rate CV or a CV
  accreting at lower mass transfer rate, but having an accretion -dominated optical
  spectrum.  From the X-ray brightness (24 counts,  $HR= 0.63 \pm
0.29$)  we obtain $L{_x} (d,N{_H}=10^{21}) = 1.9 \times 10^{31} \times
(\frac{d}{1.0~kpc})^2$ erg s$^{-1}$.  

\subsection*{CX73 = CXOGBS J174447.5-270101}

The only emission line detected in the optical spectrum of CX73
(Fig. 1) is H$\alpha$ with an intrinsic FWHM of $730 \pm 20$ km
s$^{-1}$.  As in CX70 it is difficult to confirm or reject  evidence for
absorption lines in the blue part of the spectrum. The red part of the data is strongly affected by fringing
since the observations were taken before the VIMOS detector upgrade (see Sec. 2.1)..

The Mosaic-II light curve (Fig. 4) shows aperiodic variability with an RMS
scatter of 0.08 mag and $\Delta r' = 0.19 $ mag.  This variability supports 
a significant contribution from the accretion flow to the optical light. Spectroscopic 
observations (for instance in the I-band) are necessary  to further constrain the nature 
of CX73.  An X-ray luminosity of $L{_x} (d,N{_H}=10^{21}) = 1.8 \times 10^{31} \times (\frac{d}{1.0~kpc})^2$ erg s$^{-1}$
is derived from the 23 counts ($HR= 0.22 \pm 0.25$) detected with {\it Chandra}.

\subsection*{CX87 = CXOGBS J173648.1-293639, an outbursting CV}

The VIMOS spectrum of the optical counterpart to CX87 can be found in
Fig. 2.  Due to its low signal-to-noise ratio we only detect 
H$\alpha$ in emission with FWHM = $1150 \pm 140$ km s$^{-1}$ and
EW = $24 \pm 5$ \AA. The line profile appears  double-peaked with a peak-to-peak
velocity separation of $\Delta v = 650 \pm 80$ km s$^{-1}$.   

The Mosaic-II light curve (Fig. 4) shows the source declining in
brightness by 2.9 mag,  reaching minimum brightness ($r' \sim  23$)
four days after  the first photometric observation.  Examination of
the VIMOS pre-image taken on 30 Mar 2011 UT and the spatial profile
of the optical counterpart  in the VIMOS spectrum   show that CX87 was
most likely in quiescence  during the VLT observations.

It is likely that the Mosaic-II photometry was obtained during the
final stage of a CV outburst that brightened by more than  2.9 mag at
optical wavelengths.  This  would support a dwarf nova nature
for CX87. Alternatively, CX87 could be an outbursting IP. Some IPs
have outbursts with  a 3-5 mag amplitude and a few days duration.
(see e.g. Ishioka et al. 2002).  The  X-ray  flux observed with {\it
    Chandra} (21 counts, $HR= 0.31 \pm 0.25$) implies  $L{_x}
  (d,N{_H}=10^{22}) = 3.0 \times 10^{32} \times (\frac{d}{3.0~kpc})^2$
  erg s$^{-1}$.

\subsection*{CX93 = CX153 = CXOGBS J174444.6-260328, a low accretion rate CV}

CX93 is a 0.237 d orbital period CV observed at a low accretion rate state. A complete
dynamical study of this source is presented in Ratti et al. (2013a). 
The VIMOS data for CX93 are analyzed in Ratti's work and therefore we will not discuss them further except for the following
erratum: the EWs measured for  the H$\alpha$ emission line in the VIMOS spectra  are  incorrect. Correct values can be found in 
the present work (Table 2). As a consequence the claim in Ratti et al. of a more active state of accretion at the time of the VIMOS observations 
must be disregarded. A finding chart for the field of CX93 can also be found in Ratti's work.

\subsection*{CX128 = CXOGBS J174028.3-271136, a likely short orbital period dwarf nova}

The emission line spectrum of CX128 (Fig. 2) shows very strong
H$\alpha$ and H$\beta$ emission lines with single-peaked profiles with
an intrinsic FWHM of $\sim 1400$ km s$^{-1}$.  He {\sc i}
$\lambda\lambda 5876, 6678, 7065$ are also in emission with similar broadening, though they appear
double-peaked. He {\sc i} $\lambda 5876$ is the stronger of these
three lines. At longer wavelengths CX128 shows emission lines of the
Paschen series. The lines are broad ($1500 - 2100 $ km s$^{-1}$ FWHM) with complex (multi-peaked)
profiles.  Neither photospheric lines nor interstellar bands are
detected indicating that light from the accreted material dominates and that the source 
suffers moderate reddening. The Mosaic-II
photometry (Fig. 5) shows flickering of 0.26 mag RMS and $\Delta r'$
= 0.89.

The lack in the spectrum of underlying features from the donor can be
explained if this GBS source is a high accretion rate (nova-like
  or magnetic) CV.  However, the EWs for the Balmer and Helium lines
(Table 2) seem high compare to those observed in nova-like CVs
(Rodriguez-Gil et al. 2007a,b; Dhillon et al. 1996).  Furthermore, we
do not detect emission from He{\sc ii} $\lambda 5412$, a signature
that would have been sufficient to confirm the above scenario. An
alternative possibility is that CX128 is a dwarf nova CV below the
period gap in quiescence at the time of our observations.  In this
scenario, the contribution from the small donor star to the optical
light is  veiled by the accretion disc and the spectrum shows only
emission lines. Examples of such spectra can be found in the dwarf
novae SS UMa, SU UMi and SW UMa (Connon Smith et al. 1997).  In
particular SW UMa resembles CX128 in the strength of its Balmer
lines. {\it Chandra} detected 15 counts during the
  discovery observation which yield $L{_x} (d,N{_H}=10^{21}) = 1.2
  \times 10^{31} \times (\frac{d}{1.0~kpc})^2$ erg s$^{-1}$.  If the source is at a distance of $\sim 1~kpc$,   
the X-ray luminosity is consistent with the luminosities found in quiescent dwarf novae (Bycklbycklying et al. 2010).

\subsection*{CX137 = CXOGBS J175553.2-281633, a low accretion rate CV or qLMXB?}

Based on the positional coincidence with the X-ray source, Udalski et
al. (2012) identified in their OGLE-IV data the optical counterpart to
CX137: a 0.43104 d periodic variable with I=15.11, V-I=1.32.  They
suggest that the light curve morphology is consistent with an
eclipsing binary harbouring a spotted star.  

The VIMOS spectrum of the counterpart is dominated by photospheric
absorption  features (see Fig. 2). Clearly detected is the
discontinuity in the continuum at $\lambda 5200$ caused by the Mg $b$
triplet and molecular MgH. Also visible are the metallic blends at
$\lambda\lambda 6165, 6495$. In the red, the Na {\sc i} doublet and Ca
{\sc ii} triplet are found  in absorption. Finally, TiO bands are
not detected in the spectral range covered by the spectrum. These
spectroscopic characteristics are consistent with those expected from
a late G-type/early K-type star (see e.g.  Jacoby et
al. 1984). However, the EW of the Ca {\sc ii} $\lambda 8542$ line is
$\lesssim 2$ \AA, lower than the $\gtrsim 3$ \AA~ observed in stars
with the above spectral types (see for instance Fig 23 in Carquillat
et al. 1997). Note that the line measurements are not affected of
contamination from a partially resolved field star located SE from
CX137.  The low EWs may indicate an additional source of continuum
light veiling the photospheric lines, for instance the accretion
disc. On the other hand, the discrepancy in the EWs is not likely to
be due to Paschen emission as there is no evidence of other Hydrogen
emission lines in the red part of the spectrum.  The presence of broad
H$\alpha$ in emission also supports an accreting binary nature for
CX137. From single  Gaussian fitting, we conclude that the
H$\alpha$ profile has an intrinsic FWHM and EW larger than $960$ km
s$^{-1}$ and $6.5$ \AA, respectively.  While the line position
is consistent with its rest wavelength, we measure a $-90 \pm 20 $ km
s$^{-1}$ radial velocity for the NaD and Ca {\sc ii} photospheric lines 
found in the red part of the spectrum.

The optical counterpart to CX137 is the brightest in the sample of GBS
sources discussed in this paper and it is saturated in most of the
Mosaic-II data.  As a result, no meaningful variability information is
available from the Mosaic-II images.  We therefore investigated the
publicly available OGLE-IV data sets which at the time of writing
consist on a total of 4358 data points in the I-band acquired during
the years 2010-2012. Visual inspection of the light curves suggested
small changes in brightness and/or morphology during the three years and therefore we
analyzed each year separately. We measure a mean
I-band magnitude of 15.11, 15.08 and 15.10 (with an RMS of 0.06 mag)
and $\Delta r'$ of 0.20, 0.24 and and 0.23 mag
for 2010, 2011 and 2012, respectively.  Periodograms computed for each
year have the highest peak centered at 0.21551725(15) d, where the
numbers in parentheses quotes the uncertainty in the last digits.
This is half of the orbital period since the light curves phase folded
on the 0.4 d period show a much less scattered sinusoidal-like
modulation than when folded on the 0.2 d periodicity.  Thus we confirm
the orbital period found by Udalski et al. (2012).  The folded light
curves (see Fig. 5) show that during 2010 the two maxima and two
minima have similar brightness. In contrast, during 2011 and 2012 both
minima and maxima occur at different brightness levels.  Another
characteristic of the light curve is the lack of strong flickering which supports a
stellar component as the main contributor to the optical light.

Udalski et al. (2012) interpreted the sinusoidal light curve as due to
eclipses in a contact binary where the yearly changes are due to star
spots. On the other hand, the light curve morphology can be explained as due to
ellipsoidal variability of the companion star in an accreting system
observed at low or moderate inclination.  In this scenario the
long-term variability observed in the ellipsoidal light curve can also
be associated to star spots. Other unknown mechanisms may also be at
play causing the changes in the light curve morphology (see
e.g. Thomas et al. 2012, Ratti et al. 2013a).

In an interactive binary the mean density of the Roche lobe-filling donor star can be
  determined from the orbital period by following Eggleton (1093):
  $\bar{\rho} (g~cm{^{-3}}) \cong 110  P{^{-2}} (hr) =  1.0$.  This
  indicates that the donor star  in CX137 is enlarged compare to  G-K
  main sequence stars since the latter have higher mean densities (the
  1.4 g cm$^{-3}$ of our Sun, for instance).  We adopt a K5 V
  companion (M$_V = 7.35$, M$_r' = 6.98$,M$_I = 5.73$), null reddening
  and disc contribution to the optical light to derive a crude  lower
  limit to the distance of  0.7 kpc. Additionally, we use the 15 counts
  detected with {\it Chandra} to derive an lower  limit on the
  luminosity of $L{_x} (d,N{_H} = 10^{21}) > 5.8  \times 10^{30}
  \times (\frac{d}{0.7~kpc})^2$ erg s$^{-1}$.     

The detection of photospheric lines and a broad H$\alpha$
  emission line in the optical spectrum suggest that CX137 is a  CV
  accreting at low accretion rate or a qLMXB.  High resolution
  spectroscopy will show if  CX377 is a single (accreting) or double-lined
(contact)  binary.

\subsection*{CX142 = CXOGBS J174403.7-312305, a CV with an accretion-dominated optical spectrum}

The VIMOS spectrum of CX142 (Fig. 2) shows a source rich in narrow emission
lines and lacking photospheric absorption features. DIBs are not observed. At blue
wavelengths H$\alpha$ (with intrinsic broadening of $340 \pm 20$
km s$^{-1}$) and He {\sc i} $\lambda\lambda 5876, 6678, 7065$ lines are
visible in emission. In the red, the Ca {\sc ii} triplet at
$\lambda\lambda 8542, 8662, 8498$ are prominent emission lines. They are
stronger than the nearest Hydrogen lines of the Paschen series - EW(Ca
{\sc ii}) / EW(P14) $\sim 2$.

The Mosaic-II light curve (Fig. 5) shows an RMS scatter of 0.18 mag
which is only twice the average statistical error in the photometry. Even
though $\Delta r' = 0.6 $ mag, the source is at the faint edge of what
the Blanco observations can detect making it difficult to set strong
constraints on the optical variability.

The optical counterpart to CX142 shows spectroscopic characteristics
similar to those of a CV accreting at high rate or low accretion
  rate CV with a major contribution from the disc to the total
  optical light.  The low FWHM values found for all the emission
lines (Table 2) supports a system observed at low inclination if
  the emission-line region is the accretion disc. Given the good
quality of the spectrum we speculate that CX142 was observed in
outburst, well above  the $r'=22.2$ brightness measured in the
Mosaic-II photometry. We lack other  objects in the slit to confirm
this possibility and thereby confirm a dwarf nova CV nature. We infer
$L{_x} (d,N{_H} = 10^{21}) = 1.1  \times 10^{31} \times
(\frac{d}{1.0~kpc})^2$ erg s$^{-1}$ from the 14 counts  detected with
{\it Chandra}.  Finally, note that strong emission lines from the Ca
{\sc ii} infrared triplet have been already reported in CVs. For
instance see the spectrum of the dwarf nova  system UU Aql near
quiescence (Connon Smith et al. 1997) or GW Lib in outburst (van
Spaandonk et al. 2010).

\subsection*{CX154 = CXOGBS J173838.7-283539, a low accretion rate CV or qLMXB}
 
Two epochs of spectroscopy separated by about two weeks were acquired,
with the second epoch being of higher quality. This is shown in
Fig. 2. H$\alpha$ in emission is observed in both nights, while He
{\sc i} $\lambda 6678$ is only detected in the second  observation.
The H$\alpha$ emission line is weak (EW$\sim 8$ \AA) with an
unresolved profile.  In the red, the Ca {\sc ii} $\lambda 8542$ 
  appears to be present in absorption with EW$\sim 1.0$ \AA~and a 
radial velocity of $\sim 120$ km s$^{-1}$ in both nights. This
line is most certainly a photospheric feature intrinsic to CX154 and
not due to contamination from a nearby field star.  Finally, from the
Mosaic-II data we derive $\Delta r' = 0.5$ mag and an RMS scatter of
0.1 mag.

Based on  the likely detection of one of the components of the ionized
Calcium triplet, CX154 can be tentatively classified as a low accretion rate CV or qLMXB.  
The low FWHM shown by  H$\alpha$ could be interpreted as originating in a low
inclination system if the emission arises from an accretion disc.  An alternative
scenario is that where the line is produced in the donor star by
reprocessing of X-rays from a neutron star or white dwarf primary
(e.g. Bassa et al. 2009, Ratti et al. 2012, Rodr{\'{\i}}guez-Gil et
al. 2012). From the 15 counts observed with {\it Chandra} we 
estimate $L{_x} (d,N{_H} = 10^{21}) = 1.1  \times 10^{31}
  \times (\frac{d}{1.0~kpc})^2$ erg s$^{-1}$.

\subsection*{CX207 = CXOGBS J174625.3-263133, a likely Polar}

Spectra of the optical counterpart to CX207 were obtained in two
nights separated one month.  In both nights, the spectrum is dominated
by strong H$\alpha$ in emission (EV $>80$ \AA).  This line is 
remarkable for being single-peaked despite its broadening (FWHM = 1500
- 1600 km s$^{-1}$) and because it presents a large radial velocity
offset ($|RV| > 350$ km s$^{-1}$) with respect to its rest wavelength.
Weaker He {\sc i} and  Paschen emission is also detected in the
spectrum.  H$\alpha$ and some of the other features are consistent with having
two-component line profiles.  The two epochs of spectroscopy show
clear variability in the line profiles, in particular their central
wavelengths  (see Table 2).

The optical light curve for CX207 (Fig. 5) shows an eclipse of a depth
more than 3 mag at HJD 2455389.590991 and a possible second eclipse in
progress at HJD 2455386.633003 when the source is observed 1 mag
fainter than its mean brightness.  Outside these two events, the
Mosaic-II data has $\Delta r' = 0.7$ and aperiodic variability with an
RMS of 0.2 mag.

The optical spectrum of CX207 resembles that of the neutron star X-ray
binary Circinus X-1 (Mignani 1997, Johnson et al.  2001, Jonker et
al. 2007) or the black-hole transient V4641 Sagittari (Lindstr{$\o$}m
et al.  2005).  In this case the large broadening, large radial
velocity offset and variability may be due in part to an ouflow in an
active X-ray binary harbouring a compact primary and a early-type
companion.  This model for CX207 requires the X-ray binary to be at a
large distance and suffer significant reddening in order to  explain
the faintness of the X-ray and optical counterparts (11 {\it Chandra}
counts equivalent to $L{_x} (d,N{_H} = 10^{21}) = 8.7  \times
  10^{30} \times (\frac{d}{1.0~kpc})^2$ erg s$^{-1}$,
  $r'=21.1$). However, the optical spectrum  lacks evidence for strong
  DIBs  ruling  out this scenario and favours  the possibility that
  CX207 is a nearby CV.   In this regard, broad single peaked emission
  lines are typically observed in nova-like variables.  In this class
  of  high-accretion rate CVs  even eclipsing systems do not show a
  double-peaked profile.  The optical spectra in nova-like CVs can
  show emission lines that reach profile broadenings of values similar
  to that observed in CX207 (see e.g.  Rodr{\'{\i}}guez-Gil et
  al. 2007a,b; Dhillon et al. 1992). In this scenario the large radial
  velocity offset observed in CX207 will have to reflect a combination
  of a large systemic velocity and $K_1$.  Alternatively, CX207 may be
  a member of the magnetic (Polar) class of CVs. These systems can
  exhibit  emission lines  with similar line morphology and large EWs
  as observed in the GBS source. In Polars the line profile is
  composed of (at least) a narrow and a broad component, the former
  originating in the irradiated donor star.  These components can
  exhibit radial velocities comparable or larger than those measured
  here -   see for instance the eclipsing Polars  EXO 033319 - 2554.2
  (Allen et al. 1998), MN Hya (Ramsay \& Wheatley 1998) and the
  non-eclipsing system  ET Can  (Williams et al. 2013),  Given that
  the observed systemic radial velocities in nova-like CVs do not go
  above $\sim 200$ km s$^{-1}$ (Ak et al. 2010), the magnetic CV
  scenario seems more likely as it does not require of such unusual
  high $\gamma$ to explain the radial velocities of the emission lines
  in CX207.

\subsection*{CX377 = CXOGBS J174316.5-274537, a likely low accretion rate CV or qLMXB}

Two epochs of spectroscopy separated by nearly two months were
obtained with VIMOS.  In Fig 2.~we show the spectrum obtained on
23 Jul 2011 UT with the slit oriented such to avoid spatial blending with
the nearby field stars.  H$\alpha$ is unambiguously detected in
emission. Unfortunately, instrumental artifacts 
are also apparent in the blue part of the spectra.
This makes it difficult to address the presence/absence of stellar
features and interstellar bands at wavelengths shorter than H$\alpha$.
To achieve the detection of photospheric lines we extracted and
studied the spectra of field stars in the CX377 slit. In this way we
confirm the non detection of the metallic blends at $\lambda\lambda
6165, 6495$ which are strong features in the spectra of late-type
stars.  We cannot establish the presence of He {\sc i} emission at
$\lambda 6678$.  In the red, the spectrum reveals the absorption lines
from the Ca {\sc ii} infrared triplet.  On the other hand, neither the
Na {\sc i} double nor the P14 $\lambda 8598$ line or TiO bands appear
to be present at the time of the observations.  These spectroscopic
characteristics suggest a late F-type/early G-type donor star in CX377
(see e.g. Jacoby et al. 1984, Zhou 1991, Munari \& Tomasella 1999).
We measure for the Ca {\sc ii} $\lambda 8542$ line an EW of $\sim 1.5$
\AA, lower than the values observed in F/G of any luminosity class
(Jones et al. 1984, Zhou 1991). As in CX137 this discrepancy can be
due to veiling of the lines by the accretion disc continuum
emission. Disc veiling could also explain the non detection of the
metallic lines described above. 

During the two epochs of spectroscopy the profile of the H$\alpha$
emission line is intrinsically broadened by $\sim 1200$ km s$^{-1}$
(FWHM) and asymmetrically double-peaked with the blue peak being
stronger than the red peak.  The double-peak separation is $\Delta v =
610 \pm 20$ km s$^{-1}$ and $\Delta v = 720 \pm 30$ km s$^{-1}$ during
the first and second epoch, respectively.  Given the asymmetry in the
line profiles, a better estimation of the line center than that
obtained from a Gaussian fit (Table 2) is provided by the centroid of
the line derived from the emission line peaks.  We measure in this way
radial velocities of $70 \pm 20$ km s$^{-1}$ and $60 \pm 30$ km
s$^{-1}$ for the first and second epoch, respectively. Both values are
consistent within the errors. On the other hand, we derive  from
  the Ca {\sc ii} triplet radial velocities of  $80 \pm 6$ km s$^{-1}$
  and $90 \pm 10$ km s$^{-1}$ for  epoch 1 and 2, respectively. 
 These velocities are consistent with those  measured from H$\alpha$. 
Note here that the emission line is weak (EW$\sim 7$ \AA)  and thereby its double-peak morphology 
may be caused by the underlying photospheric H$\alpha$.

The VIMOS spectra also show the
$\lambda 8620$ DIB with an EW = $0.7 \pm 0.1$ \AA.  We derive $E(B-V)
= 1.9 \pm 0.2$ according to the relation between reddening and EW for
this interstellar band (Munari 2000). This value is equal to the $1.9
\pm 0.3 $ mag derived from the reddening maps of the Galactic bulge
(Gonzalez et al. 2012; these maps have a spatial resolution of $2
\times 2$ arcmin$^2$ resolution).

The Mosaic-II light curve of CX377 (Fig 5) shows photometric
variability with an RMS scatter of 0.05 mag and $\Delta r'$ = 0.22
mag. There is no evidence for a periodic modulation.

The detection of the donor star in the VIMOS spectrum, the H$\alpha$
profile broadening and the high reddening towards the source suggest
that CX377 is a (bulge) low accretion rate CV observed at a
  moderate inclination or a bulge qLMXB.  Using the
  relationship of Bohlin et al. (1978) $N{_H} = 5.8 \times
  10^{21} \times E(B-V)$  cm$^{-2}$ and adopting as before an
  absorbed  power-law spectrum  with index $\Gamma = 2$, we derive
  from the detect 7 {\it Chandra} counts  an unabsorbed flux of $1.0
    \times 10^{-13}$ erg cm${^{-2}}$ s$^{-1}$ (0.5 - 10 keV) and 
a  luminosity of $L{_x}  (d,N{_H} = 1.1 \times 10^{22}) = 7.6  \times 10^{32} \times
  (\frac{d}{8.0~kpc})^2$ erg s$^{-1}$.

\subsection*{CX446 = CXOGBS J174627.1-254952, a high inclination CV or qLMXB}

The spectrum of the optical counterpart to CX446 can be found in
Fig. 3.  Due to its low signal-to-noise ratio we only detect strong
H$\alpha$ in emission with EW = $50 \pm 4$ \AA. The line profile
appears single-peaked despite its large intrinsic broadening (FWHM =
$2200 \pm 50 $ km s$^{-1}$).  This appearance may be caused both by
noise and/or the presence of a narrow emission line component
filling-in the region between the two peaks.  The large line
broadening is suggestive of a high inclination system.  In fact, the
Mosaic-II light curve (Fig. 5) shows an eclipse-like event at HJD
2455387.824233 with a 0.4 mag depth with respect to a mean 21.16 $r^\prime$
band magnitude.  Our photometric sampling does not allow
us to set constraints on the duration of this event.  Other decreases
in brightness observed in the light curve are consistent either with
photometric variability and/or another eclipses. We performed a
periodicity analysis of this light curve but this yielded no
significant period.

In summary, the broadening of the line and the eclipse-like event
observed in the light curve suggest a high inclination (eclipsing) CV or qLMXB
nature for CX446. From the 6 counts detected by {\it Chandra} we 
estimate $L{_x} (d,N{_H} = 10^{21}) = 4.7  \times 10^{30}  \times (\frac{d}{1.0~kpc})^2$ erg s$^{-1}$.

\subsection*{CX522 = CXOGBS J175432.5-282918, a CV with an accretion-dominated optical spectrum}

The optical counterpart to this GBS source was misidentified in
Udalski et al. (2012) with a close bright field star. The coordinates
for the correct optical counterpart to CX522 can be found in Table 2.
The VIMOS spectrum (Fig. 3) shows , H$\beta$ and He {\sc i}
$\lambda\lambda 5876, 6678, 7065$  ith some evidence for 
Paschen emission in the red part of the spectrum. H$\alpha$
is moderately broad with $850 \pm 10 $ km s$^{-1}$ FWHM.   
Neither photospheric lines from the donor star nor strong DIBs are 
detected.

The optical counterpart to CX522 is too close to the bright field star
mentioned above to allow for a reliable photometric study on the
Mosaic-II data.  From the VIMOS pre-images obtained on 8 Apr
  2011 UT  we derive $R \sim 22$ by performing differential
  photometry respect to the USNO B1.0 object 0615-0617155 (R1 =
  18.68). On the other hand, examination of the spatial profile of both the optical
  counterpart spectrum and the spectrum of a field star partially
  contained in the slit suggests that CX522 may have been brighter at
  the time of the spectroscopic observations .

On the basis of the  spectroscopic  characteristics described above,
CX522 is a high accretion rate CV or a low accretion rate CVV
  with an accretion-dominated optical continuum.   The 5 counts
detected  with {\it Chandra} imply $L{_x} (d,N{_H} = 10^{21}) = 4
\times 10^{30} \times (\frac{d}{1.0~kpc})^2$ erg s$^{-1}$.

\subsection*{CX772 = CXOGBS J174405.0-263159}

Only broad H$\alpha$ in emission with an intrinsic FWHM of $900 \pm
180$ km s$^{-1}$ is apparent in the VIMOS spectrum (Fig. 3).  On the
other hand, it is not possible to address the presence of line
features in the red part of the spectrum  given that the spectrum is affected by
fringing - the observations were taken before the upgrade of the VIMOS
detectors (see Sec. 2.1).

We measure $\Delta r' = 1.5$ and an RMS scatter of 0.28 mag for
the light curve of CX772 (see Fig. 6). However, the brightness of the optical counterpart 
($r' = 22.2$) is at the faint end of the range of detected objects in the Mosaic-II
data. Thus the photometry has on average a large uncertainty of 0.18 mag. 

With the  current  data we cannot establish the nature of CX207 further than 
it is a CV or a qLMXB. The 4 counts detected with {\it Chandra} yield
$L{_x}  (d,N{_H} = 10^{21}) = 3.2  \times 10^{31} \times  (\frac{d}{1.0~kpc})^2$ erg s$^{-1}$.

\subsection*{CX781 = CXOGBS J174311.1-271621, a CV with an accretion-dominated optical spectrum}

The VIMOS spectrum of CX781 (Fig. 3) exhibits H$\alpha$ and He {\sc i}
$\lambda 6678$ emission with unresolved line profiles.  The H$\alpha$ line
profile is weak with EW = $5.0 \pm 0.2$ \AA. Instrumental artifacts 
are evident in the blue part of the spectrum making difficult to identify
stellar features and interstellar bands at wavelengths shorter than H$\alpha$.
In the red  part of the spectrum (right panel) absorption lines from the donor star are not
detected, in particular, there is no evidence of  the Ca{\sc ii} triplet in absorption. Only H{\sc i} $\lambda9229$ (P11) seems to be present, but  in emission.   

The Mosaic-II light curve for the optical counterpart to CX781
(Fig. 6) exhibits clear variability:  the source brightness decreases
by $0.25$ mag four days after the start of the observations  to return
slowly to full brightness of $r'=17.9$ mag approximately four days
later.  Long-term erratic variability usually observed in accreting
binaries can explain this behaviour.  There is also variability at
shorter time scales (flickering)  of  $0.05$ mag amplitude.  We derive
$\Delta r' = 0.28$ mag and RMS of 0.09 mag from the whole light curve.

The lack of photospheric features in the VIMOS  spectrum, the
detection of flickering together with possible erratic long-term photometric
variability support both a high accretion rate CV nature or a low accretion rate 
CV where the accretion flow is the dominant source of optical continuum - a short orbital period 
IP or dwarf nova in quiescence for instance.  The narrow and weak H$\alpha$ emission
line is arising either in the disc of a low inclination binary or
in the irradiated hemisphere of the donor star. We estimate 
$L{_x}  (d,N{_H} = 10^{22}) = 3.2  \times 10^{30} \times  (\frac{d}{1.0~kpc})^2$ erg s$^{-1}$.from
the 4  counts detected with {\it Chandra}.

\subsection*{CX794 = CXOGBS J174142.3-275829, an eclipsing nova-like} 

Udalski et al. (2012) reported an I=19.4 optical counterpart to CX794
and found in the OGLE light curve a possible 0.11786 d period together
with evidence for eclipses.  Dimming events of up to 2 mag are
apparent in the Mosaic-II photometry (Fig. 6) as well as in the public
OGLE light curve.  A Lomb-Scargle periodogram analysis of our data
shows a $0.1179$ d period with another peak at $0.1052$ d, adding
weight to the period reported by Udalski et al. (2012).  The long
($\sim 2$ years) interval spanned by the OGLE observations allows us
to assess the eclipsing nature of CX794 and establish its orbital
period by using an O-C curve analysis. For this we have taken the
minimum brightness in the OGLE light curve to be the event closest to
or at the time of a mid-eclipse. This yields the following ephemeris:
T$_{mid-eclipse}$ = HJD $ 2455768.57874 + 0.11786 \times E$.  Next the
times for eleven events with a $>0.6$ mag drop in brightness were
identified in both the OGLE and Mosaic-II data.  The O-C curve (not
shown) displays $< 0.1$ cycles offsets between these times and the
eclipse times derived from the ephemeris. The same analysis shows that
the periodicities found in the Mosaic-II data gave larger
residuals. Fig. 6 shows the Mosaic-II light curve folded on the above
ephemeris.

Our optical spectrum (Fig. 3) shows H$\alpha$, H$\beta$ and He {\sc i}
$\lambda\lambda 5876, 6678, 7065$ lines in in emission. The FWHM of H$\alpha$ (1050 km
s$^{-1}$) is on the low side for a high-inclination system.
Furthermore, the line profiles of both lines are single-peaked.  Lines from the
Paschen series are also detected in emission with P9 being the strongest. On the other hand, there is no evidence for
absorption features from the donor star or the interstellar medium.

These above properties suggest that CX794 is a high accretion rate CV within the 2-3 hr
period gap and with spectroscopic similarities to the SW Sextantis nova-like variables.
We derive $L{_x}  (d,N{_H} = 1.1 \times 10^{21}) = 3.2  \times 10^{30} \times  (\frac{d}{1.0~kpc})^2$ erg s$^{-1}$
from the 4 {\it Chandra} counts. At present we lack an infrared counterpart to obtain a reliable constraint on the distance 
towards the source by using the method by Knigge (2006).

\subsection*{CX1004 = CXOGBS J174623.5-310550, a low accretion rate eclipsing CV or qLMXB}
  
The most remarkable feature in the spectrum of CX1004 ( Fig. 3) is a
broad double-peaked H$\alpha$ emission line with FWHM = $2120 \pm 20$
km s$^{-1}$ and a peak-to-peak velocity separation of $\Delta v = 1170
\pm 10$ km s$^{-1}$.  This is the single emission line detected in the
spectrum. The centroid of the H$\alpha$ line profile measured using
the red and blue peak positions yields a radial velocity of $-170 \pm
20$ km s$^{-1}$, slightly lower than the value obtained by fitting a
single Gaussian to the line profile (Table 2).  Absorption bands are
prominent in the spectral range $5850-7350$ \AA~ and are caused by the
TiO $\alpha,\beta,\gamma$ and $\gamma^{'}$ band systems (see
e.g. Coelbo et al. 2005 for their identification in a broad-band
spectrum).  Redward of 8000 \AA~the $\delta$ and $\epsilon$ systems
are not detected.  In this region the Na {\sc i} doublet at
$\lambda\lambda 8183, 8195$ is in absorption together with the Ca {\sc
  ii} and Hydrogen atomic lines. No interstellar features are found,
except perhaps for   the NaD doublet at $\lambda\lambda 5890, 5896$
unresolved from their stellar counterparts.  We measure a radial
  velocity of $-170 \pm 20$ km s$^{-1}$ from fits to the metallic
  lines in the red part of the spectrum. This value is in agreement
  with  the radial velocity obtained from the H$\alpha$ emission
  line. 

On the other hand,  the Mosaic-II light curve (Fig. 6)  does not
exhibit  significant photometric variability for this source in 35
observations.  The RMS scatter and $\Delta r'$  in the photometry is
0.04 mag and 0.17 mag, respectively. 

The large FWHM and peak-to-peak velocity separation in the H$\alpha$
profile in CX1004 is similar to that observed in eclipsing dwarf novae CVs in
quiescence or qLMXBs harboring a black-hole. Examples are
% Z Cha (P$_{orb}$=1.36 hr, K$_2$ =  375 km s$^{-1}$;   $\Delta v$ = 1180   km s$^{-1}$; Marsh, Horne \& Shipman 1987),
OY Car (P$_{orb}$=1.51 hr, $\Delta v$ = 1160 km s$^{-1}$; Schoembs \&
Hartmann 1983), U Gem (P$_{orb}$=4.22 hr, $\Delta v$ = 920 km
s$^{-1}$; Echevarria et al. 2007, Naylor et al. 2005), XTE J1118+480
(P$_{orb}$=4.08 hr, $\Delta v$ = 1800 km s$^{-1}$; Torres et al. 2004)
and A0620-00 (P$_{orb}$=7.68 hr, $\Delta v$ = 1270 km s$^{-1}$; Marsh
et al. 1994).  

As shown above, the detection of photospheric lines and the large
  broadening of the H$\alpha$ emission line are properties that
  suggest that CX1004 is an eclipsing  CV  accreting at low accretion
  rate or a  high inclination qLMXB with a black hole primary.  However, the lack of
  significant photometric variability is difficult to explain as we
  would expect to see eclipses, an ellipsoidal modulation and/or
  flickering.Finally, the lack of DIBs suggests a nearby source for
which we derive  from the  3 {\it Chandra} counts an X-ray luminosity
of  $L{_x} (d,N{_H}=10^{21}) = 2.4 \times 10^{31} \times
(\frac{d}{1.0~kpc})^2$ erg s$^{-1}$. As a reference an M2 V
($M_r'=9.14$) unreddened star with the same apparent magnitude  as
CX1004 will be at a 2.1 kpc distance.
 
\subsection*{CX1011 = CXOGBS J174604.7-311722, a nova-like variable}

A strong feature in the VIMOS and IMACS spectrum (Fig. 3 and 7) is 
H$\alpha$ with a narrow and weak emission line profile contained within a broad absorption
component.  Redward of H$\alpha$,  He{\sc i} $\lambda 6678$ is in
emission as well as several Paschen lines with P9 being the
strongest of them. In addition to the saturated interstellar NaD doublet,
DIBs at $\lambda 5780$ and $\lambda 6284$
are prominent with EWs of $0.40 \pm 0.03$ \AA~ and $1.28 \pm 0.05$ \AA,
respectively. Photospheric lines from the donor star are not
detected, in particular, there is no evidence of  the Ca{\sc ii}
triplet in absorption or emission.  

The spectral resolution and blue coverage provided by the IMACS
observation (Fig. 7) allows us to measure an intrinsic FWHM of $578
\pm 2$ km s$^{-1}$ for H$\alpha$ and we find that the other Balmer
lines are composed by a broad absorption component filled-in
with core emission.  We have measured the full-width zero intensity
(FWZI) of the absorption components to be $7522 \pm 555$, $4964 \pm
898$ and $4593 \pm 950$ km s$^{-1}$ for H$\beta$, H$\gamma$ and
H$\delta$, respectively. The large uncertainties are due to the
difficulty of defining the continuum. Broad He{\sc ii} $\lambda 4686$
and the C{\sc iii}/N{\sc iii} Bowen blend near $\lambda 4648$ are
obvious in the IMACS spectrum. We also detect He{\sc i} $\lambda 4471$ with emission
and absorption components as observed for the Balmer lines.  

This source is near the saturation limit of the Mosaic-II data.  There
are 28 observations which do not saturate. As shown in Fig. 6, the photometry
is characterized by  low-amplitude flickering with an RMS of 0.01
mag and $\Delta r' = 0.04$. This variability is not significantly correlated with the seeing,
which suggests it is real and not a product of the non-linear CCD
response at high count rates.

We derive an E(B-V) = 0.8 mag from the calibration between reddening
and the EW for the $\lambda 5780$ DIB (Herbig 1993). This value is
smaller than the $1.6 \pm 0.2$ mag obtained from the reddening maps of
the Galactic bulge (Gonzalez et al. 2012).   Using $N{_H} = 5.8 \times
10^{21} \times E(B-V)$  cm$^{-2}$ (Bohlin et al. 1978)  and an
absorbed power-law spectrum with photon index $\Gamma = 2$, we
calculate from the detected 3 {\it Chandra} counts an unabsorbed flux
of $5.7 \times 10^{-14}$ erg cm${^{-2}}$ s$^{-1}$ (0.5 - 10 keV) and
X-ray luminosity of $L{_x} (d,N{_H} = 4.6 \times 10^{21}) = 6.1
\times 10^{31} \times (\frac{d}{3.0~kpc})^2$ erg s$^{-1}$. 

The presence of high excitation emission lines and broad absorption lines
of H{\sc i} and He{\sc i} partially filled-in by emission cores is
frequently  observed in nova-like variables (non magnetic CVs in a
state of high accretion rate).  The  narrow emission cores observed in
CX1011 suggests a low-inclination system.   However, we cannot exclude
an origin for the emission in the irradiated donor star or hot spot
instead of the disc as observed in the nova-like IX Vel  (Beuermann \&
Thomas 1990).  The optical spectrum of CX1011 resembles that of IX
  Vel  except that in our spectrum we do not detect He{\sc i} $\lambda
  4922$ in emission and the Bowen blend appears stronger than the
  He{\sc ii} $\lambda 4686$ line.

\section{Conclusions and Remarks}

In this paper and in Britt et al. (2013) we have utilized optical
spectroscopy to identify the emission line optical counterparts to twenty eight GBS
sources and to classify them as accreting binaries.  To derive 
sample of accreting binaries  from the optical counterparts to
GBS sources as clean as possible of unrelated objects,we have adopted three discriminators, viz. H$\alpha$ in
emission with i) EW $> 18$ \AA, ii) FWHM $> 400$ km s$^{-1}$ and/or
the presence of He{\sc i} $\lambda\lambda 5876,6678$ emission with EW
$> 3$ \AA~and  EW(He {\sc i})/EW(H$\alpha$) $>0.2$.  Such a selection
towards strong H$\alpha$ emitters permits to derive a sample wealky or not
contaminated by chromospherically active stars/binaries which are
expected to represent $\sim 40\%$ of the X-ray sources found in the
GBS area (Jonker et al. 2011).  In fact we have identified with VIMOS
a large number of optical counterparts to GBS sources that show
spectra characteristic of this type of sources. A disadvantage of
using this approach is that accreting binaries with an emission line
spectrum not fulfilling any of the above criteria will be
missed. Furthermore, a selection based only on the detection of
objects with emission lines, blue colours or/and a composed disc --
donor spectra will neglect sources in quiescence with spectra fully or
mostly dominated by the donor star. For instance, the black hole LMXBs
GRO J1655-40 (Orosz et al. 1997; Shahbaz et al. 1997), 4U 1543-47
(Orosz et al. 1998) and SAX J1819.3-2525 \footnote{For the sake of
  completeness and comparison, we provide orbital parameters for
  several sources examined in this section. GRO J1655-40:
  P$_{orb}$=2.62 d, $\gamma = $ -142 km s$^{-1}$, K$_2$ = 228 km
  s$^{-1}$, q=0.33; F3-6 IV companion. 4U 1543-47: P$_{orb}$=1.12 d,
  $\gamma = -82$ km s$^{-1}$, K$_2$ = 124 km s$^{-1}$, A2 V
  companion. SAX J1819.3-2525: P$_{orb}$=2.82 d, $\gamma = 107$ km
  s$^{-1}$, K$_2$ = 211 km s$^{-1}$, q=0.67, B-type companion.  FIRST
  J102347.6+0038: P$_{orb}$=4.75 hr, $\gamma = 1 $ km s$^{-1}$, K$_2$
  = 268 km s$^{-1}$, mid-G V companion.  IGRJ19308+053: :
  P$_{orb}$=0.61 d, $\gamma = -18 $ km s$^{-1}$, K$_2$ = 91 km
  s$^{-1}$, q=1.78, F-type companion. Cyg X-2; P$_{orb}$=9.84 d,
  $\gamma = -210$ km s$^{-1}$, K$_2$ = 86 km s$^{-1}$, q=0.34 A9 III
  companion. Cen X-4: P$_{orb}$=15.1 hr, $\gamma = 184$ km s$^{-1}$,
  K$_2$ = 150 km s$^{-1}$, q=0.17 K3-5 V companion.}  (Orosz et
al. 1998) would have been overlooked since during quiescence the
emission lines may become apparent as residual features only after
subtracting the donor star spectrum.  The same selection caveat can
affect systems harboring less massive primaries.  This is the case of
IGRJ19308+053 (Ratti et al. 2013b) where light from the F-type
companion dominates over both accretion disc and white dwarf
emission. Another interesting example is the millisecond radio pulsar
FIRST J102347.6+003841 (Thorstensen \& Armstrong 2005, Wang et
al. 2009) where occasionally only photospheric features from the
G-type dwarf donor are detected during quiescence.  While low in
number, such objects are of interest for two reasons: first, A and F
donors are visible at larger distances and reddening than cooler
companions.  Second, their optical and infrared counterparts are
partially or totally free of variable emission from the accretion
disc. As a consequence inclination and distance measurements become
less dependent on uncertainties in the disc contribution to the total
light. The selection effect also applies to LMXBs in
  hierarchical triple systems in which the main source of optical
  light during quiescence is the  star orbiting around the inner LMXB.
  Considering only the two confirmed triple systems among  the current sample of
  recycled millisecond pulsars, we expect a $\sim 10\%$ of the LMXBs  to be triple system.

In order to increase the chances of finding accreting binaries in
sources with no or only weak signs of on-going accretion in their
optical spectra, spectroscopic surveys should also employ radial
velocity measurements.  Stellar-like objects with high radial
velocities (or proper motions) should be re-observed to confirm if
they are binaries.  The reason for this is that, except for runaway
stars, field stars in the direction of the bulge should show only
small radial motions with respect to the Sun.  A radial velocity
search for accreting binaries will be sensitive to LMXBs with cool (G
to M-type) donors and black hole LMXBs with cool (G to M) and hot (A
to F-type) donors since in these cases $q \lesssim 1$ and thereby $K_2
\gtrsim 100$ km s$^{-1}$ for orbital periods less than a day.
Detection of neutron star LMXBs and CVs with hot donors in a single
radial velocity measurement is more unlikely since $q \gtrsim 1$ and
$K_2 \lesssim 100$ km s$^{-1}$ (however, see Ratti et
al.~2013b). Furthermore, some neutron star LMXBs may have received a
significant natal kick that may result in a high systemic radial
velocity for the system. Examples of systems with a high $\gamma$ are
Cyg X-2 (Casares et al. 1998, Kolb et al. 2000) and Cen X-4 (Torres et
al.  2002, Gonz{\'a}lez Hern{\'a}ndez et al. 2005).  Accreting
  binaries  in the Bulge may also have a high radial velocity
  component caused by the kinematics in this region. Bulge members
  can be identified by the  presence of strong DIBs in their optical
  spectra. The strengths in many of these features correlate well with
  the extinction towards the source and thereby they can be used
  together with 2D/3D extinctions maps of the GBS areas (e.g. Gonzalez
  et al. 2012, Chen et al. 2013). The GBS, with a selection of sources based on their X-ray detection, a
spectroscopic search accounting for radial velocities and a photometric scrutiny to identify ellipsoidal variables, will allow us
to investigate the frequency and nature of accreting binaries lacking
signs for accretion activity in their optical spectra.

\section*{Appendix A: finding charts}
Finding charts (only on-line) for the optical counterparts to 20 of
the 23 accreting binaries presented in this paper.  North is up and
East is to the left. The charts are 20 arcsec square with the optical
counterpart in the center. The images have been created by using
R-band images obtained with VIMOS (except for CX44 for which Mosaic-II
r$^\prime$-band imaging was used).  Finders for CX28 and CX93 can be
found in Britt et al. (2013) and Ratti et al. (2012), respectively.

\section*{Acknowledgments} \noindent We are thankful to Marina Rejkuba
(ESO) for supporting our service mode observations with VIMOS.  We are
grateful to Eva Ratti, Oliwia Madej and Marianne Heida and for
helping in the preparation of the VIMOS masks.  We also thank Victoria
Gabb and Monique Villar for assistance with Mosaic-II data analysis
and Lauren Gossen for assistance at the Blanco telescope.  PGJ
acknowledges support from a VIDI grant from the Netherlands
Organisation for Scientific Research. DS acknowledges support from STFC through an Advanced Fellowship
(PP/D005914/1) as well as grant ST/I001719/1 RIH, CTB and CBJ acknowledge support from the National
Science Foundation under Grant No. AST-0908789.

\newpage

\begin{figure*}
\includegraphics[width=7.0in, angle=0.0]{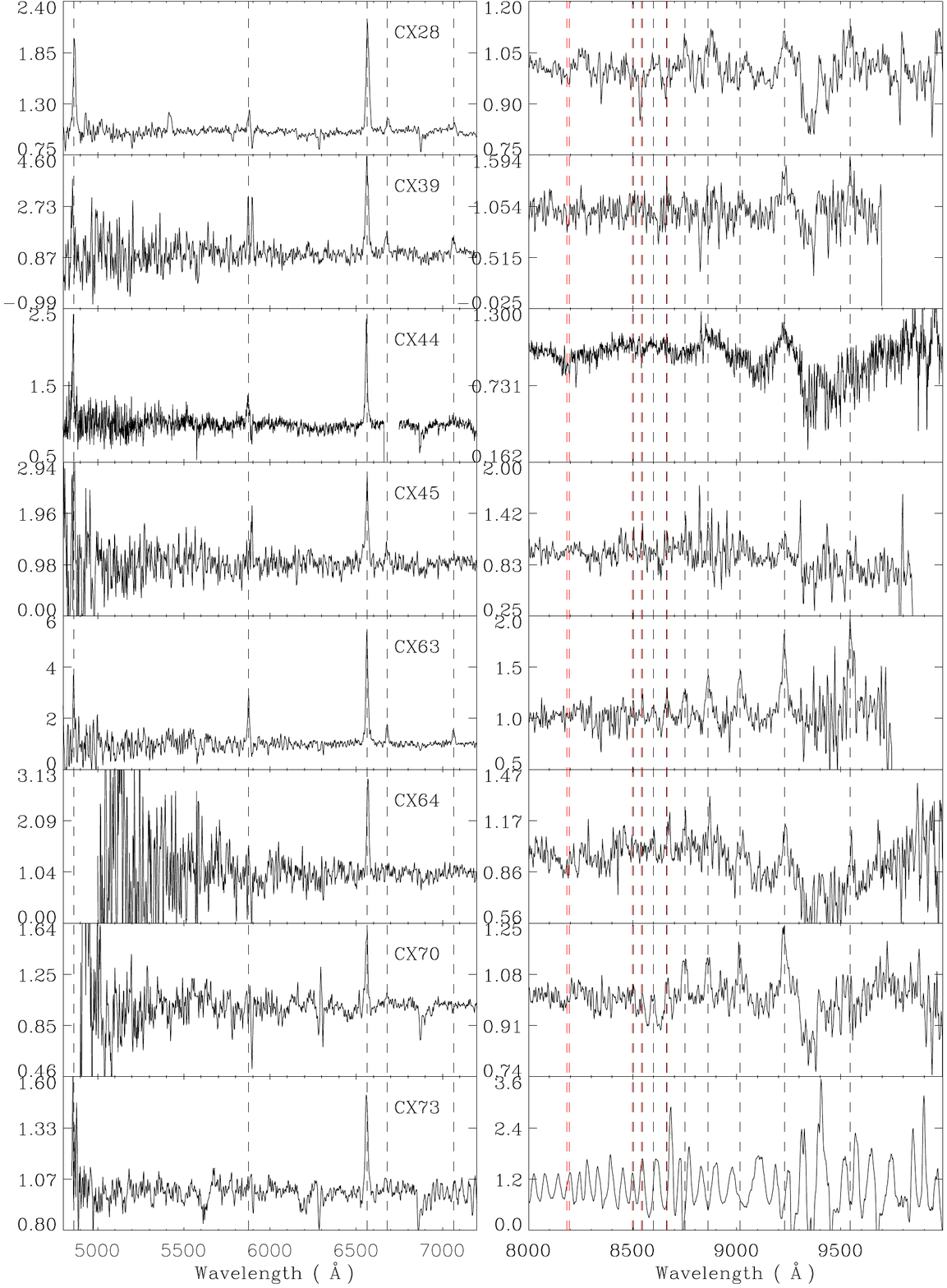}
\caption{Continuum-normalized spectra of GBS sources in two wavelength
  ranges. The rest wavelenths for the lines given in Sec 3.1 are
  marked with black dashed lines except the Ca {\sc ii} IR triplet
  (red dashed lines). Note that the P13, 15 and 16 Paschen lines are
  blended with the Ca triplet.}
\label{licus}
\end{figure*}

\newpage

\begin{figure*}
\includegraphics[width=7.0in, angle=0.0]{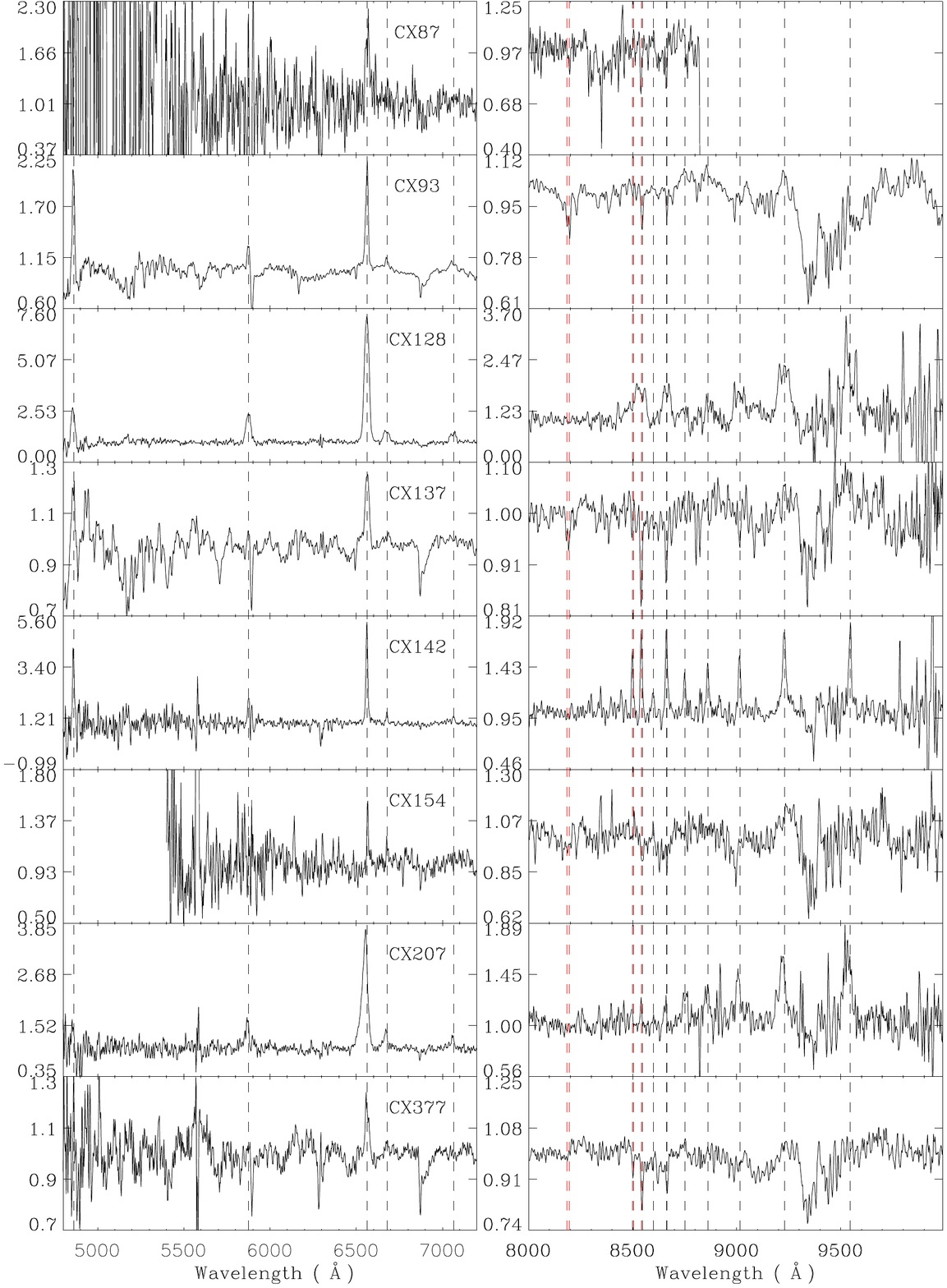}
\caption{Normalized spectra of GBS sources (continued).}
\label{licus}
\end{figure*}

\newpage

\begin{figure*}
\includegraphics[width=7.0in, angle=0.0]{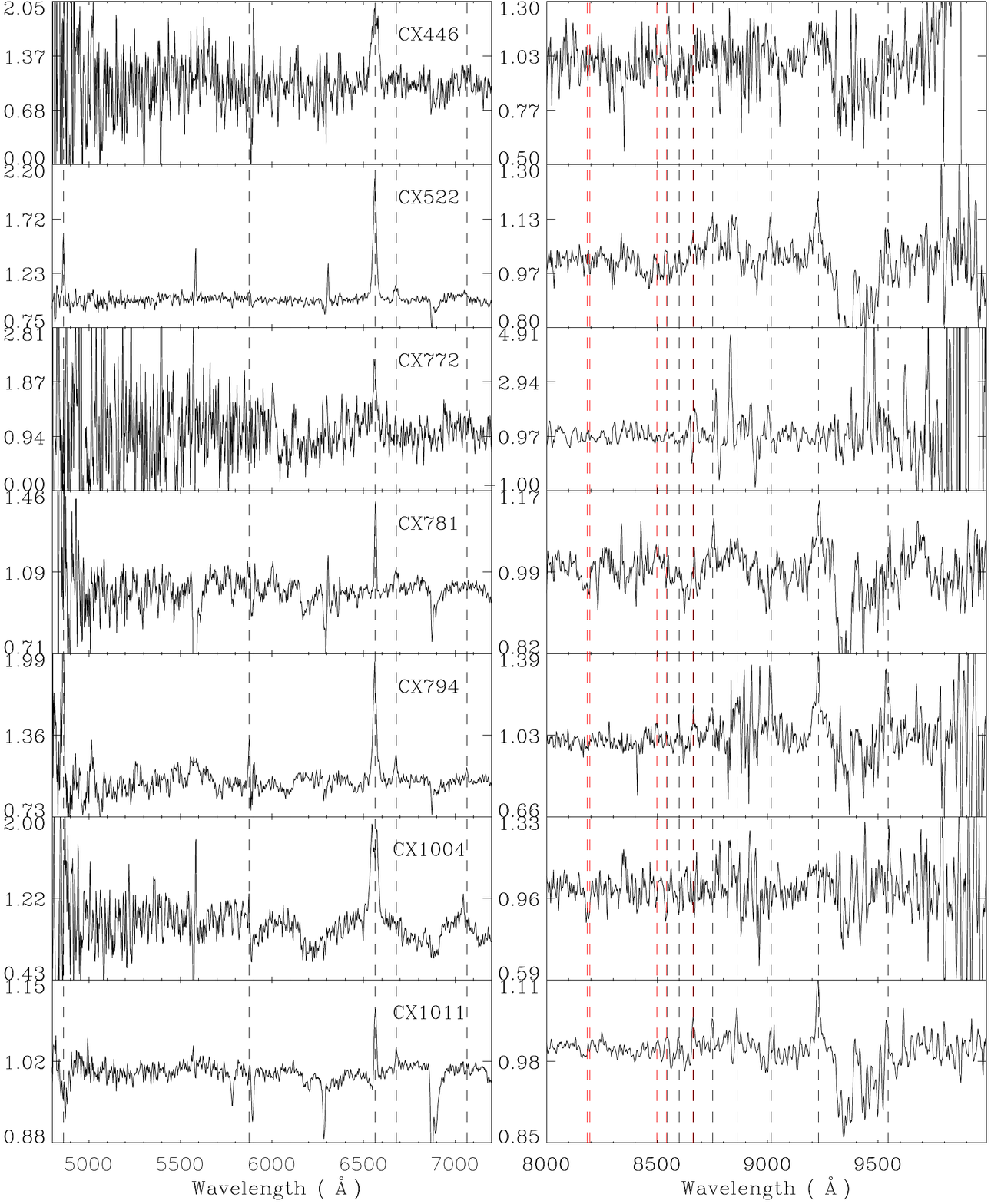}
\caption{Normalized spectra of GBS sources (continued).}
\label{licus}
\end{figure*}

\newpage

\newpage

\begin{figure*}
\begin{center}$
\begin{array}{ccc}
\includegraphics[width=2.2in, angle=90.0]{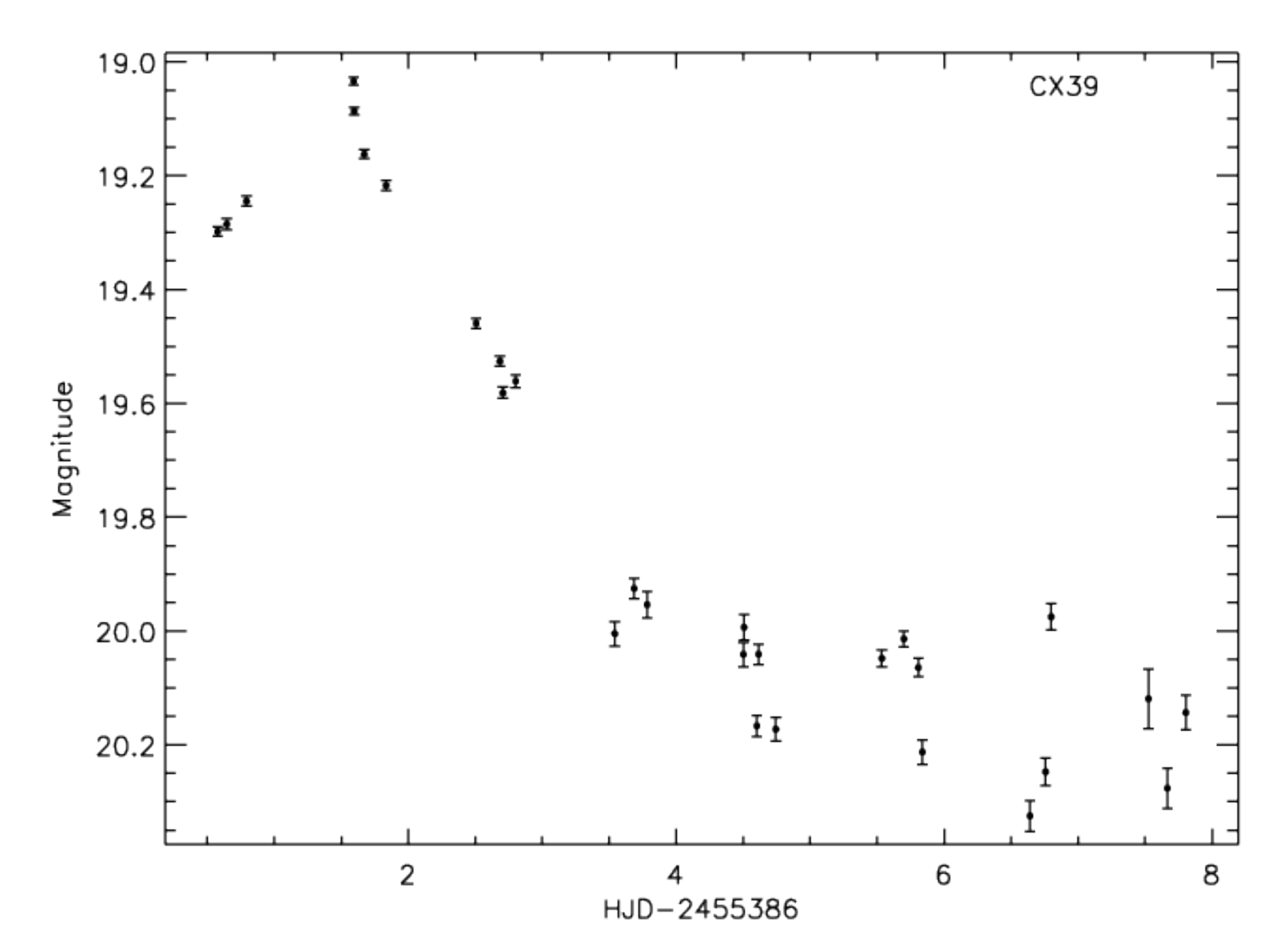} &
\includegraphics[width=2.2in, angle=90.0]{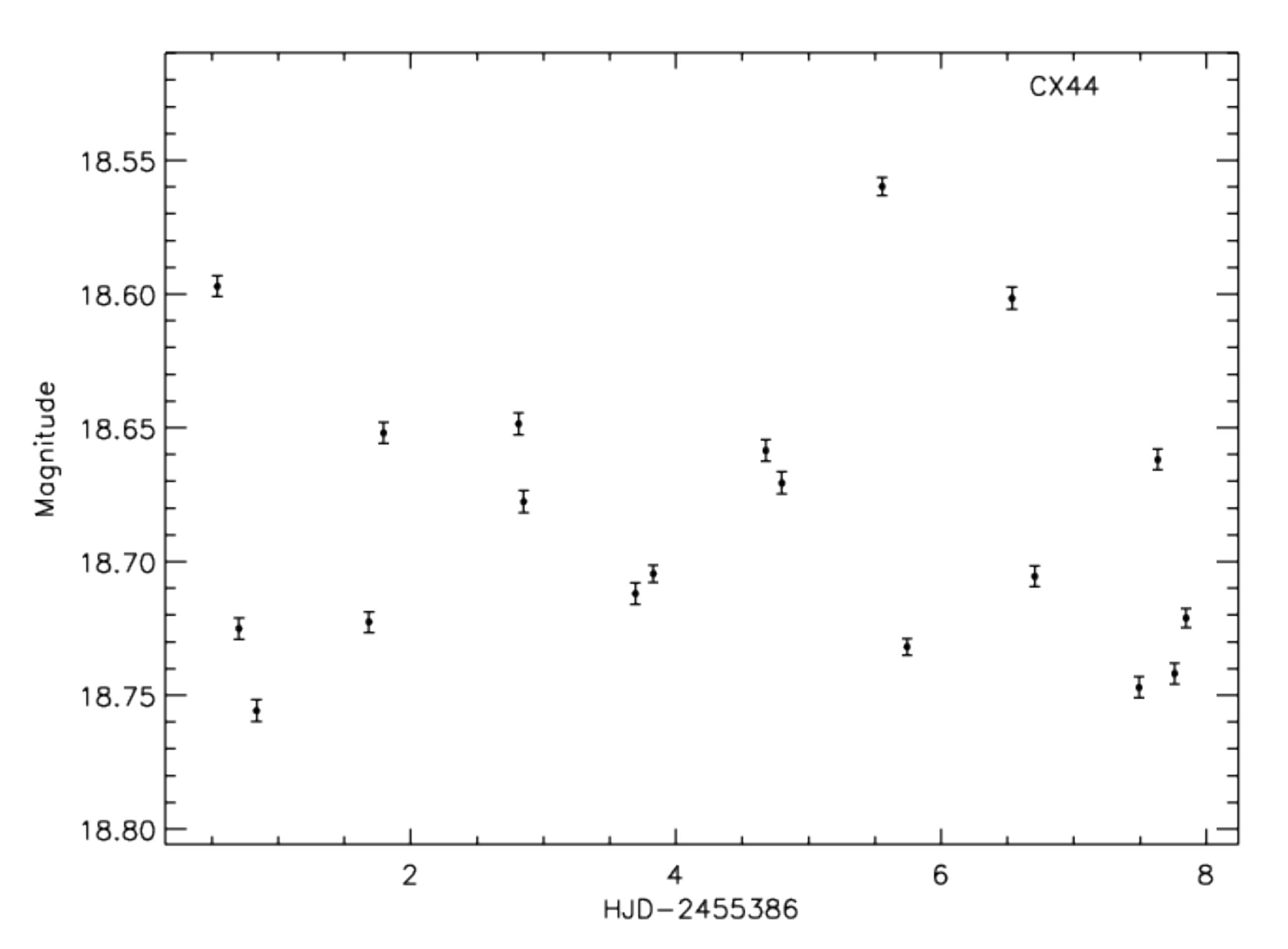} \\
\includegraphics[width=2.2in, angle=90.0]{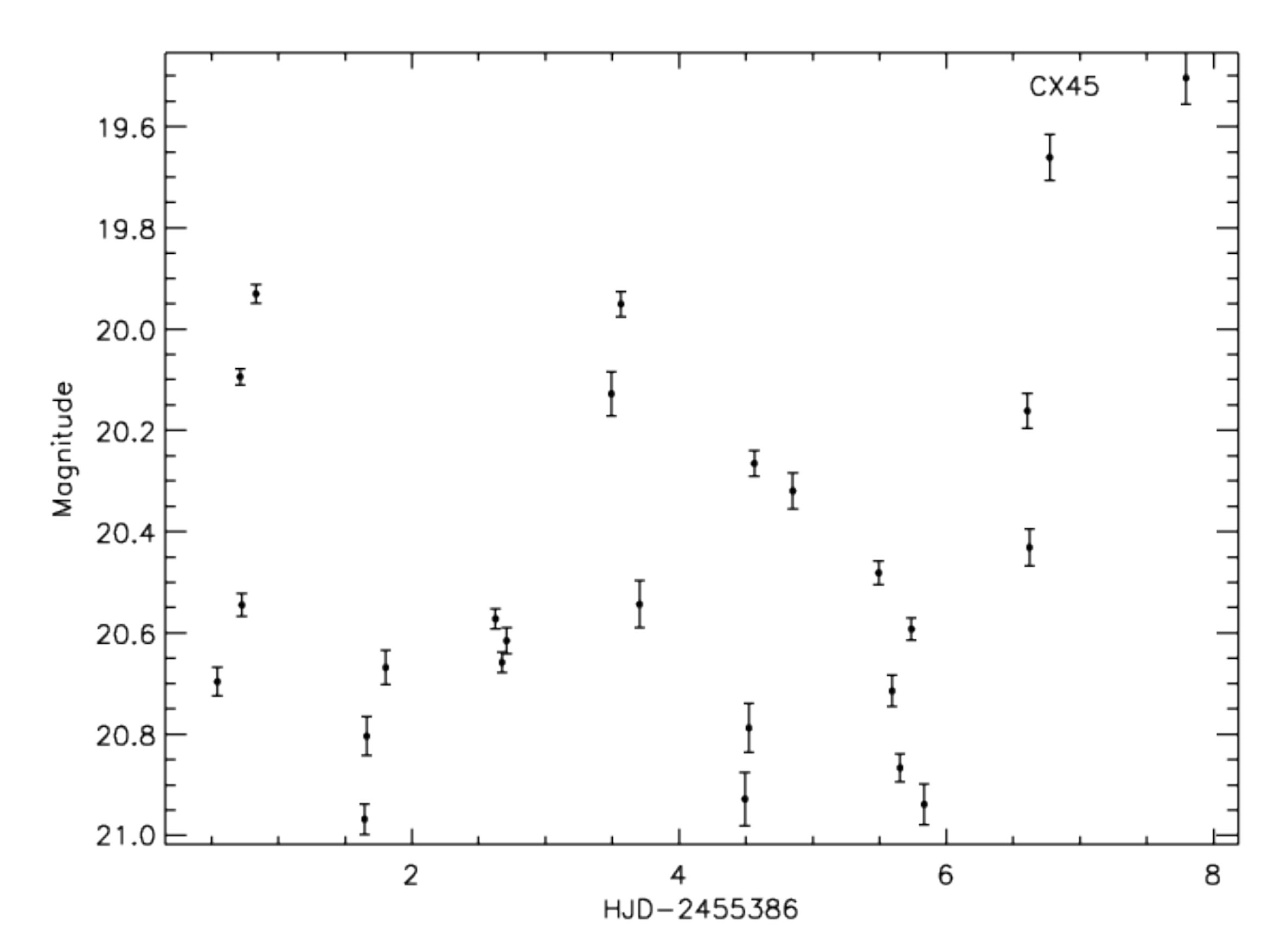} &
\includegraphics[width=2.2in, angle=90.0]{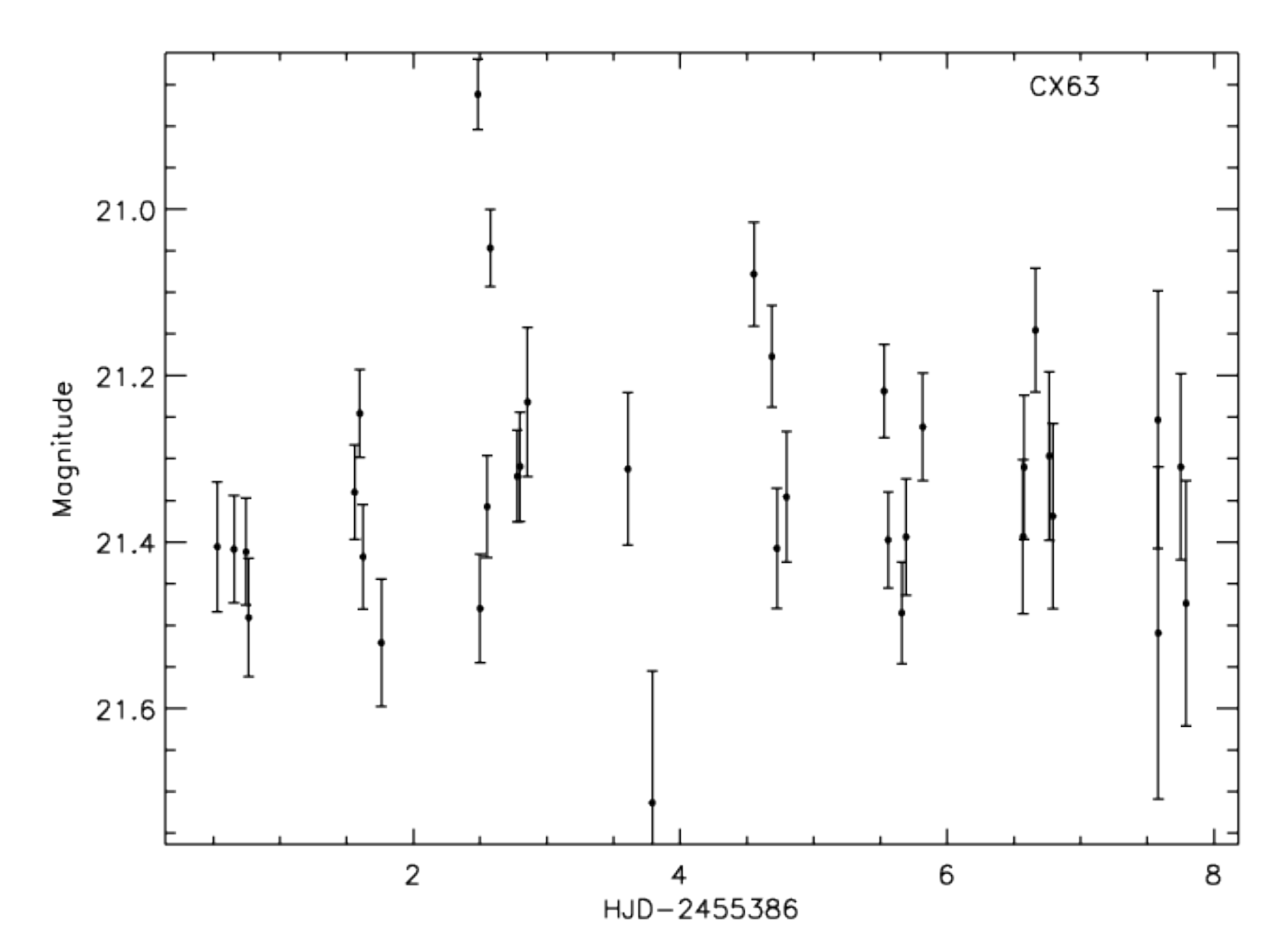} \\
\includegraphics[width=2.2in, angle=90.0]{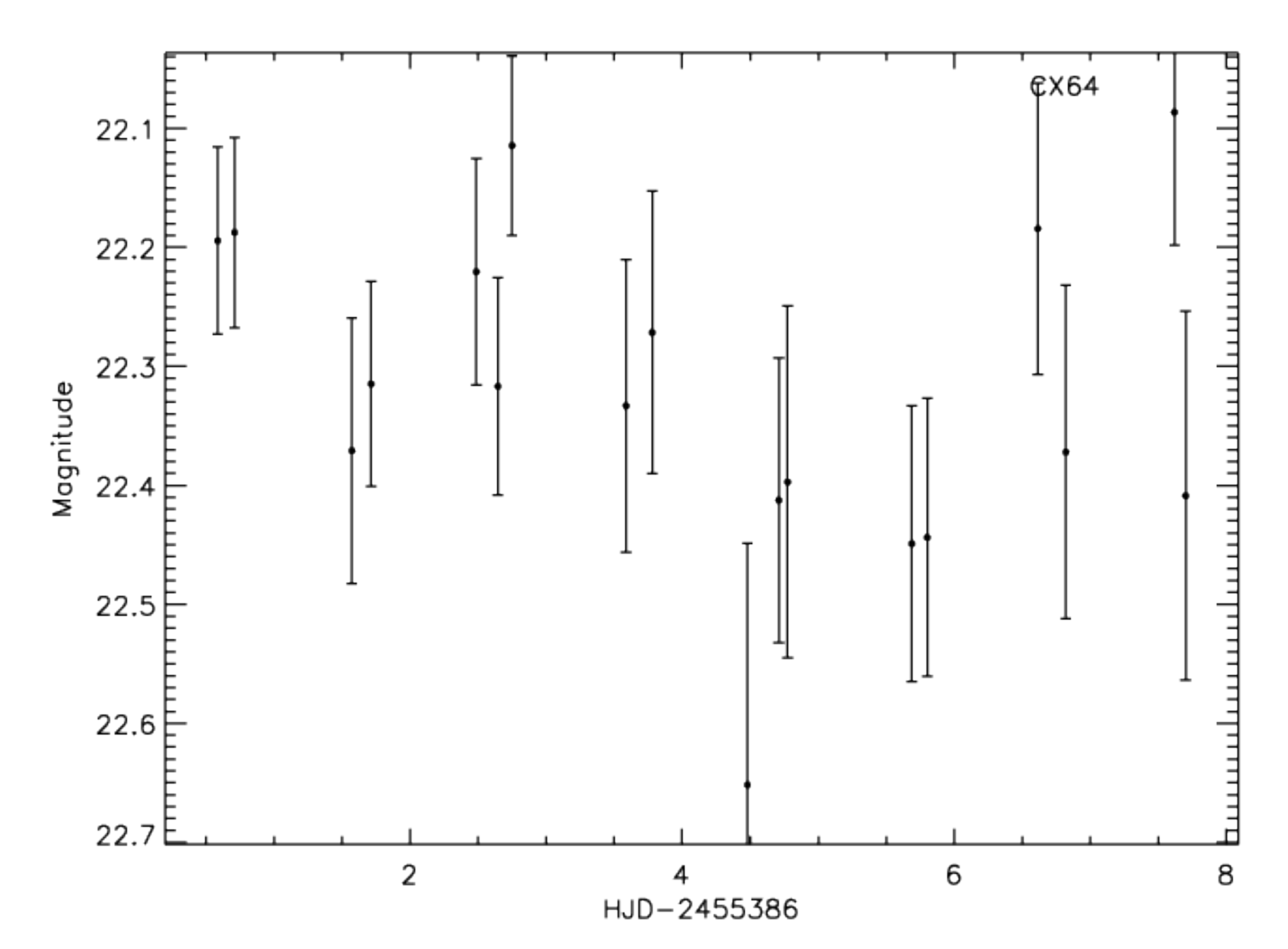} &
\includegraphics[width=2.2in, angle=90.0]{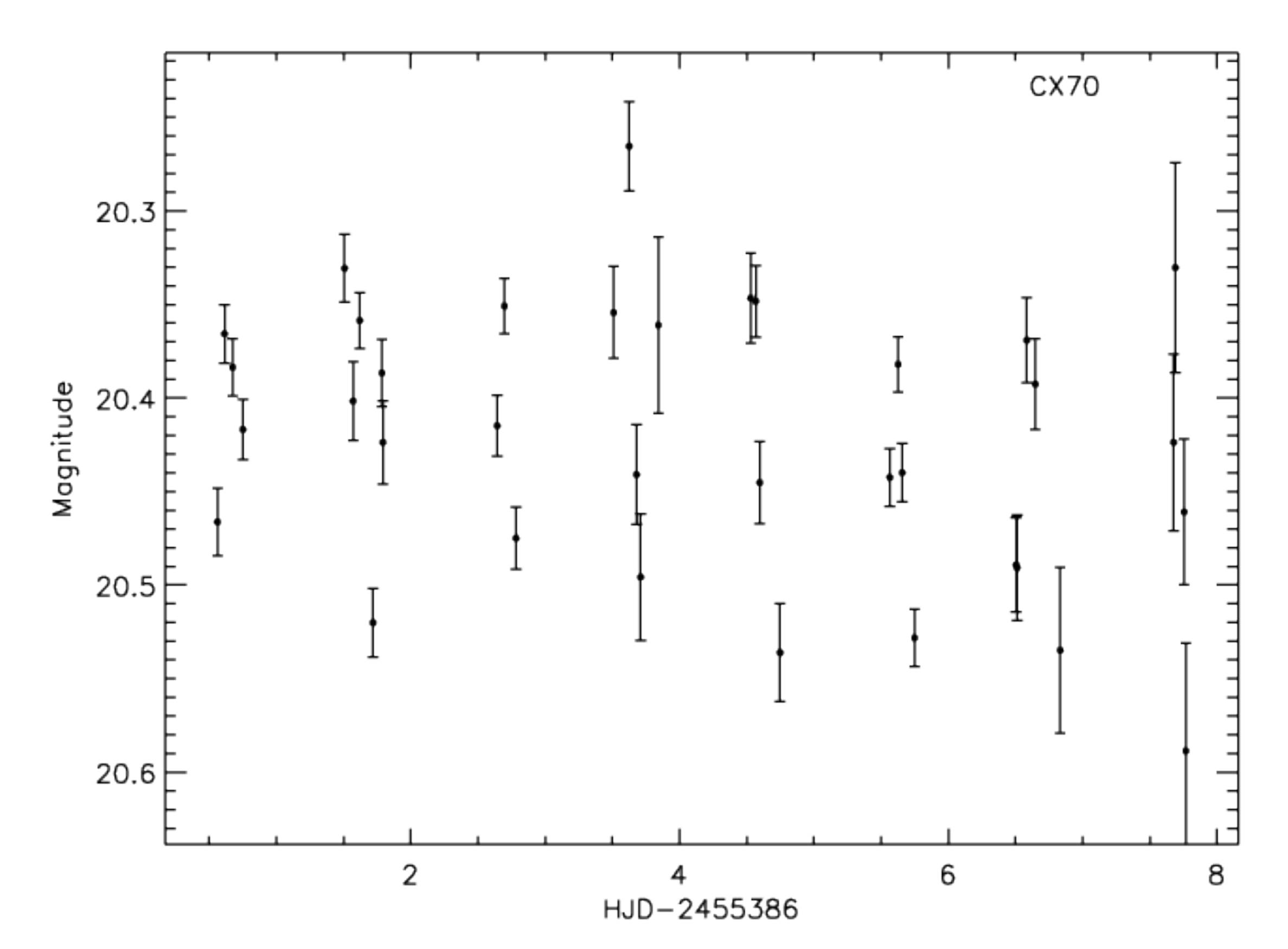} \\
\includegraphics[width=2.2in, angle=90.0]{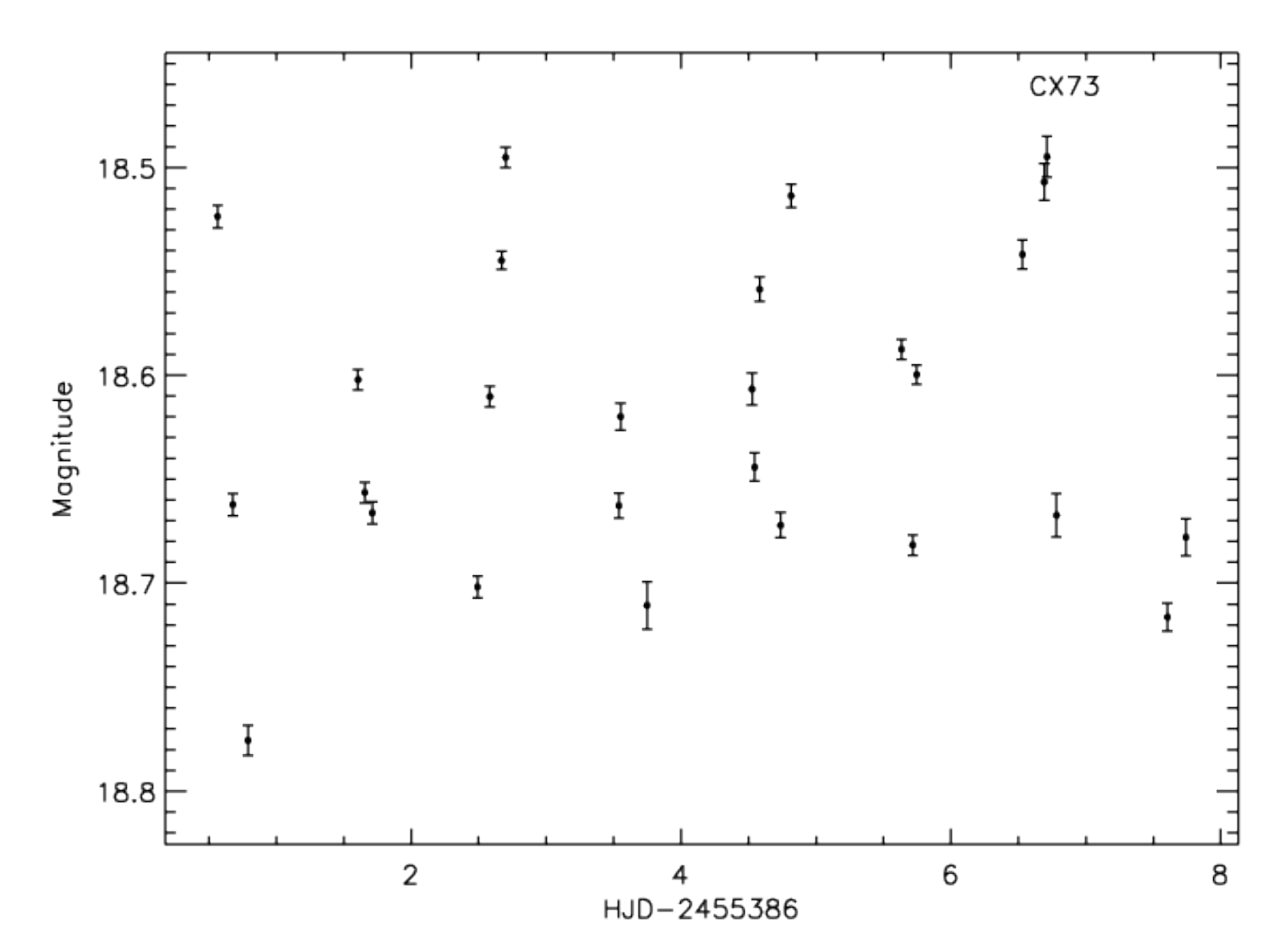} &
\includegraphics[width=2.2in, angle=90.0]{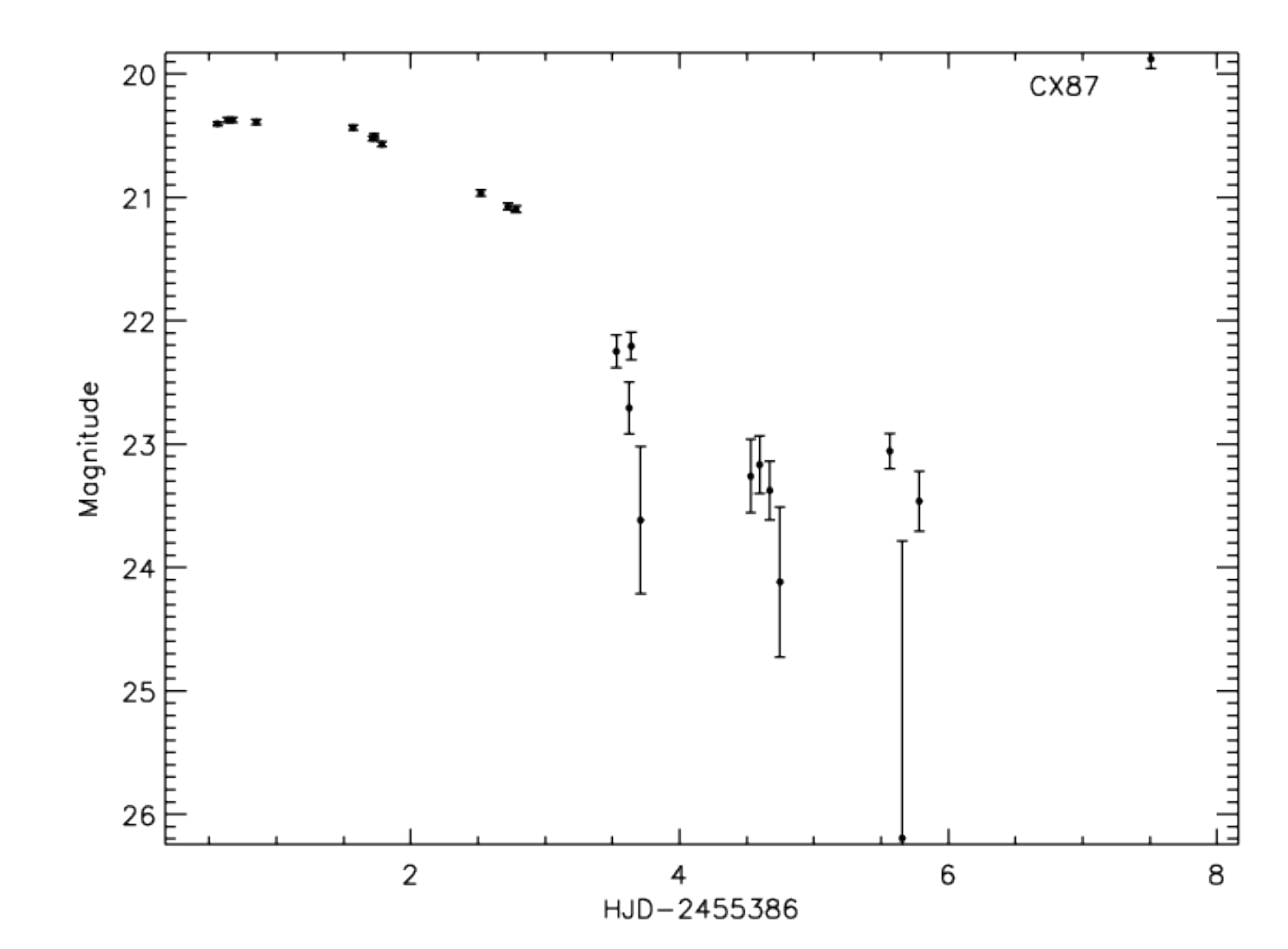} \\
\end{array}$
\end{center}

\caption{Optical light curves for 20 of the 23 accreting binaries
  reported in this paper. The light curves for CX28 and CX93 can be
  found in Britt et al. (2013) and Ratti et al. (2012a). No suitable
  data was available to obtain a light curve for CX522.}
\label{licus}
\end{figure*}

\newpage

\begin{figure*}
\begin{center}$
\begin{array}{ccc}

\includegraphics[width=2.2in, angle=90.0]{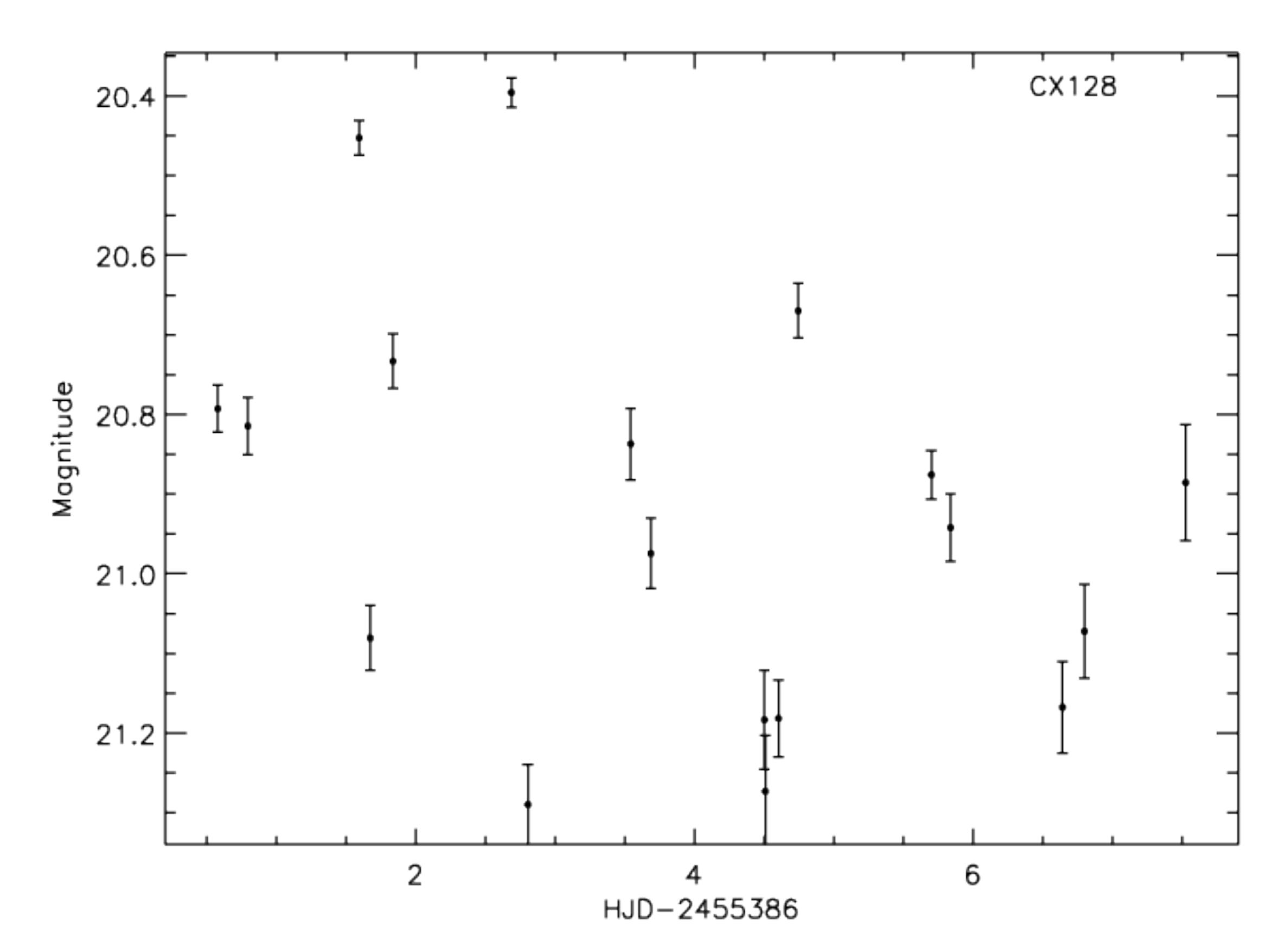} & 
\includegraphics[width=2.2in, angle=90.0]{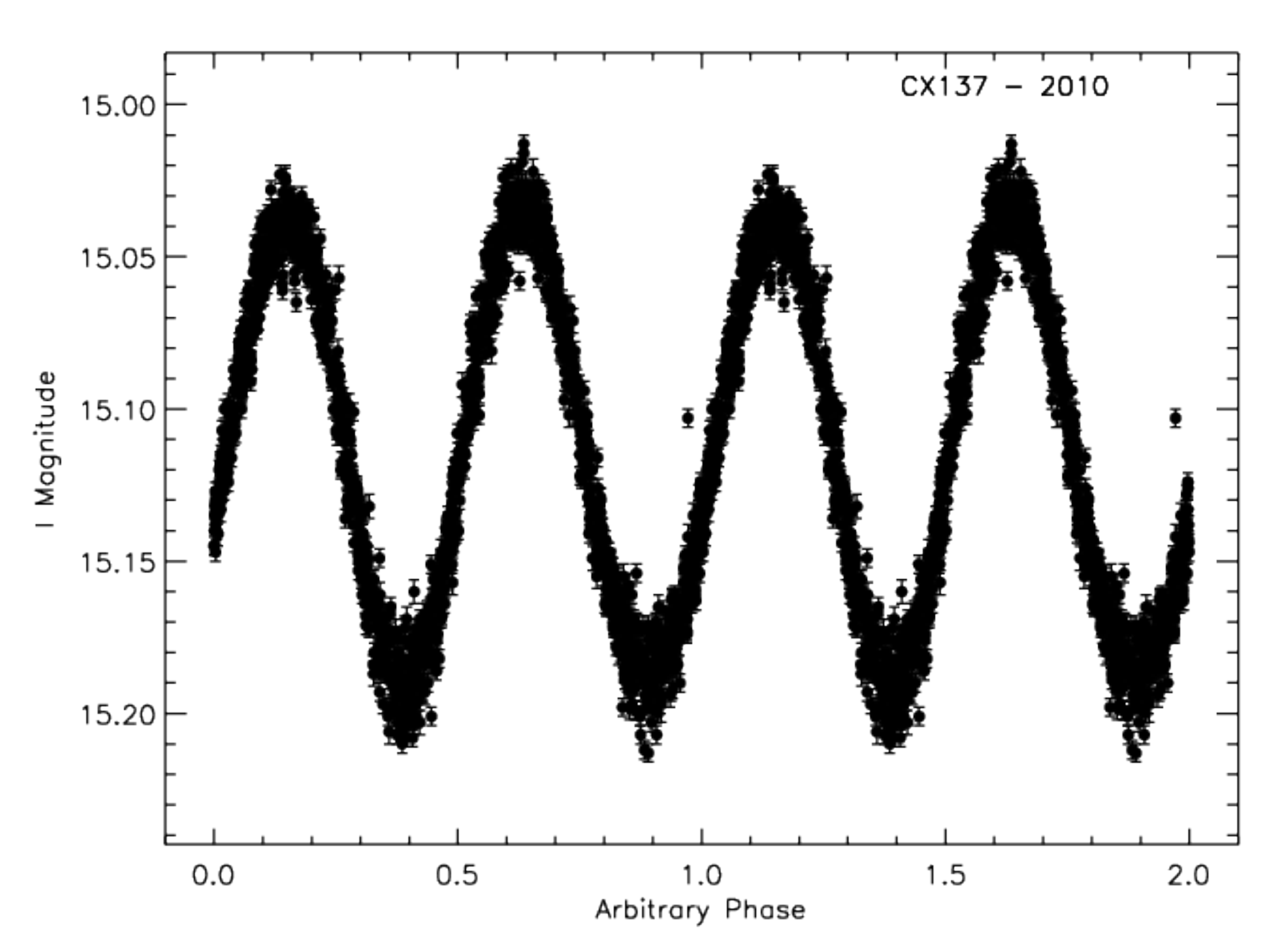} \\

\includegraphics[width=2.2in, angle=90.0]{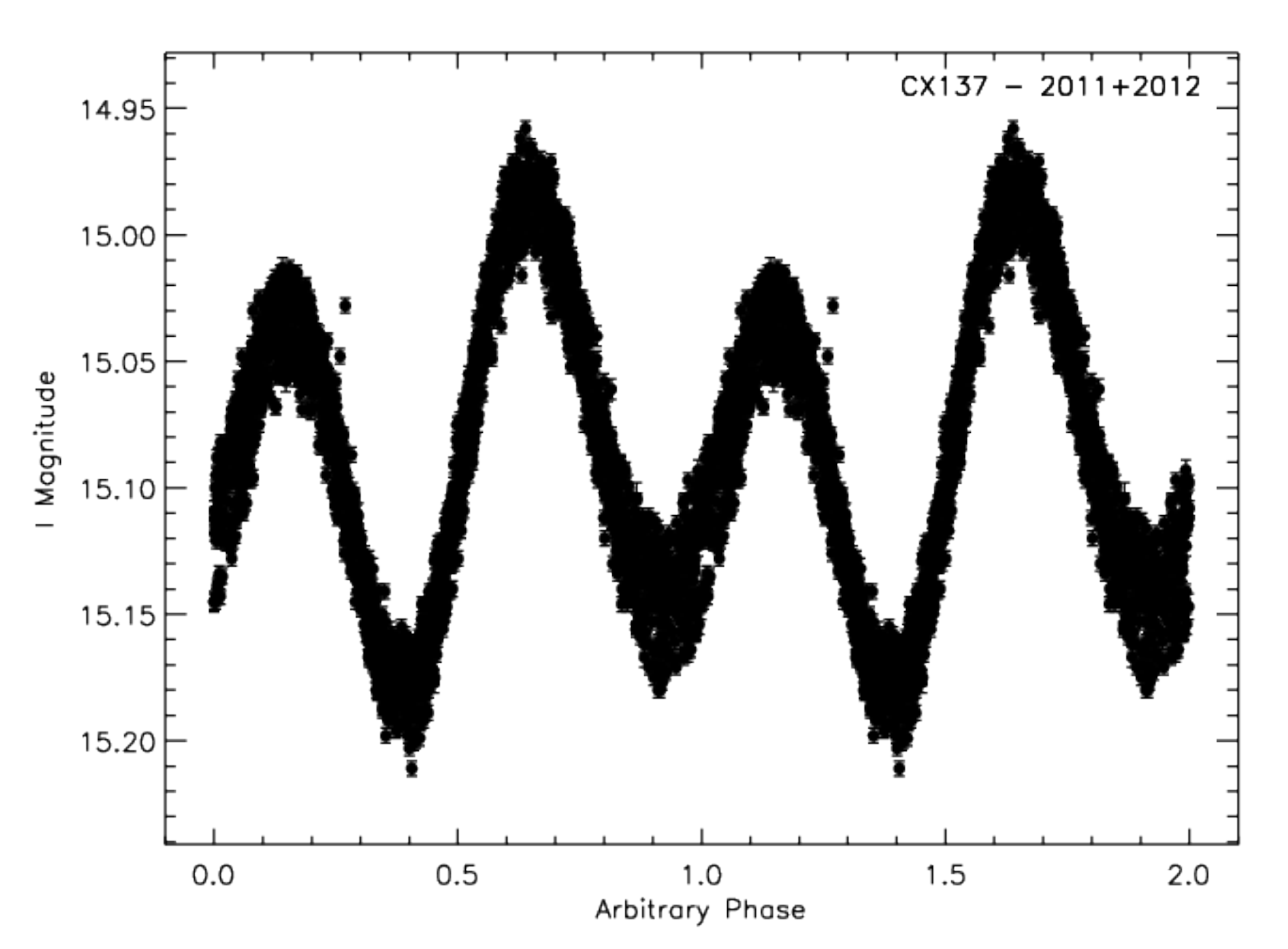} &
\includegraphics[width=2.2in, angle=90.0]{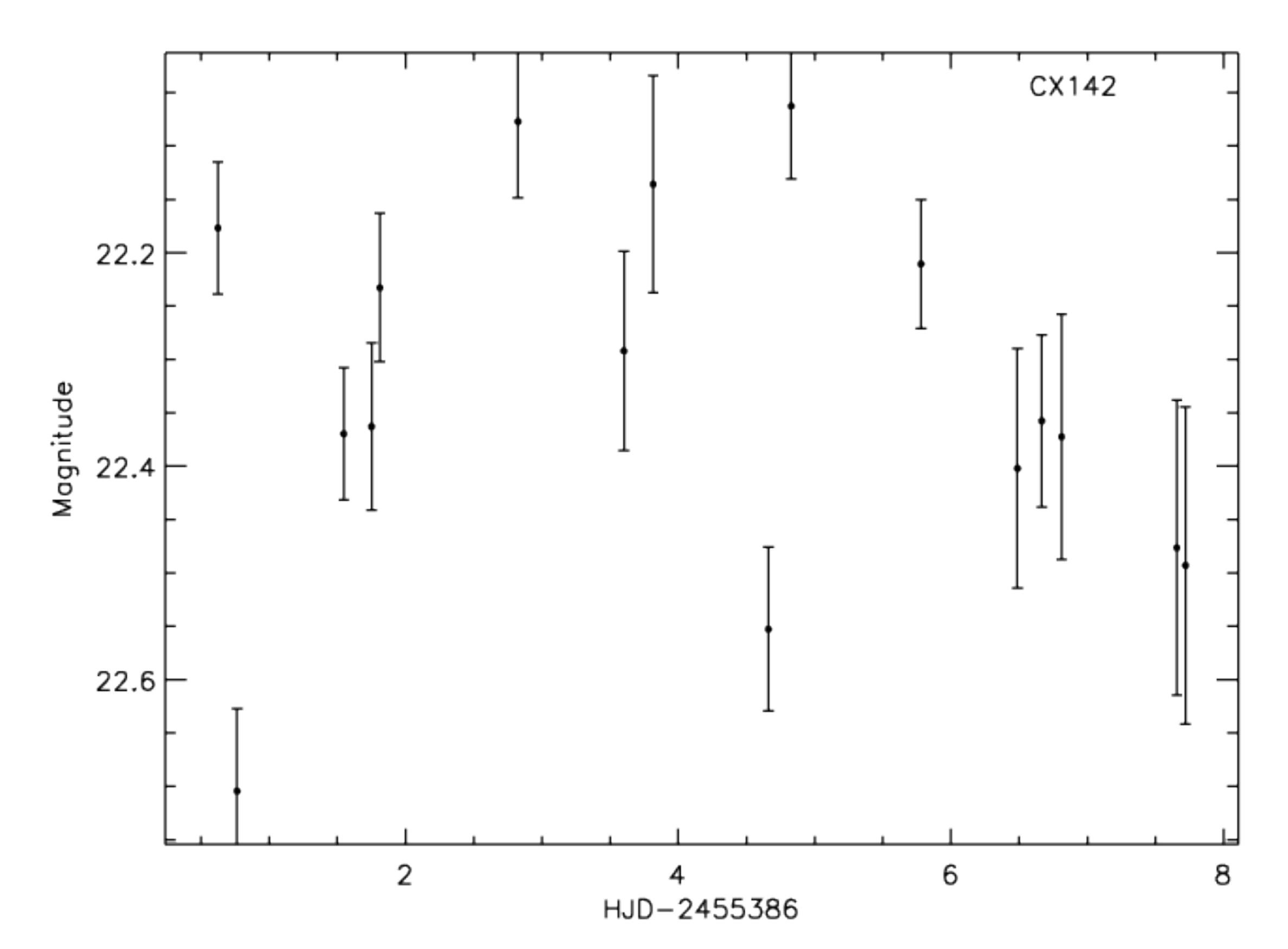} \\

\includegraphics[width=2.2in, angle=90.0]{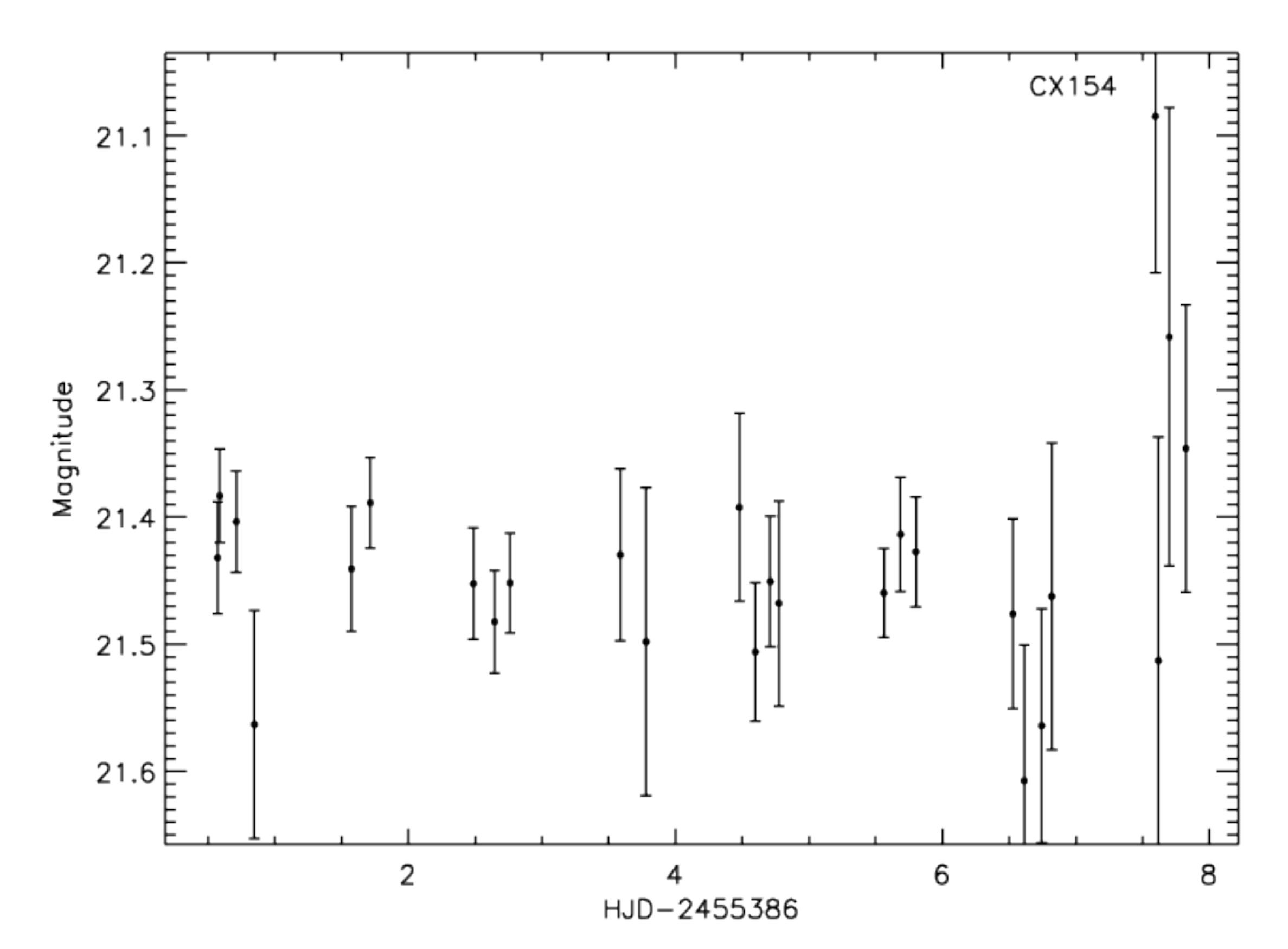} &
\includegraphics[width=2.2in, angle=90.0]{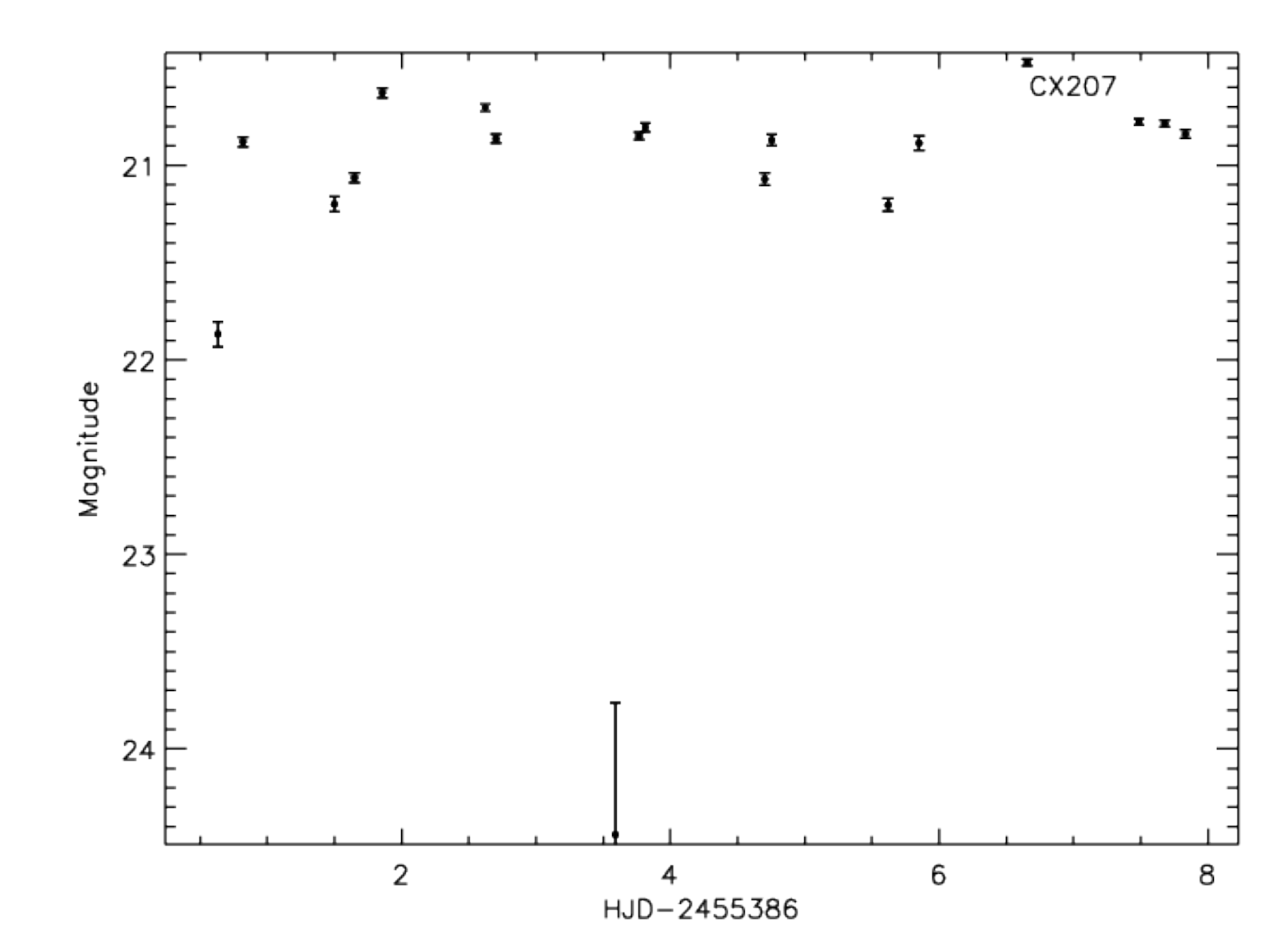} \\

\includegraphics[width=2.2in, angle=90.0]{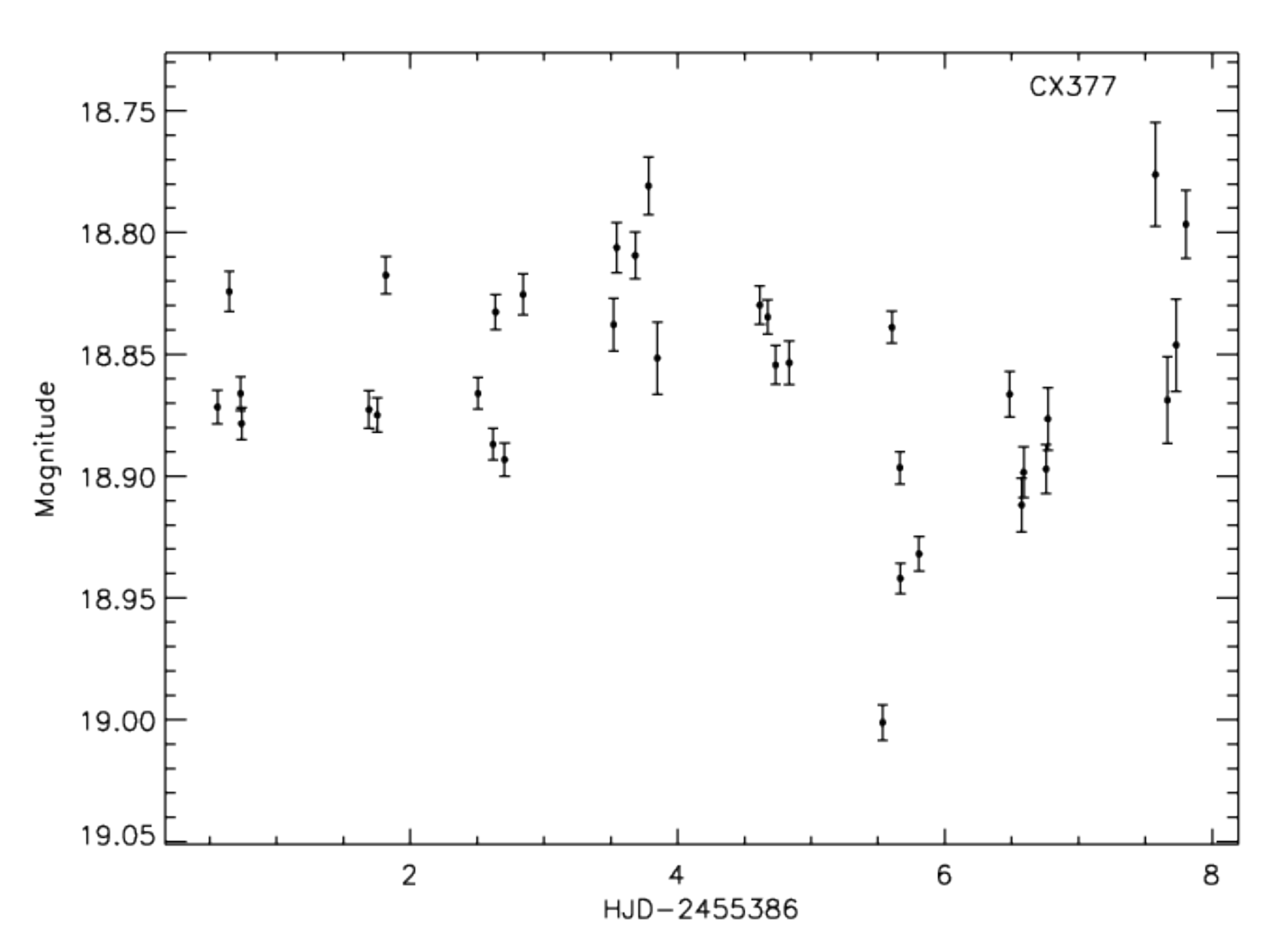} &
\includegraphics[width=2.2in, angle=90.0]{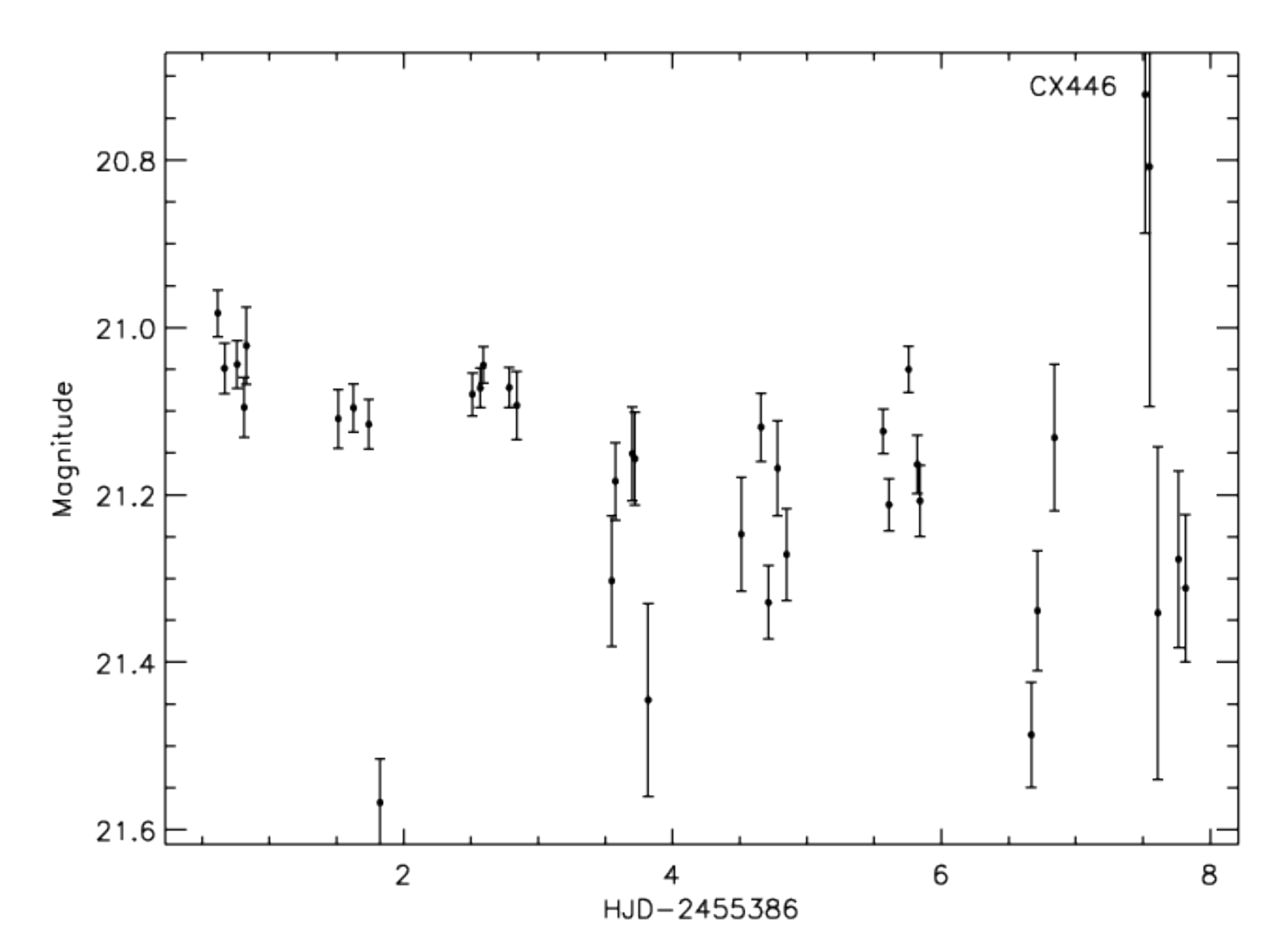} \\

\end{array}$
\end{center}
\caption{Optical light curves (continued). The CX137 light curves are based on OGLE-IV data.}
\label{licus}
\end{figure*}

\begin{figure*}
\begin{center}$
\begin{array}{ccc}

\includegraphics[width=2.2in, angle=90.0]{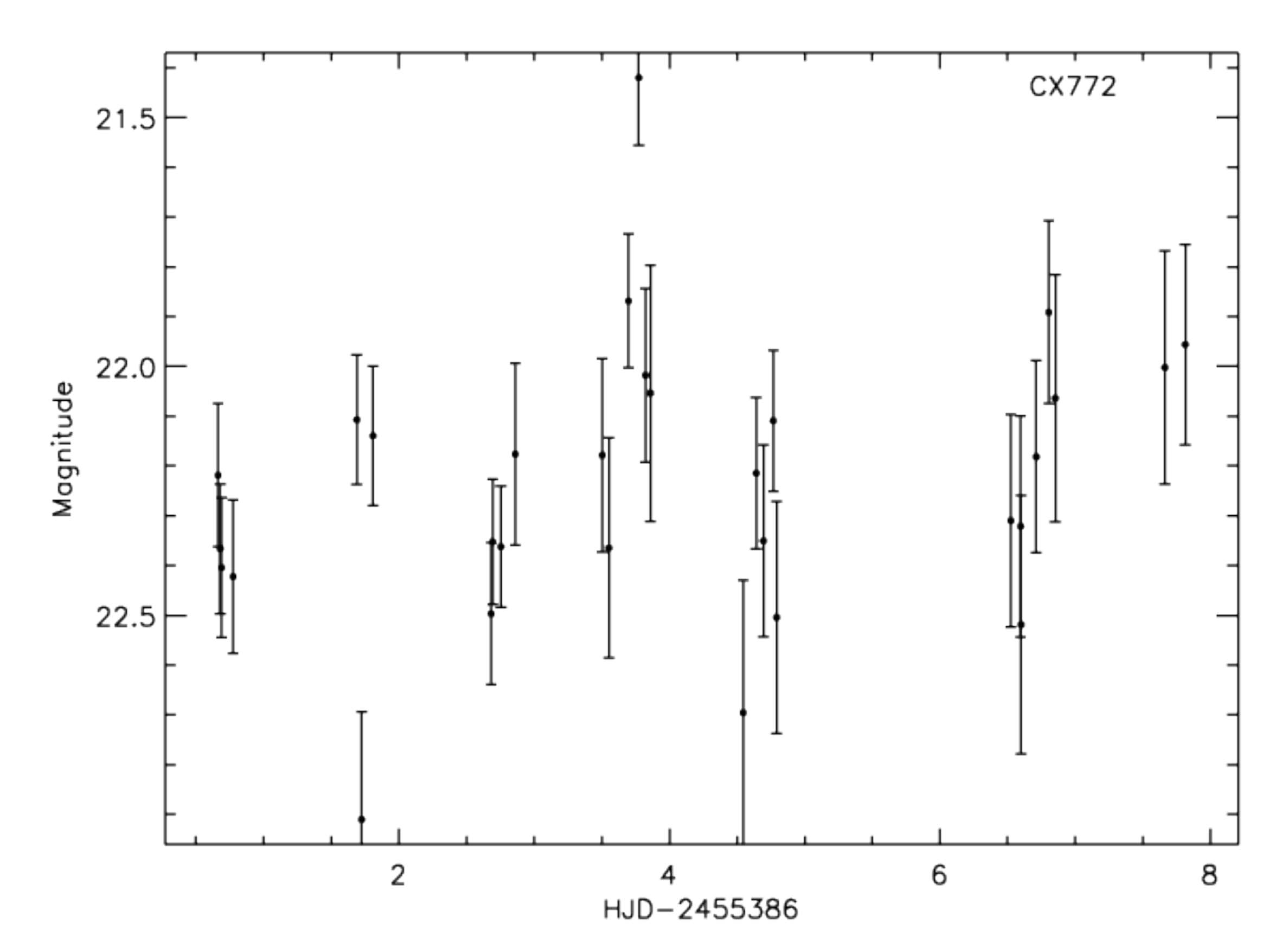} &
\includegraphics[width=2.2in, angle=90.0]{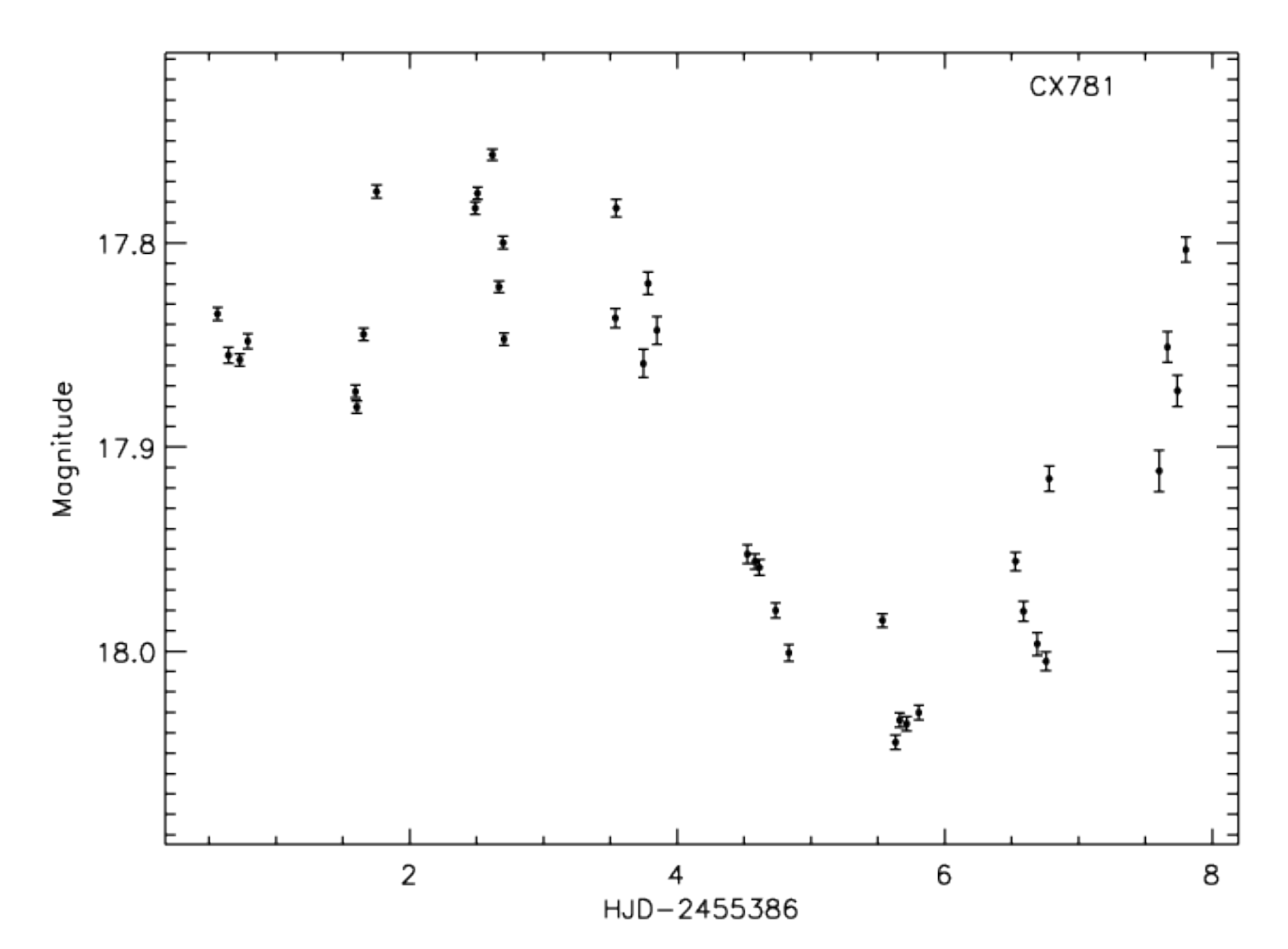} \\

\includegraphics[width=2.2in, angle=90.0]{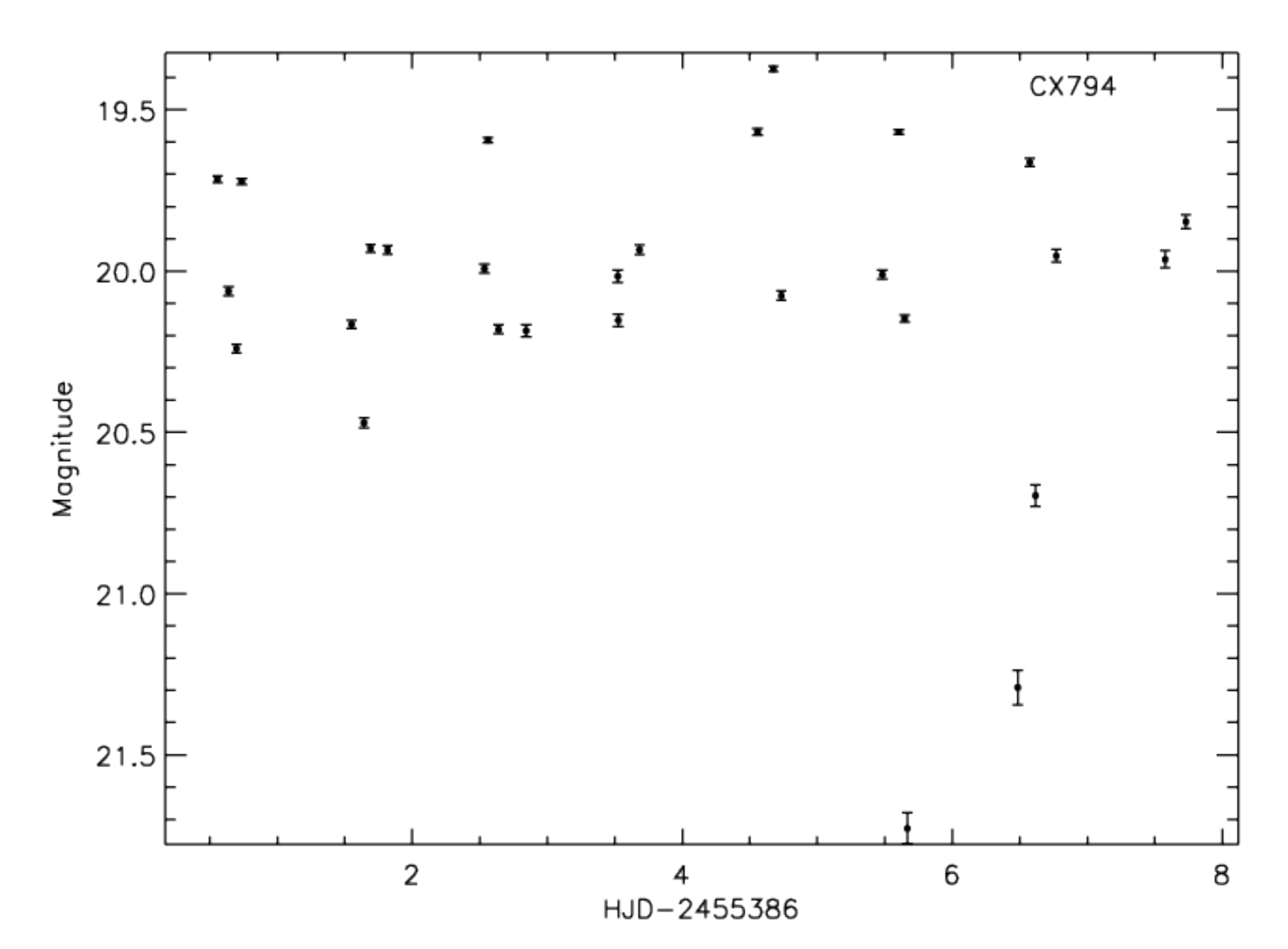} &
\includegraphics[width=2.2in, angle=90.0]{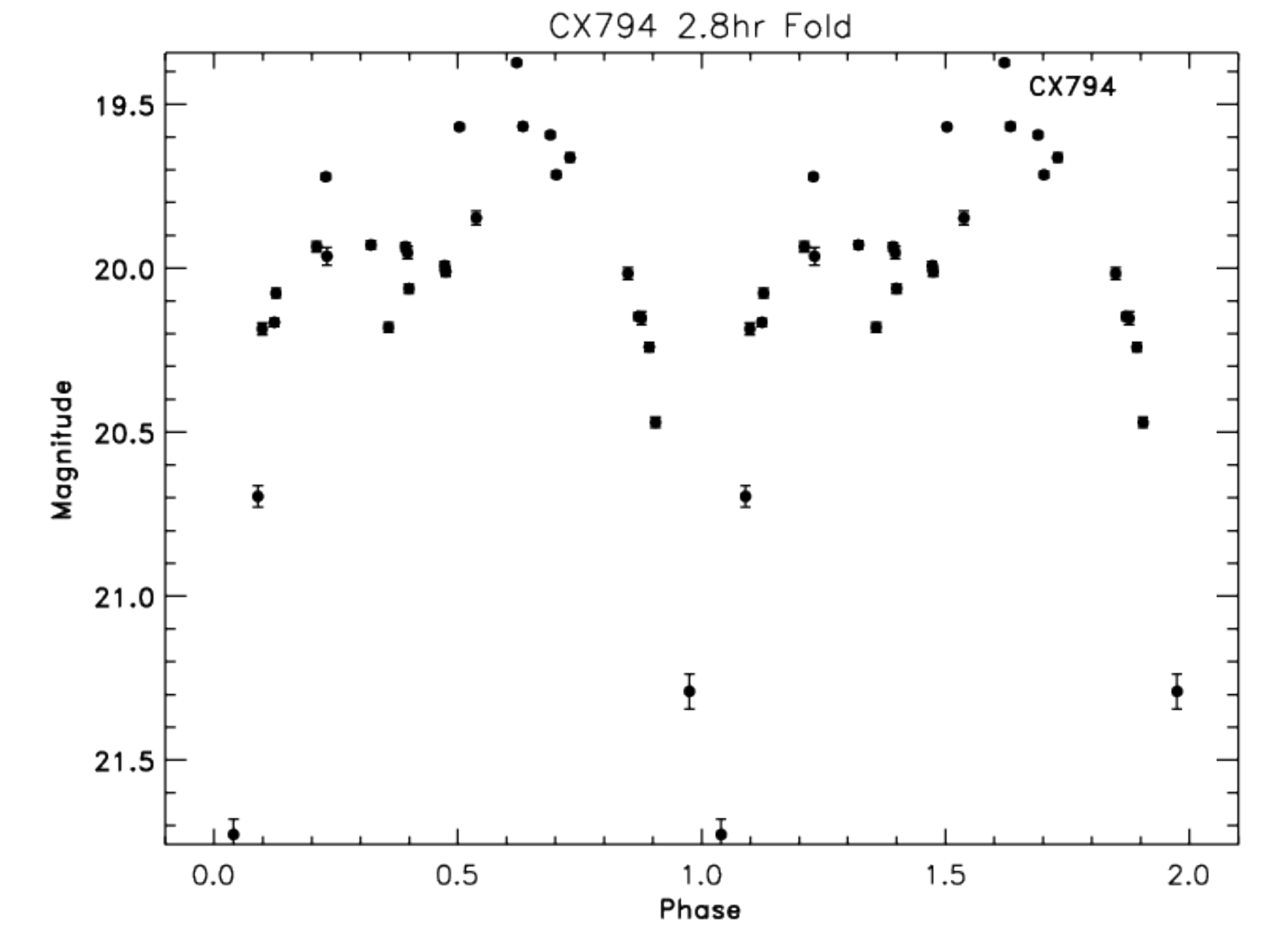} \\

\includegraphics[width=2.2in, angle=90.0]{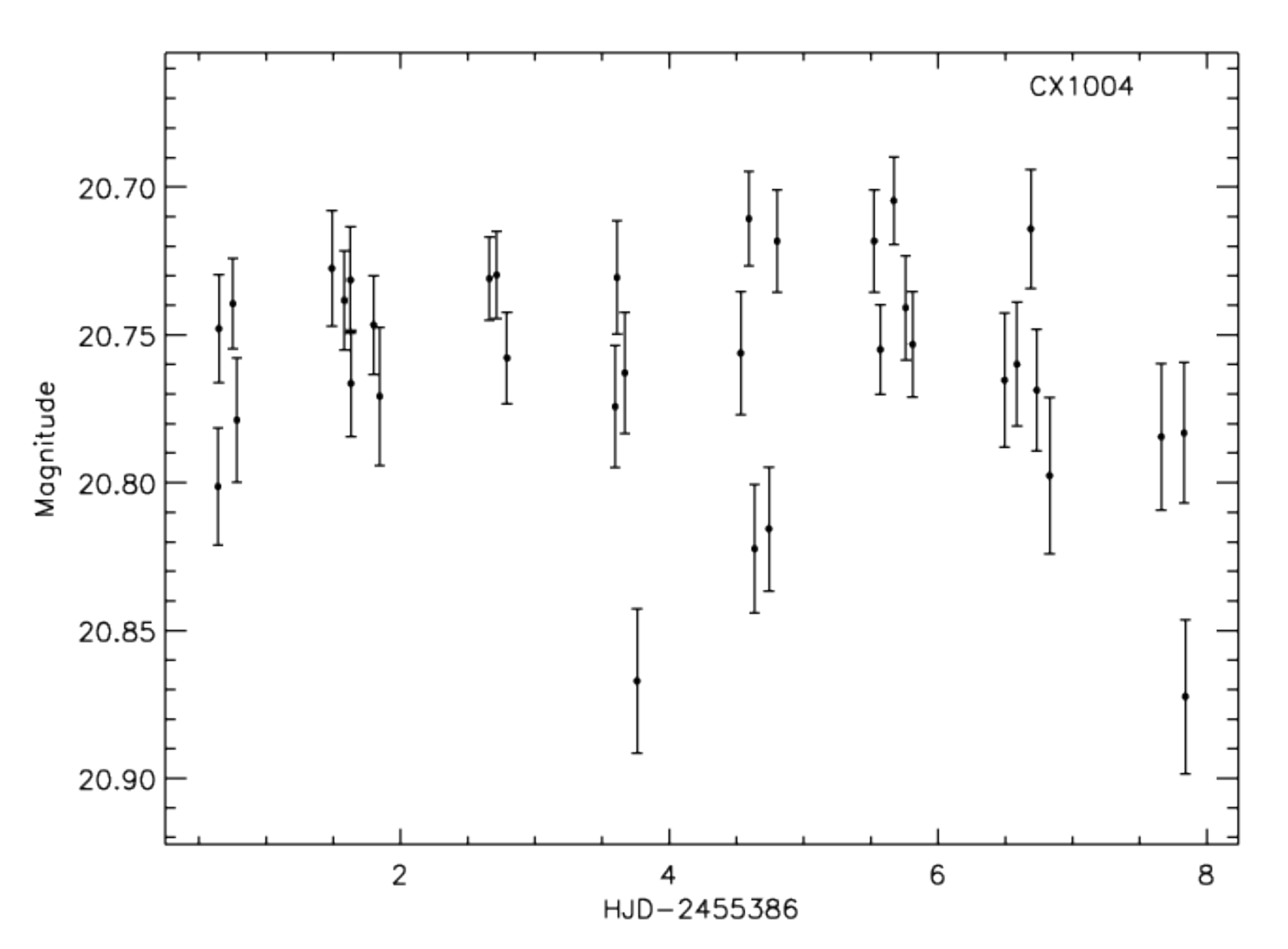} &
\includegraphics[width=2.2in, angle=90.0]{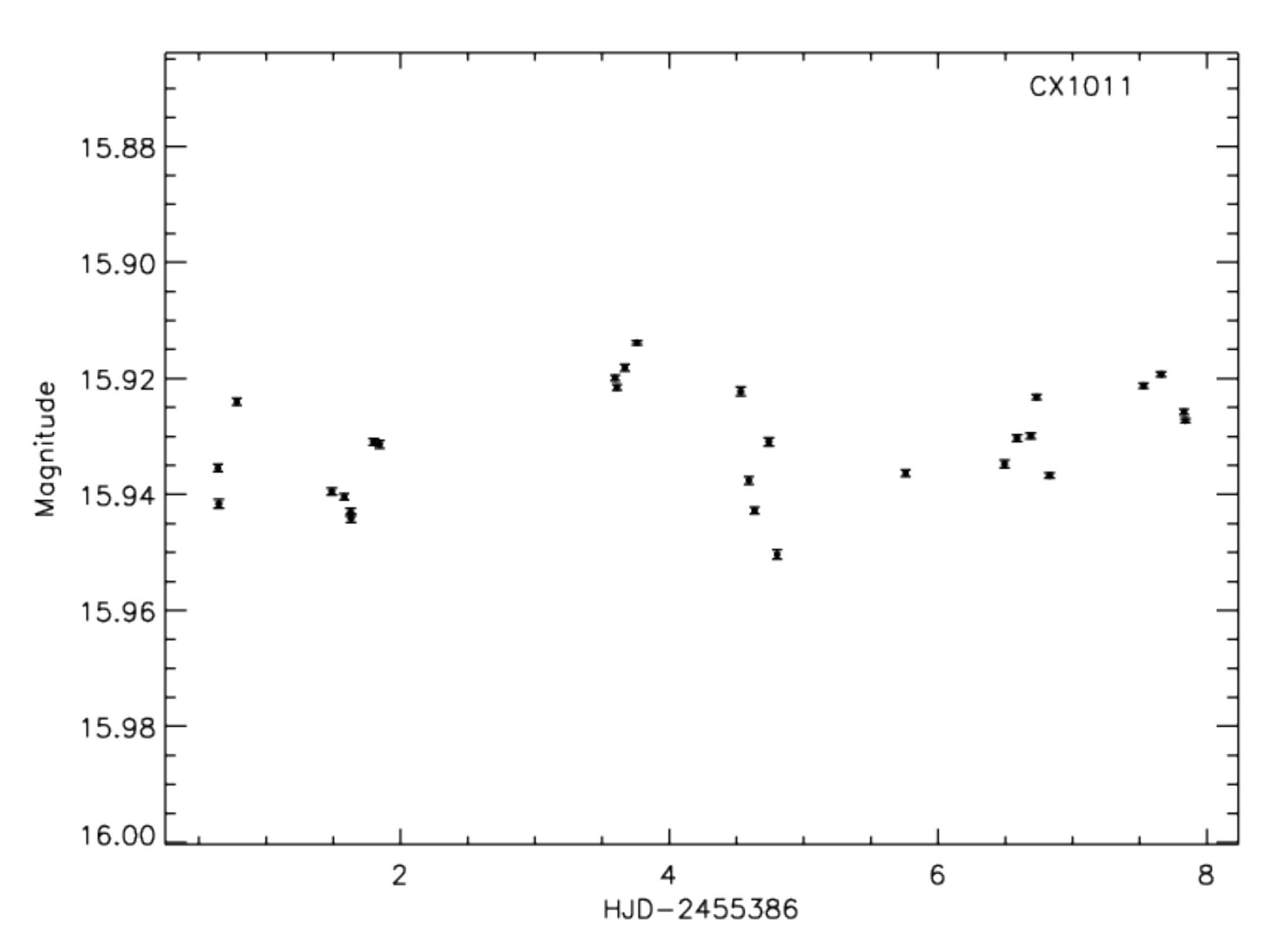}                                                                                                                \\
\end{array}$
\end{center}
\caption{ Optical light curves (continued). The Mosaic-II light curve
  of CX794 is also shown folded on the ephemeris of Section N.}
\label{licus}
\end{figure*}

\begin{figure*}
\includegraphics[width=4.0in, angle=180.0]{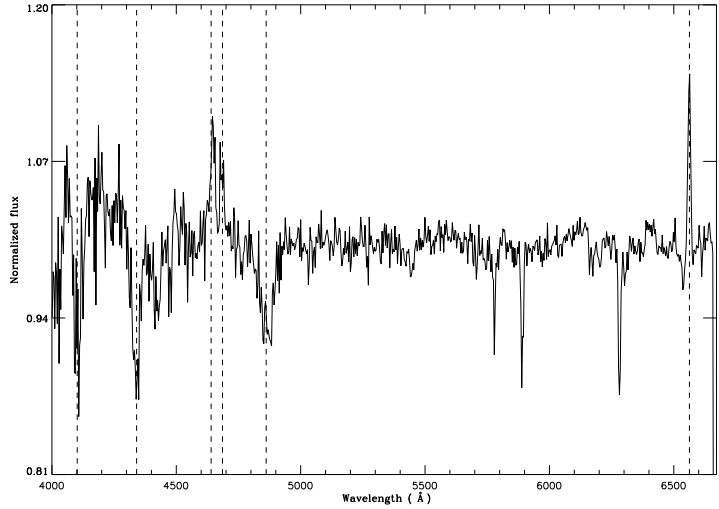}
\caption{Normalized IMACS spectrum of CX1011. Markers are for H$\delta$, H$\gamma$, the Bowen blend, He{\sc ii} $\lambda 4686$, H$\beta$ and H$\alpha$.  The three prominent absorption features in the region $\lambda\lambda 5800-6300$ are interstellar in origin.}
\label{licus}
\end{figure*}

\newpage

\begin{table*}
\caption{A   log of the VIMOS  and Mosaic-II observations.}
\label{log}
\begin{center}
\begin{tabular}{cccccc}
\hline

GBS        & VIMOS OB ID     & VIMOS OB ID   &  VIMOS   & Date UT              & Mosaic-II          \\
No           &  (spectroscopy)    & (pre-imaging)       & Quadrant  & (spectroscopy)   & N frames        \\
\hline
\hline
       &                  &                   &     &                          &     \\
CX28  & 453466     & 509361      & 4 & 2011-04-02    & 28        \\     
CX39  & 508808     & 453460      & 1 & 30-03-2011    & 29 \\  
CX45  & 575808    &    453494   & 1 & 04-07-2011     & 26 \\         
CX63  & 575585    &    453506   & 2 &  30-06-2011    & 35     \\
CX64  & 575148    &    453474   & 3 & 09-06-2011     & 18 \\
CX70  & 577698    &    453498   & 1 & 01-07-2011     & 35      \\
CX73  & 484756    &    453436   & 2 & 09-05-2010 & 27     \\
CX87  & 575612    &   453490    & 2 &  03-07-2011  & 21     \\
CX93   &  509202    &    453434     & 2 &  29-04-2011 & 35     \\
CX128 &  509361    &    453466      & 2 & 02-04-2011 & 18          \\
CX137 &  577805    &    544378     & 1 & 22-07-2011 & 4          \\
CX142 &  577743    &    544353     & 4 & 26-06-2011 &  16          \\
CX154 & 575008     &    453472     & 3  & 26-05-2011 &  26         \\
        & 575148    &    453474     &  2 & 09-06-2011 &        \\
CX207 &  508997    &    453430    &  2 & 02-05-2011 &  35        \\
        &  575272    &    453430     &  2 & 06-06-2011 &         \\
CX377  &  575026    &    453452     &  1  & 28-05-2011  & 35      \\
         &  575799    &    453452     &  1  & 23-07-2011  &   \\
CX446  & 508787     &    453428     &  1 & 02-05-2011  &  37     \\
CX522  & 575918     &    544379     &  4 & 04-08-2011   & ---     \\
CX772  & 484978     &    453440     &  2 & 11-05-2010  & 30   \\
CX781  &  575403    &    453446     &  2 & 27-07-2011   & 38      \\
CX794  & 509817     &    453456     &  1 &  03-04-2011   &  28   \\
        & 575275     &    453456      &  1 &  26-05-0211  &      \\                      
CX1004  & 577836     &    544354      &  2 &  28-06-2011   & 35     \\
CX1011  & 577836     &    544354      &  4  & 28-06-2011    & 28     \\
          &                    &                       &     &                         &     \\
\hline
\end{tabular}             
\end{center}
\end{table*}

\newpage

\begin{table*}
\caption{Spectroscopic measurements. An * indicates that the measured EW is most likely a lower limit due to contamination of unresolved star/s during the extraction of the spectra.}
\label{log}
\begin{center}
\begin{tabular}{lcccl}
\hline

GBS No       & $\alpha$ (deg)     & $\delta$ (deg)  & $r'$             &  Remarks        \\
\hline
            Specie $\lambda_0$           & RV  (km/s)       & EW (\AA)  & FWHM (\AA)      \\ 
\hline
\hline
                     &                               &                         &                     &                                \\       
%                    &   H{\sc i}                    &   $   \pm   $              &   $  \pm  $ &      $  \pm  $   &    &    \\      

CX28$^{*}$   &  264.94581                &  -27.302481           &  $16.89 \pm 0.20$             &                              \\  
\hline
                   He{\sc ii} 4686.75   &   $  420 \pm  30 $   &   $  23 \pm  3 $      &    $ 19.2 \pm  0.7 $   & High accretion rate CV    \\        
                   H{\sc i} 4861.327    &   $  320  \pm 20  $  &   $ 13.4  \pm 0.5$ &     $ 14.3 \pm  0.5 $  &           \\      
                   He{\sc ii} 5411.551   &   $  330 \pm  20 $   &   $  5.1  \pm 0.3 $ &     $ 22.3  \pm 0.7 $   &      \\      
                   He{\sc i} 5875.618  &    $  150 \pm  20  $ &   $  3.9 \pm  0.1$  &     $ 16.1  \pm  0.3$  &          \\      
                   H{\sc i} 6562.76      &   $  150 \pm  10 $   &   $ 22.5 \pm 0.9$ &      $ 19.8 \pm 0.2 $  &            \\      
                   He{\sc i} 6678.149  &   $  300  \pm  20  $ &   $  3.4 \pm  0.5 $ &     $  21.4 \pm  0.2$  &            \\      
                                                      &                                  &                              &                                     &           \\                                            
CX39$^{*}$  &    265.4167                & -27.293809              & $19.8 \pm 0.4$  &              \\    
\hline
                  H{\sc i} 6562.76     &   $ -5 \pm 4   $   &   $86 \pm 4$          &      $20 \pm  1$  & Outbursting CV                    \\          
                  He{\sc i} 6678.149   &   $ 151 \pm 9 $   &   $ 15 \pm  1 $        &     $21.2  \pm 0.8$  &             \\      
                  He{\sc i} 7065.188   &   $ -50  \pm  20 $ &   $  11.4  \pm 0.5 $ &     $ 19.4  \pm 0.7$  &            \\      
                                                     &                                &                                 &                                      &             \\          
CX44  &   268.92844                 &  -28.302561          &  $18.68 \pm 0.05$ &                               \\  
\hline
         H{\sc i}  4861.327            & $ -190  \pm 20  $  &   $  11 \pm 1 $        &      $  9 \pm 2 $    & Quiescent neutron star LMXB          \\      
         He{\sc i} 5875.618            &  $ -170  \pm 40  $ &   $  5.8 \pm 0.5 $  &      $  21.0 \pm  0.7 $   & or low accretion rate CV          \\      
          H{\sc i} 6562.76              &  $-240 \pm 9$     &  $20 \pm 1$            &      $13.4 \pm  0.1$     &                         \\            
                 &                                    &                                &                                 &                                                \\         
CX45  &   263.9106     &      -28.881169             &   $20.4 \pm 0.3$        &                               \\  
\hline
          H{\sc i} 6562.76     &  $-48 \pm 4 $      &  $40 \pm 3$            &      $21.6  \pm  0.3$  &  CV with accretion-dominated                   \\            
         He{\sc i} 6678.149  &   $  -2  \pm  16  $ &   $ 8.5  \pm 0.2   $ &     $  20.9 \pm  0.9$  &  optical spectrum         \\    
                                            &                              &                                 &                                     &        \\            
%        &                        &                                &                   &                    &    & 22-07-2011    \\
CX63$^{*}$  &  263.54742    &     -29.521676            &  $21.3 \pm 0.2$    &                               \\  
\hline
              H {\sc i} 4861.327   &   $ -63  \pm 4   $  &  $\gtrsim 35 \pm 2 $       &  $\gtrsim 9 \pm 1$   & CV with accretion-dominated    \\        
              He{\sc i} 5875.618 &   $ -13  \pm 5      $ &   $ 31  \pm 8  $ &     $ 12.1  \pm 0.9  $  & optical spectrum           \\      
              H{\sc i} 6562.76     & $ -45 \pm 2 $   &  $65 \pm 2$            &      $14.8  \pm  0.1$   &         \\     
              He{\sc i} 6678.149 & $ 13  \pm  4  $ & $9.8   \pm 0.1   $ &     $ 13.4  \pm  0.5 $  &            \\      
              He{\sc i} 7065.188 & $-60  \pm  10$ & $9.0   \pm 0.3  $ &     $  14.6 \pm  0.5 $  &            \\      
      
      H{\sc i} 8665.019 (P13)  &   $-30  \pm  10  $ &   $ 4.7  \pm 0.9   $ &     $ 20.8  \pm 0.9 $  &            \\      
      H{\sc i} 8750.473  (P12)  &   $-2 \pm 7   $ &   $ 8.8  \pm 0.4  $ &     $ 23.5  \pm 0.9 $  &            \\      
      H{\sc i} 8862.784 (P11)&   $ 50 \pm 30   $ & $ 14.6  \pm 2.4  $ &     $ 32   \pm 4 $  &            \\      
      H{\sc i}  9014.911 (P10) &   $ 50  \pm 20    $ &   $ 9.6  \pm 1.7  $ &     $ 22.0  \pm  0.9$  &            \\      
      H{\sc i}   9229.015 (P9) &   $-30  \pm 6    $ &   $ 18.1  \pm 0.7  $ &     $ 28.1   \pm 0.5  $  &            \\      
%       H{\sc i}  9545.972  (P8) &   $ ---    $ & $  --- $ &     $  ---  $  &            \\      
                                                  &                                  &                            &                                     &           \\                   
CX64 &   264.51153   &      -28.523984            &   $22.3 \pm 0.1$      &                               \\  
\hline
               H{\sc i} 6562.76     &  $ 212 \pm 2$                       &  $33 \pm 1$            &      $14.8  \pm  0.1$  & CV in a low state of accretion   \\
                                                 &                                  &                              &                                     & or CV in the Bulge region          \\                

                                                  &                                  &                              &                                     &           \\              
CX70  &   263.89611   &      -29.994897            &    $20.42 \pm 0.07$                            &                               \\  
\hline
        H{\sc i} 6562.76     &   $ -133 \pm 2 $            &  $15 \pm 1$            &      $20.3  \pm  0.2$ &  CV with accretion-dominated   \\    
        H{\sc i}  8598.392 (P14) &   $ -70 \pm 20  $ &   $ 2.9   \pm 0.1  $ & 25.0    $   \pm 0.2  $  &  optical spectrum            \\     
        H{\sc i} 8665.019 (P13)  &  $  61 \pm 7    $ &  $ 3.6  \pm  0.3  $ &  $ 23  \pm 2 $  &         \\       
        H{\sc i} 8750.473 (P12) &  $ -90 \pm 10  $ &   $ 4.0   \pm 0.2    $ & $ 27.6   \pm 0.8  $  &            \\       
       H{\sc i} 8862.784 (P11)   &  $ 30  \pm 40  $ &   $ 3.6  \pm 0.4  $ &     $ 25  \pm 1 $  &            \\      
                                                     &                                  &                              &                                     &           \\                        
CX73       &   266.19797   &      -27.017144             &    $18.62 \pm 0.08$                            &                               \\  
\hline
               H{\sc i} 6562.76     &   $ -116 \pm 4 $                     &  $11.4 \pm 0.2$            &      $18.8  \pm  0.4$    &   Unknown nature             \\            
                                      &                                  &                              &                                     &           \\                
                                    &                                  &                              &                                     &           \\                
\hline
\hline
\end{tabular}		      
\end{center}
\end{table*}

\newpage

\begin{table*}
\caption{Spectroscopic measurements (continued).}
\label{log}
\begin{center}
\begin{tabular}{llcccl}
\hline

GBS No     & & $\alpha$ (deg)     & $\delta$ (deg)  & $r'$             &  Remarks        \\
\hline
            Specie  $\lambda_0$         &   & RV  (km/s)       & EW (\AA)  & FWHM (\AA)      \\ 
\hline
\hline
                   &  &                               &                         &                     &                                \\       
CX87        &  & 264.20005             & -29.611037    &  $\sim 23.3$            &                             \\  
\hline
               H{\sc i} 6562.76     & &  $ 50  \pm 50   $          & $ 24  \pm 5   $    & $ 27  \pm 3   $   & Outbursting CV                       \\        
                                                 &  &                             &                      &           &           \\            

CX93  & &   266.18661   &      -26.058373              &  $17.07  \pm 0.03$             &                               \\  
\hline
          H{\sc i} 4861.327   & & $ -70  \pm 20   $ &  $ 27 \pm 2 $   &      $ 16.7   \pm  0.2 $   &  Low accretion rate CV               \\    
          He{\sc i} 5875.618 & & $ -60  \pm 10   $ &   $ 5.0  \pm  0.5  $ &     $ 15.5  \pm 0.4 $  &            \\      
          H{\sc i} 6562.76     & & $-17  \pm 4 $  &  $18.4 \pm 0.3$            &      $16.7  \pm  0.3$   &                 \\    
          He{\sc i} 6678.149 & & $ -64  \pm 2   $ & $1.6   \pm 0.2   $ &     $ 18.2  \pm 0.4 $  &            \\       
                                                  & &                                  &                              &                                     &           \\                   
CX128$^{*}$   & &   265.11822   &      -27.193451             &   $20.9 \pm 0.3$                             &                               \\  
\hline
            H{\sc i} 4861.327    & &   $ -180  \pm  60  $   &  $ 91 \pm 5 $                    &      $ 26.2     \pm  0.8$  & CV with disc-dominated optical      \\       
            He{\sc i} 5875.618    & &   $ -30   \pm 20   $   &  $ 56 \pm 5 $                    &      $ 33     \pm 1 $   & spectrum, likely short orbital                   \\            
            H{\sc i} 6562.760      & &   $ -70 \pm 4$           &  $197 \pm 7$                    &      $31.0  \pm  0.2$   & period dwarf nova                  \\             
            He{\sc i} 6678.149    & &   $ -180  \pm 10   $  &  $ 20 \pm 1$                      &      $ 35.5     \pm  0.4 $    &              \\           
            He{\sc i} 7065.188    & &   $ -110  \pm 40   $  &  $ 20 \pm 1 $                    &      $ 44     \pm 4 $     &                   \\    

          H{\sc i}  8545.383 (P15) & & $ -180  \pm 10 $ &   $ 69  \pm 7  $ &     $ 61  \pm 4 $  &            \\      
          H{\sc i} 8665.019 (P13)  & & $ -20 \pm 10    $ &   $ 48  \pm 4  $ &     $ 46.6  \pm 0.7 $  &            \\      
          H{\sc i}  9014.911 (P10)  & & $ -130 \pm 30 $  &   $ 45  \pm 2  $ &     $ 58  \pm 3 $  &            \\          
          H{\sc i}   9229.015 (P9)  & &  $ -40 \pm 20   $ &   $100   \pm 30  $ &  $  65 \pm  3$  &            \\      
     
                                                  & &                                  &                              &                                     &           \\            

CX137$^{*}$  & &   268.97196   &      -28.276046            &   $16.14 \pm 0.01$                             &                               \\  
\hline
       H{\sc i} 6562.76     & &   $ -9 \pm 9$  &   $\gtrsim 6.5 \pm 0.2$            &      $\gtrsim 23.4  \pm  0.4$    & Low accretion rate CV or       \\
                                                  & &                                  &                              &                                     &  qLMXB?, P$_{orb}$ = 0.431 d         \\                       
CX142         & &   266.01566   &     -31.384838            &  $22.3 \pm 0.2$                              &                               \\  
\hline
          H{\sc i} 4861.327    & & -90   $  \pm 10  $                       &  $ 120 \pm 80 $            &      $14   \pm 2  $   & CV with accretion-dominated                 \\    
            He{\sc i} 5875.618  & &  ---       &     --- &  ---  &  optical spectrum         \\      
            H{\sc i} 6562.76      & &   $ -69 \pm 1 $   &  $59 \pm 4$             &  $12.4  \pm  0.2$    &                \\       
            He{\sc i} 6678.149  & & $-28  \pm 9$ &   $ 4.9  \pm 0.3  $ &     $ 9  \pm 1 $  &            \\          
            He{\sc i} 7065.188  & & $-42  \pm 8   $        &  $6.2 \pm 0.4$   &      $ 16.0     \pm 0.9 $     &                   \\    

           Ca{\sc ii}  8498.02 + P16   & &   $0  \pm 10   $ &  $ 8.0  \pm 0.4  $ &     $10.4   \pm 0.3 $  &            \\      
           Ca {\sc ii} 8542.09 + P15   & &$-39  \pm 2   $  &   $ 10.8  \pm 0.3  $ &     $ 1.1  \pm 0.3 $  &            \\      
           H{\sc i}  8598.392 (P14)    & & $ -10  \pm 20   $ &   $ 4.2  \pm 0.6   $ &     $ 13.3  \pm 0.5 $  &            \\     
           Ca{\sc ii}  8662.14  + P13  & &   $ -7 \pm 2   $ &   $ 9.2  \pm 0.3  $ &     $ 11.1   \pm 0.4 $  &            \\       
           H{\sc i} 8750.473 (P12)     & &   $-37  \pm 7    $ &   $ 4.8  \pm 0.2  $ &     $ 11.0   \pm 0.3 $  &            \\      

          H{\sc i} 8862.784 (P11) & &   $ -68 \pm  7   $ &   $ 7  \pm  1 $ &     $ 14  \pm 2 $  &            \\      
          H{\sc i}  9014.911 (P10) & &   $-54  \pm  7   $ &   $ 8.0  \pm 0.4  $ &     $12.8   \pm 0.7 $  &            \\      
          H{\sc i}   9229.015 (P9) & &   $-31  \pm 7   $ &   $ 11  \pm 1  $ &     $ 17   \pm 1 $  &            \\      
          H{\sc i}  9545.972  (P8)  & &   $-44  \pm 7   $ &   $ 10.2  \pm 0.6   $ &     $ 13.6  \pm 0.2 $  &            \\      

                                                  & &                                  &                              &                                     &           \\            
CX154$^{*}$   & &  264.66119    &     -28.594338           &   $21.4 \pm 0.1$                             &                               \\  
\hline

            H{\sc i} 6562.76 & 1 & $ 95 \pm 9$       &   $   8  \pm 2 $          &  $ 11.3 \pm  0.8 $  & Low accretion rate CV          \\
                                         & 2 & $142 \pm 4$     &  $  7.6  \pm 0.4 $     &  $ 8.5 \pm  0.5 $    & or qLMXB            \\
            He{\sc i} 6678.149   & 2 &   $ 10 \pm 9 $ &   $ 2.8  \pm 0.4  $ &   $ 12.7  \pm 0.4 $  &   (1,2: 2011 May 26, Jun 9)       \\          
%                                               & &   $  \pm    $ &   $   \pm   $ &     $   \pm  $  &            \\                    
                                                  & &                                  &                              &                                     &           \\            
CX207$^{*}$    & &   266.60619   &      -26.526381            &  $21.1 \pm 0.9$                              &                               \\  
\hline
            He{\sc i} 5875.618 & 1 & $ -370  \pm 20 $ &   $ 14  \pm 2  $ &  $  22 \pm 1 $  &  Eclipsing high accretion rate CV,          \\      
                                             & 2 &  $ -210  \pm 50 $ &   $ 21  \pm 2  $ &  $  41 \pm  4 $  & most likely a magnetic (Polar) CV           \\      
            H{\sc i} 6562.76    & 1 &  $ -558 \pm 4$                     &  $107  \pm  8$            &      $ 37.7 \pm  0.7 $   & (1,2: 2011 May 2, Jun 6)       \\    
                                             & 2 & $ -358 \pm 4$                       &  $  83  \pm 9 $            &      $ 35.1 \pm  0.6 $  &           \\    
              He{\sc i} 6678.149  & 1 &   $  -330 \pm 10  $ &   $ 10   \pm 1  $ &     $ 22.0  \pm 0.8 $  &            \\      
                                                 & 2 &   $ -200 \pm 100 $ &   $ 7  \pm 1  $  &     $ 32  \pm 5 $  &            \\       
               He{\sc i} 7065.188 & 1 &   $  -250 \pm 10  $  &  $ 6.8 \pm 0.6 $ & $ 22 \pm  2 $     &                   \\    
                                                 & 2 &   $  -190 \pm 30  $  &  $ 6.5 \pm 0.3 $ & $ 38  \pm 2 $     &                   \\    
\hline
\end{tabular}		      
\end{center}
\end{table*}

\begin{table*}
\caption{Spectroscopic measurements (continued).}
\label{log}
\begin{center}
\begin{tabular}{lccccl}
\hline

GBS No   &       & $\alpha$ (deg)     & $\delta$ (deg)  & $r'$             &  Remarks        \\
\hline
            Specie  $\lambda_0$ &         & RV  (km/s)       & EW (\AA)  & FWHM (\AA)      \\ 
\hline
\hline
                     &  &                               &                         &                     &                                \\       
CX207$^{*}$    & &    (continued)  &                         &                     &                               \\  
\hline
%               H{\sc i}  8545.383 (P15) & & $ -214  \pm 11 $ &   $ 69  \pm 7  $ &     $ 61  \pm 4 $  &            \\                    
               H{\sc i} 8665.019 (P13)   &1  &   $ -280  \pm 10   $ &   $ 3.2  \pm 0.4  $ &     $ 16  \pm 1 $  &            \\       
%                                                        & 2  &  $       \pm    $                         &  $ \pm $                    &      $      \pm  $     &                   \\          
               H{\sc i} 8862.784 (P11)   & 1  &  $ -200  \pm 30    $ &   $ 8   \pm 3  $ &     $ 29  \pm 3 $  &       \\      
%                                                        & 2 &  $       \pm    $                         &  $ \pm $                    &      $      \pm  $     &                   \\    
              H{\sc i}  9014.911 (P10)   & 1 &  $ -280  \pm 20  $ &   $ 7.7  \pm 0.9  $ &     $  19.5 \pm 0.8 $  &            \\      
%                                                       &  2 &  $       \pm    $                         &  $ \pm $                    &      $      \pm  $     &                   \\    
              H{\sc i}   9229.015 (P9)   & 1   &  $ -330 \pm 30   $ &   $ 23  \pm 6 $ &     $39   \pm 8 $  &            \\      
                                                          & 2   &  $ -180 \pm 30   $ &   $ 33  \pm 3 $ &     $ 55  \pm 3 $  &            \\      
             H{\sc i}  9545.972  (P8)    & 1   &    $ -460  \pm 30 $ & $ 37  \pm 9  $ &     $ 46  \pm 2 $  &            \\      
%                                                       & 2  &  $  \pm    $ &   $   \pm   $ &     $   \pm  $  &            \\      
                                          & &                                   &                              &                                     &           \\            
CX377$^{*}$  & &     265.8189     &      -27.760333              &         $18.86 \pm 0.05$                       &                               \\  
\hline
             H{\sc i} 6562.76     & 1  &         $ 30 \pm 30$                     &  $  6.6 \pm  0.3 $        &      $ 29 \pm  2 $        & Low accretion rate CV      \\    
                                               & 2 &          $ -39 \pm 4$                   &  $  6.4 \pm  0.2 $         &      $ 27.3 \pm  0.6 $   &  or qLMXB,    \\       
                                               &  &                                   &                              &                                     &  (1,2: 2011 Jul 23, Aug 28)           \\                                              &  &                                   &                              &                                     &           \\                    
CX446  & &     266.61317   &      -25.83124                &      $21.2 \pm 0.2$                          &                             \\  
\hline
               H{\sc i} 6562.76     & &          $ -10\pm 20$                    &  $ 50 \pm 4 $               &      $ 49 \pm  1 $     & High inclination CV   \\        
                                                 &  &                                  &                              &                                     & or qLMXB           \\            
CX522  & &      268.63525  &       -28.488483              &  ---                              &                               \\  
\hline
              H{\sc i} 4861.327   & &   $ 42.0 \pm 0.1  $  &  $ 6 \pm 1 $  &      $ 10  \pm 1  $   & CV of unkown type                \\    
              He{\sc i} 5875.618 & &   $ 30 \pm 70  $      &   $ 1.2  \pm 0.2  $ &  $ 16  \pm 3 $  &           \\      
              H{\sc i} 6562.76     & & $ -84 \pm 1$               &  $  23.3 \pm 0.9 $        &      $ 21.1 \pm  0.2 $    &        \\        
              He{\sc i} 6678.149  &  & $ -80 \pm 40   $ &   $ 3.2  \pm 0.3   $ &     $ 25  \pm 2  $  &            \\      
%              H{\sc i}   9229.015 (P9) & &     $  \pm    $ &   $   \pm   $ &     $   \pm  $  &            \\      
%              H{\sc i}  9545.972  (P8) & &     $  \pm    $ &   $   \pm   $ &     $   \pm  $  &            \\      
                                                  & &                                   &                              &                                     &           \\

CX772  & &     266.02105  &       -26.533196             &   $22.2 \pm 0.3$                        &                               \\  
\hline
              H{\sc i} 6562.76     & &        $ -40 \pm 20$    &  $  24 \pm 4 $            &      $ 22 \pm  4 $     & Unknown nature                       \\
                                                & &                                    &                              &                                     &           \\            
CX781  & &      265.79638  &       -27.272705              &      $17.89 \pm 0.09$                           &                               \\  
\hline
              H{\sc i} 6562.76     & &   $ 102 \pm 1  $  &  $  5.0 \pm 0.2 $  &   $ 10.1 \pm  0.2 $   & CV with accretion-dominated   \\         
              He{\sc i} 6678.149 & &   $  10 \pm 30 $ &   $ 2.3  \pm 0.2  $ &   $ 24  \pm 2 $  & optical spectrum          \\      
%              H{\sc i}   9229.015 (P9) & &     $  \pm    $ &   $   \pm   $ &     $   \pm  $  &            \\      
                                                  & &                                    &                              &                                     &           \\            
CX794  & &      265.42652  &       -27.975019            &  $20.1 \pm  0.5$                              &                               \\  
\hline
              H{\sc i} 4861.327   & &      $ -10 \pm 10  $  &  $ 13 \pm 1 $ &      $ 9.8  \pm 0.7  $   & High accretion rate CV                \\    
%              He{\sc i} 5875.618 & &    $  \pm    $ &   $   \pm   $ &     $   \pm  $  &            \\      
              H{\sc i} 6562.76     & &       $-69 \pm 4$                      &  $ 23 \pm 1  $            &      $ 23.0 \pm 0.4  $ & eclipsing, P$_{orb}$ = 0.11786 d          \\  
              He{\sc i} 6678.149  & &     $ -80 \pm 20   $ &   $ 5  \pm 2  $ &     $ 19  \pm 2 $  &            \\      
%              He{\sc i} 7065.188 & &   $   \pm    $ & $   \pm    $ &     $   \pm  $  &            \\      
%              H{\sc i}   9229.015 (P9) &   &   $  \pm    $ &   $   \pm   $ &     $   \pm  $  &            \\      
%              H{\sc i}  9545.972  (P8) & &    $  \pm    $ &   $   \pm   $ &     $   \pm  $  &            \\      
                                                                & &                                    &                              &                                     &           \\            
CX1004  & &     266.59799  &       -31.09717               &   $20.75 \pm 0.04$                             &                               \\  
\hline
               H{\sc i} 6562.76     & &      $ -210 \pm 20 $        &  $ 32.9 \pm 0.4 $        &      $ 46.4 \pm  0.4 $     & Low accretion rate eclipsing CV   \\     
                                                 & &                                    &                              &                                     & or qLMXB           \\           
CX1011   & &     266.51968   &      -31.289612            &    $15.93 \pm 0.01$                            &                               \\  
\hline 
              H{\sc i} 6562.76     & 1 &      $ 18 \pm 6$             &  $ 2.36  \pm 0.04 $     &      $  13.59 \pm 0.04 $           & High accretion rate CV,       \\
                                                & 2 &     $ 93 \pm 5 $             &  $ 2.14  \pm  0.07 $    &      $ 15.1 \pm  0.2 $     &  nova-like       \\    
              He{\sc i} 6678.149 & 2 &     $ 40.2\pm  10   $ &   $ 0.46  \pm 0.04  $ &     $ 13.3  \pm 0.5  $  &    (1,2: 2011 May 13, Jun 28)        \\      
              H{\sc i}   9229.015 (P9) & 2  & $ 50  \pm  10   $ &   $ 1.5  \pm 0.1  $ &     $ 16.7  \pm 0.5  $  &            \\                    
\hline
\end{tabular}		      
\end{center}
\end{table*}

\begin{figure*}
\begin{center}$
\begin{array}{ccc}
\vspace{1cm}
%\hspace{-2.0cm}
\includegraphics[width=2.5in, angle=0.0]{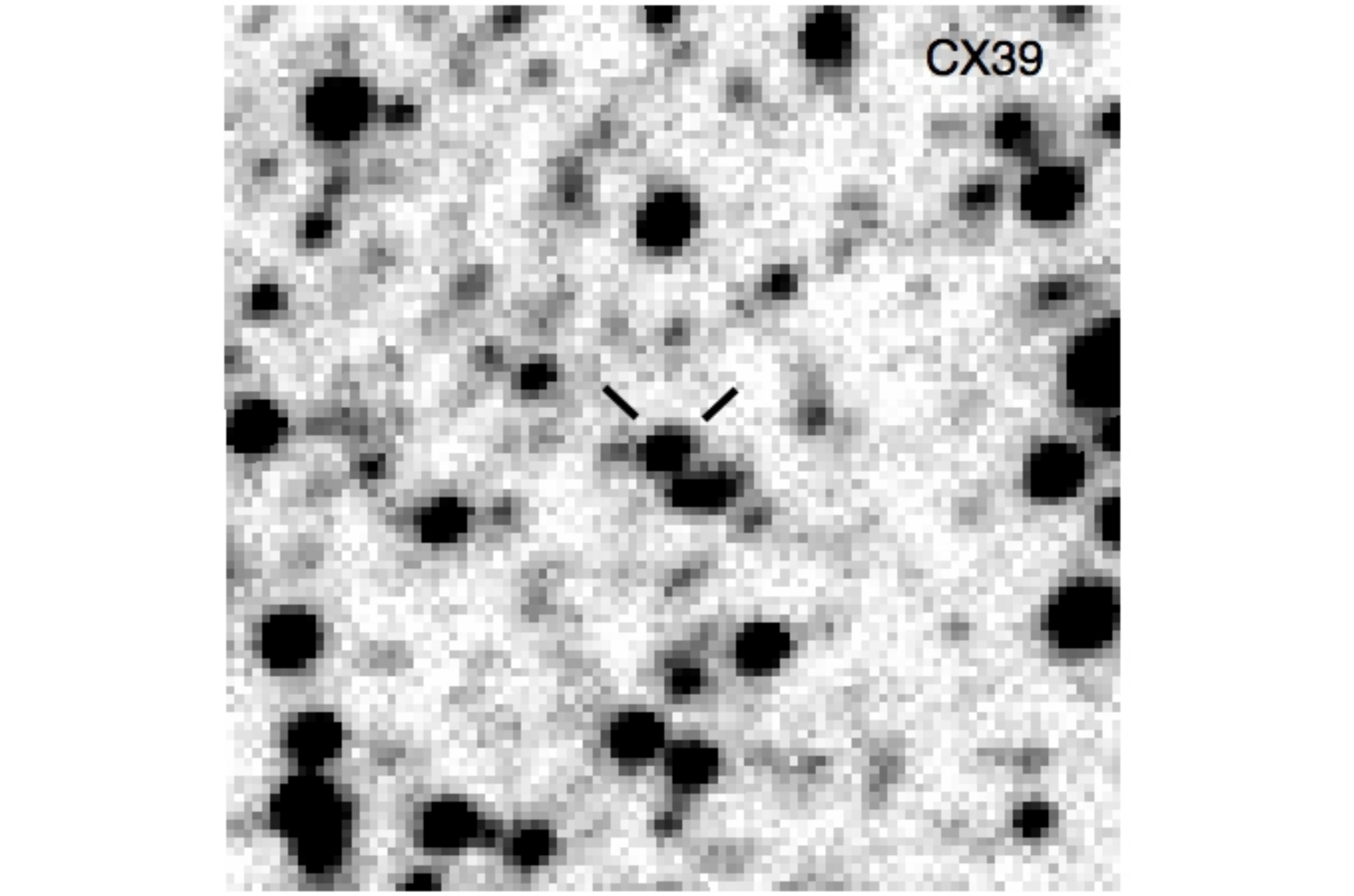} &
%\hspace{-5.0cm}
\includegraphics[width=2.5in, angle=0.0]{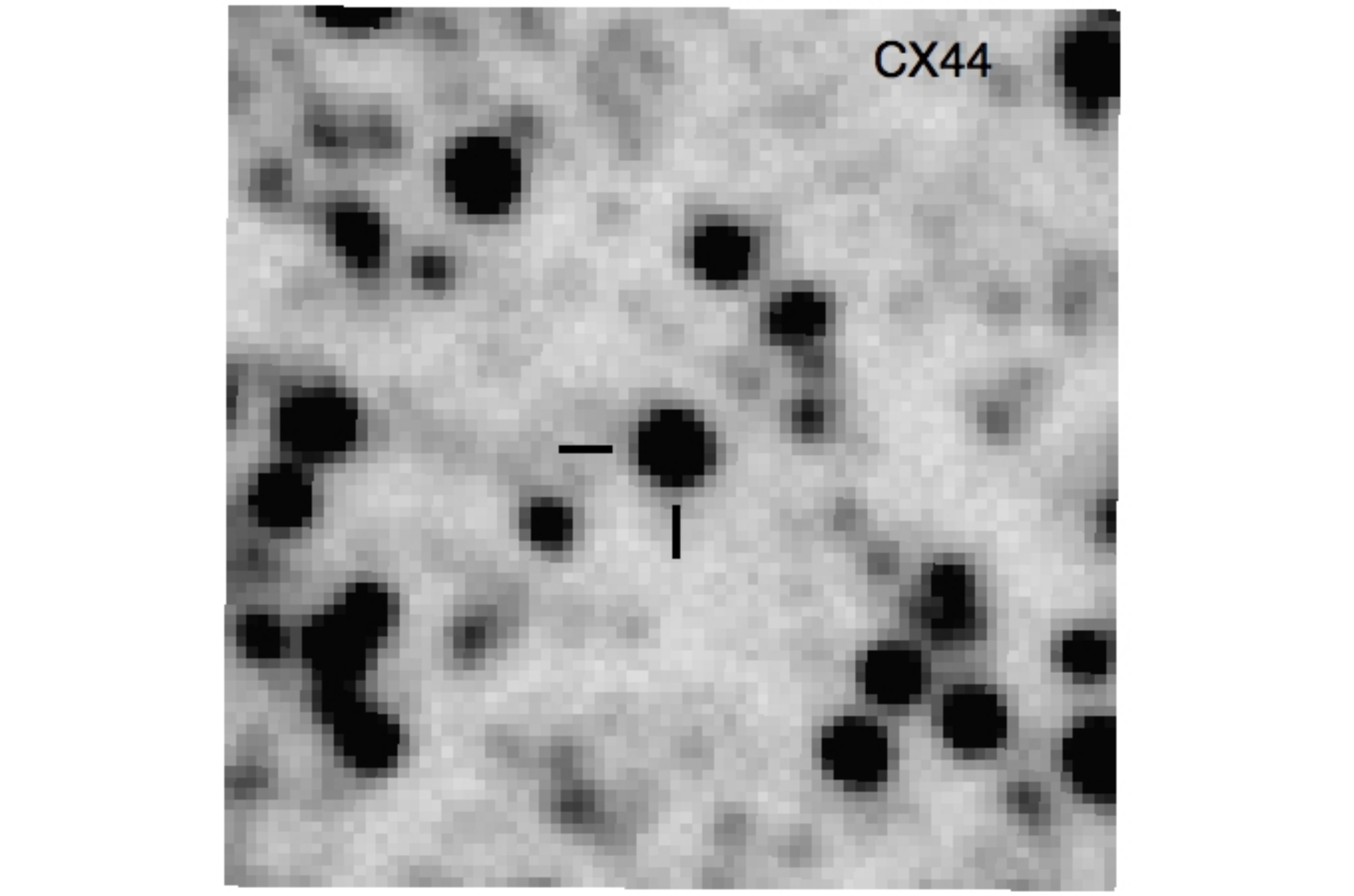} &
%\hspace{-8.0cm}
\includegraphics[width=2.5in, angle=0.0]{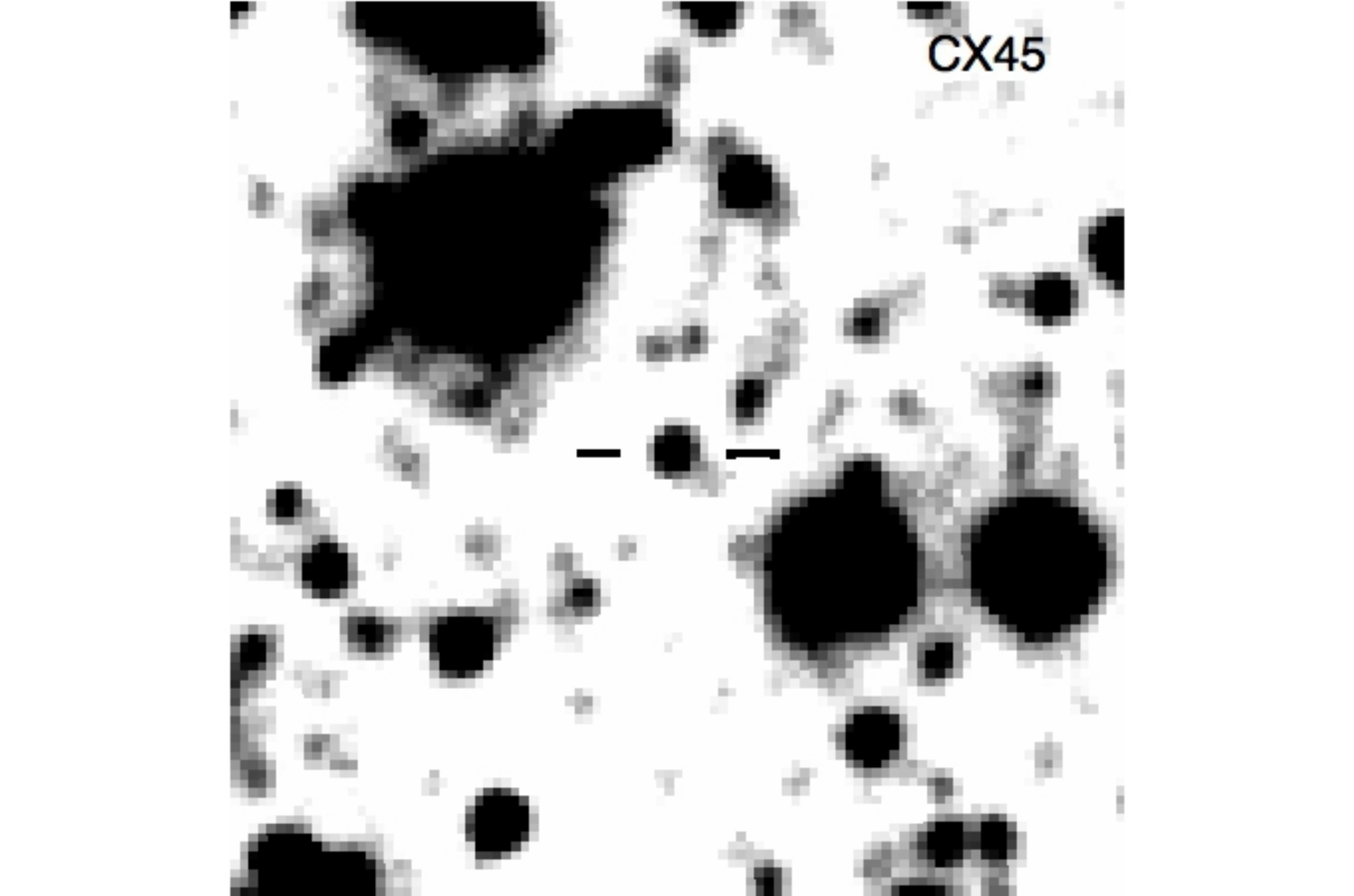} \\
\vspace{1cm}
%
%\hspace{-2.0cm}
\includegraphics[width=2.5in, angle=0.0]{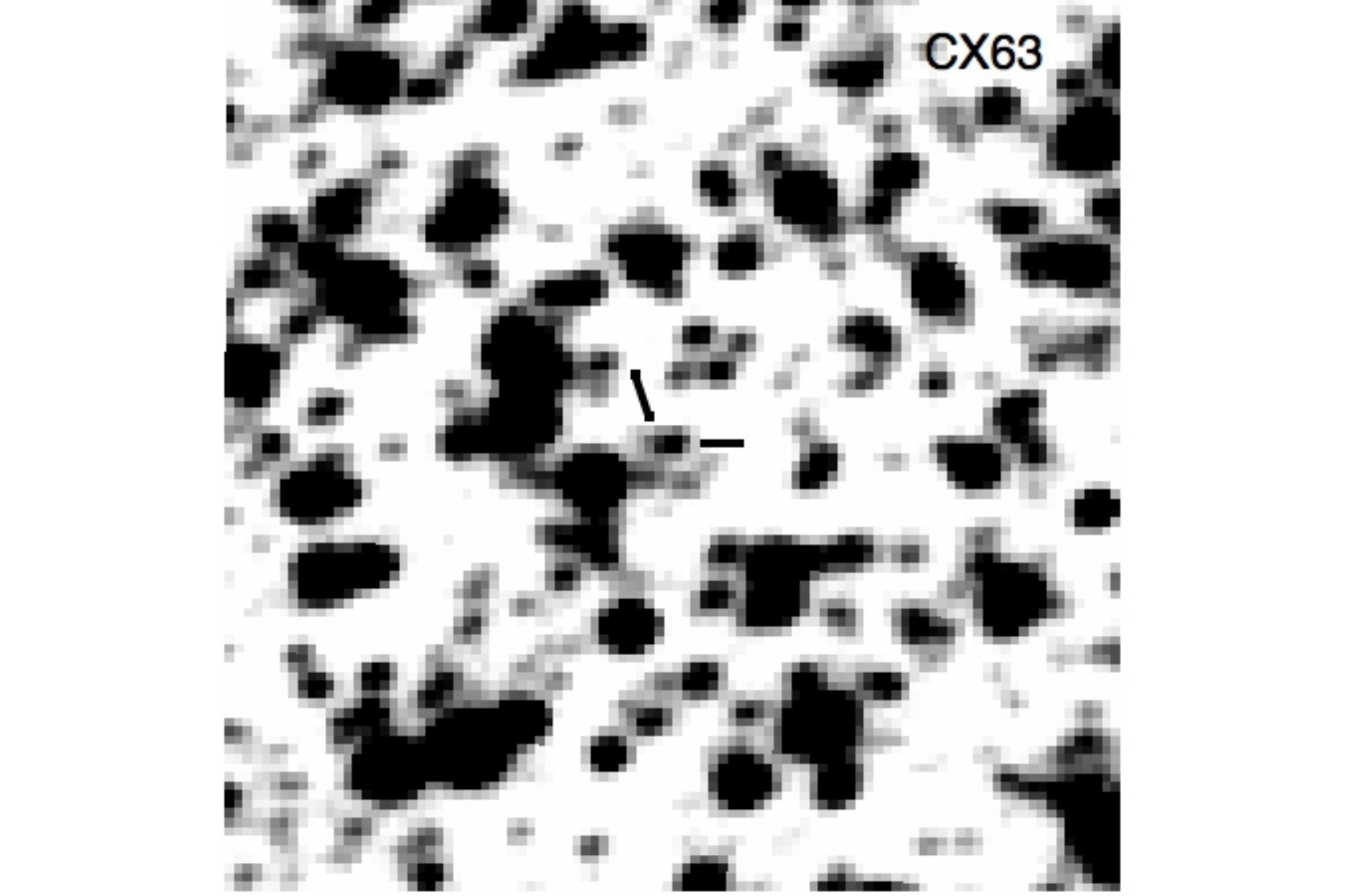} &
%\hspace{-5.05cm}
\includegraphics[width=2.5in, angle=0.0]{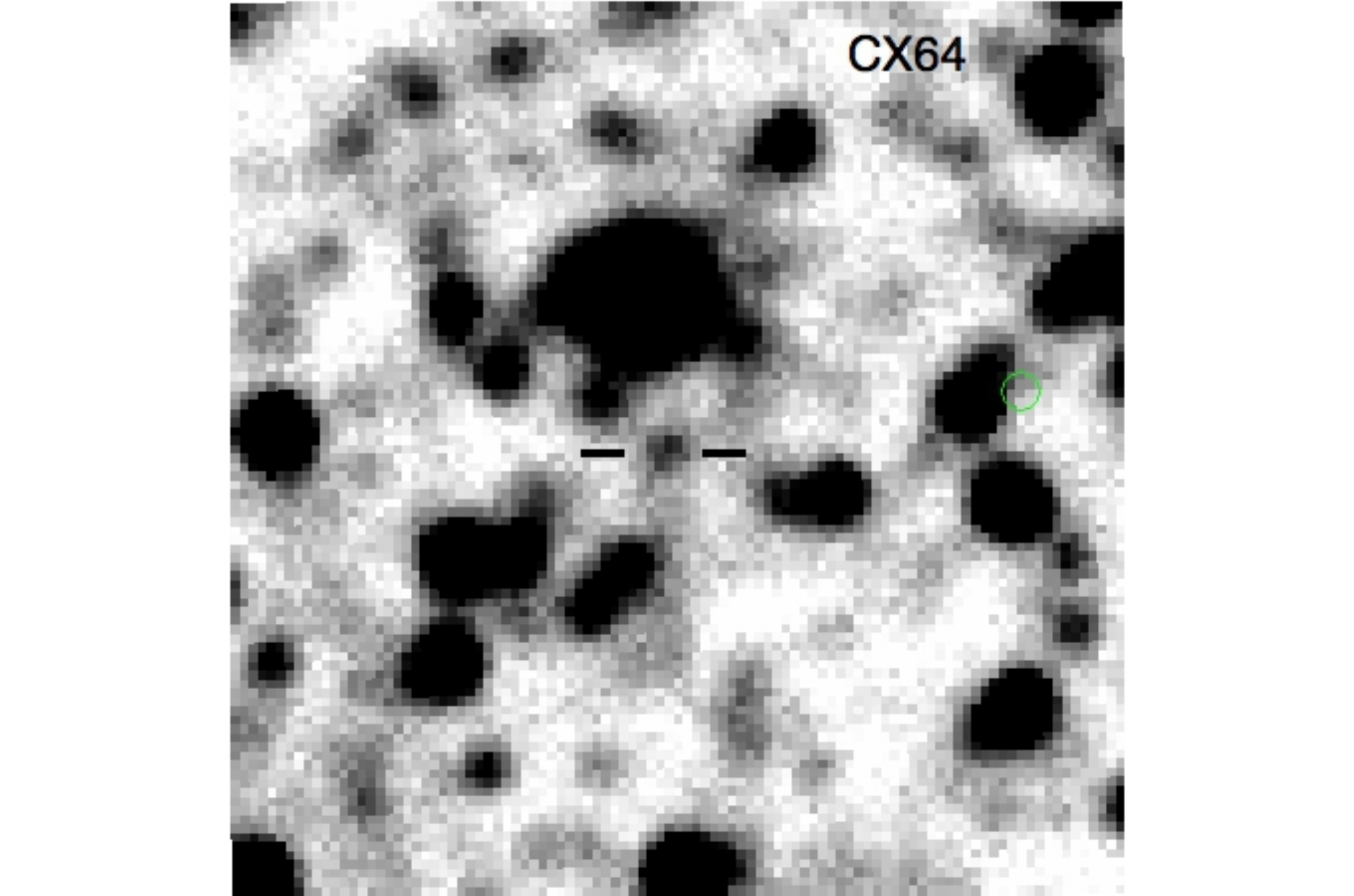} &
%\hspace{-8.0cm}
\includegraphics[width=2.5in, angle=0.0]{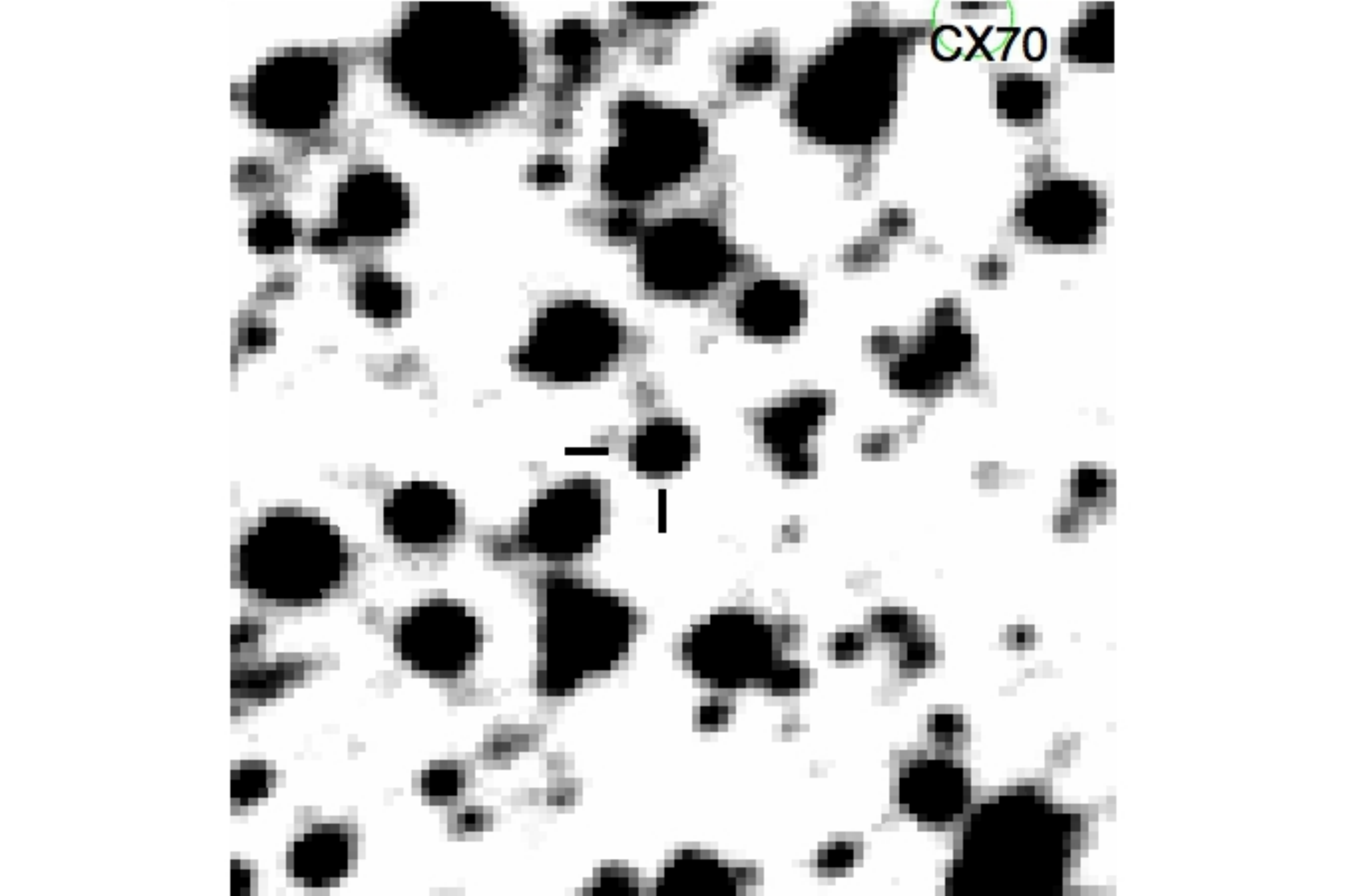} \\
\vspace{1cm}
%
%\hspace{-2.0cm}
\includegraphics[width=2.5in, angle=0.0]{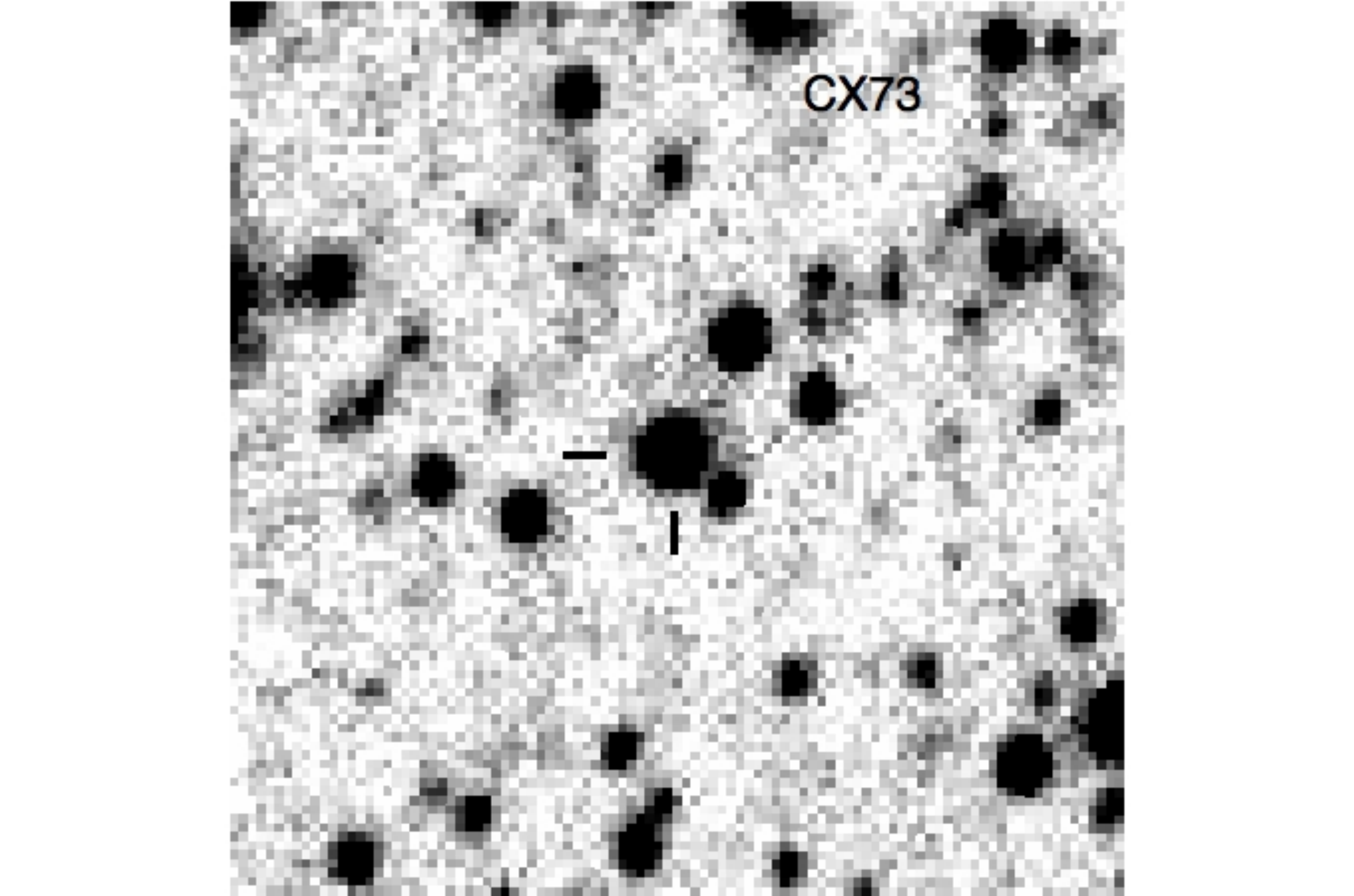} &
%\hspace{-5.0cm}
\includegraphics[width=2.5in, angle=0.0]{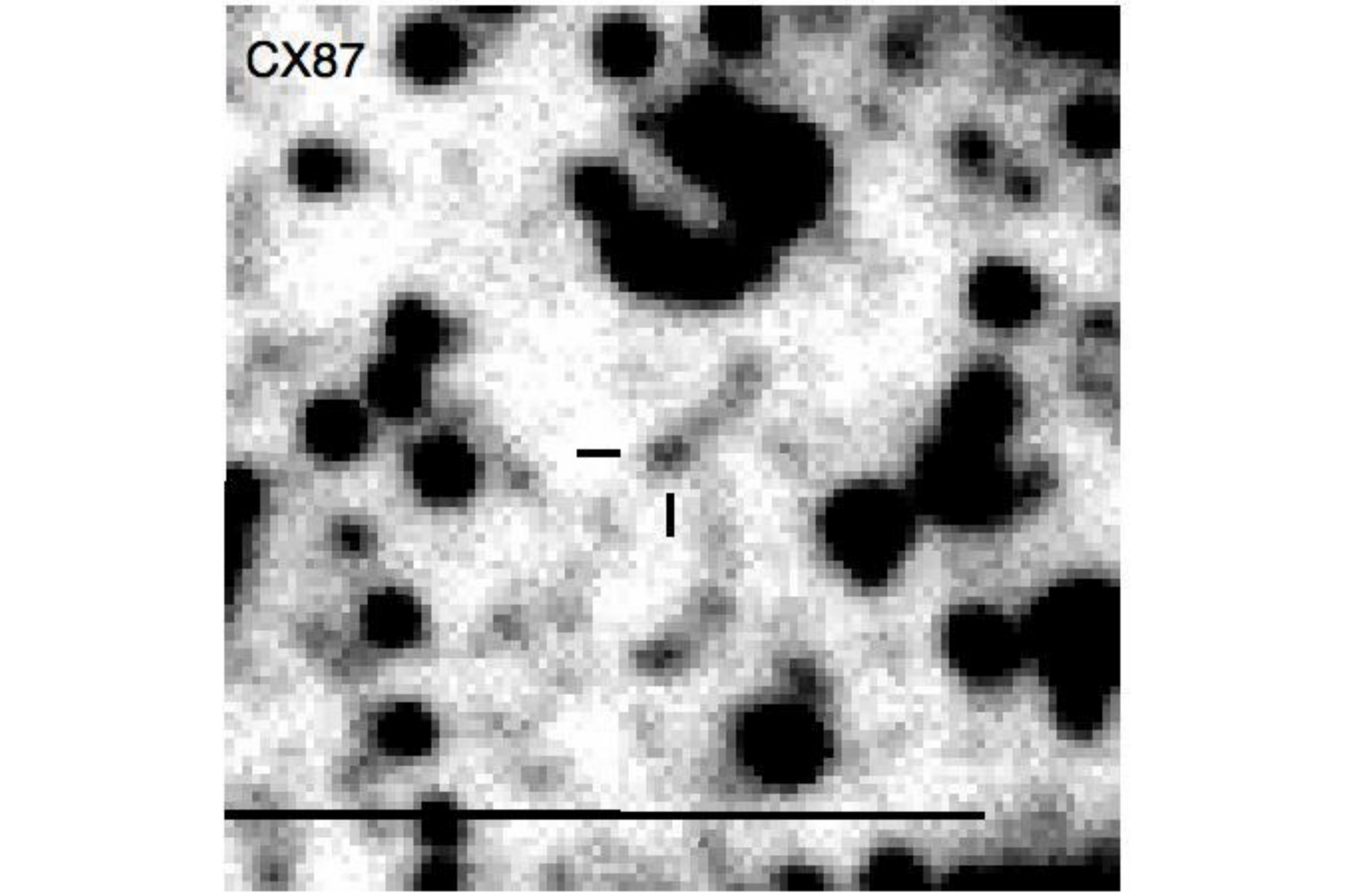} &
%\hspace{-8.0cm}
\includegraphics[width=2.5in, angle=0.0]{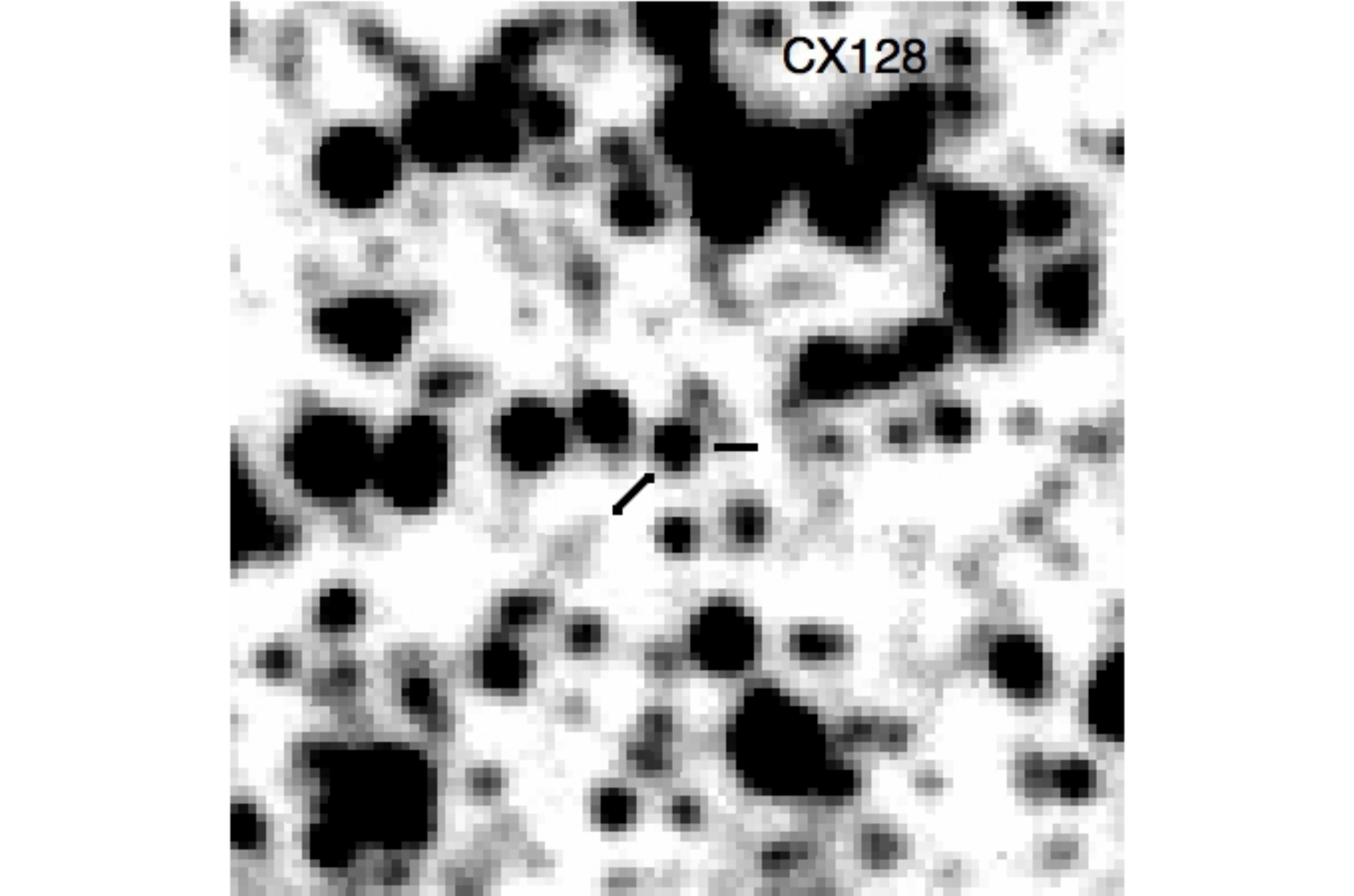} \\
\vspace{1cm}
%
%\hspace{-2.0cm}
\includegraphics[width=2.5in, angle=0.0]{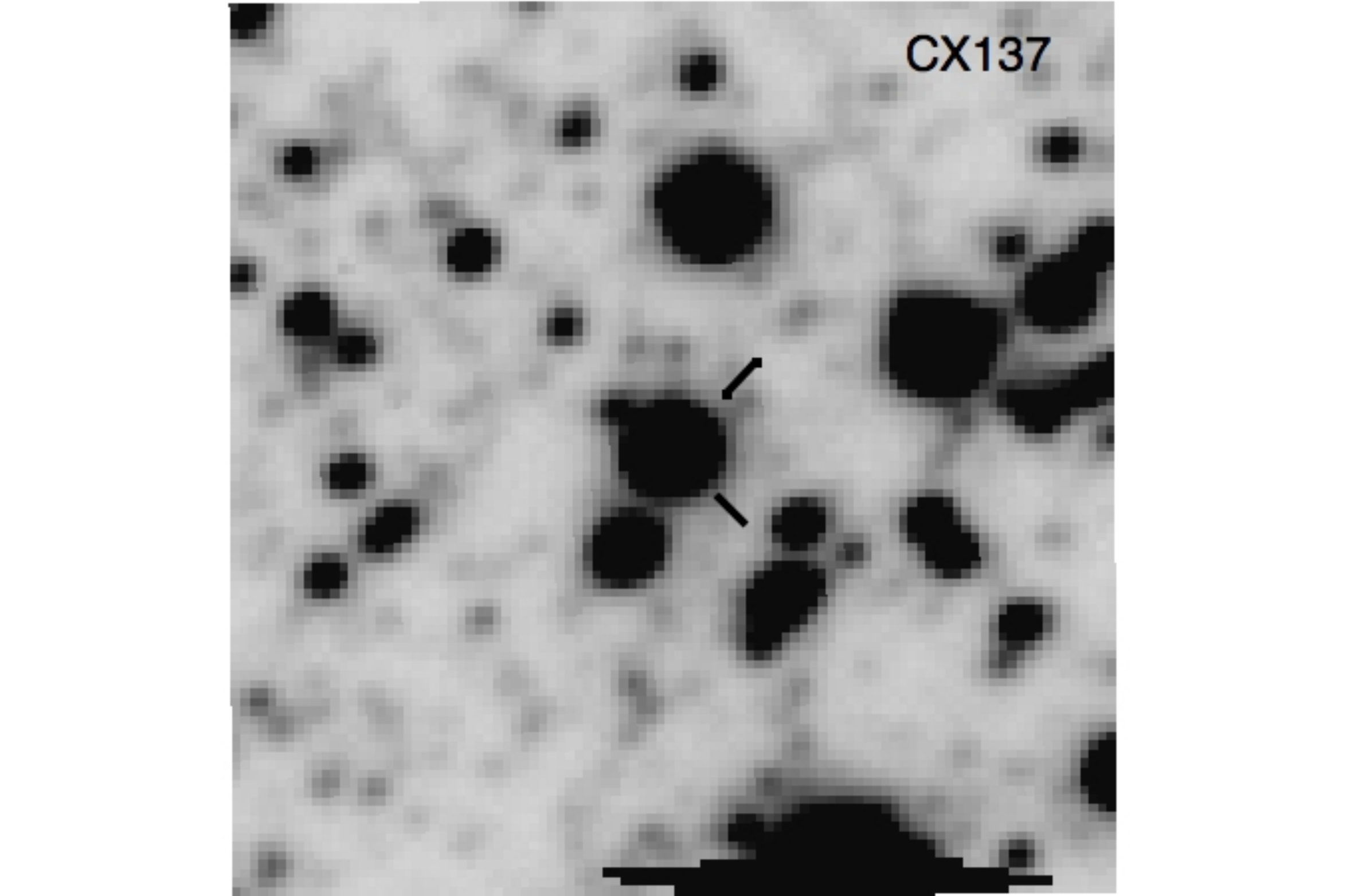} &
%\hspace{-5.0cm}
\includegraphics[width=2.5in, angle=0.0]{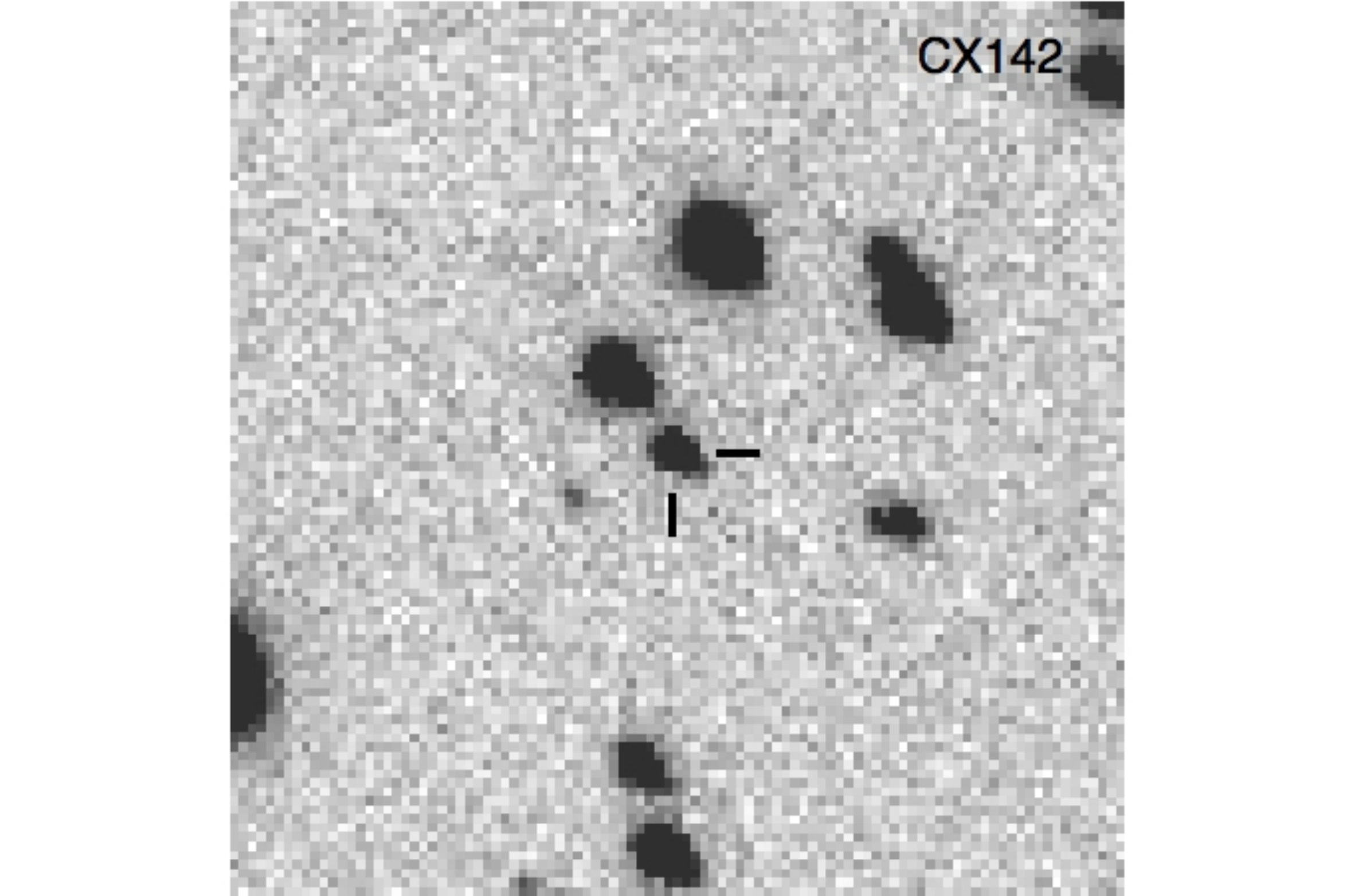} &
%\hspace{-8.0cm}
\includegraphics[width=2.5in, angle=0.0]{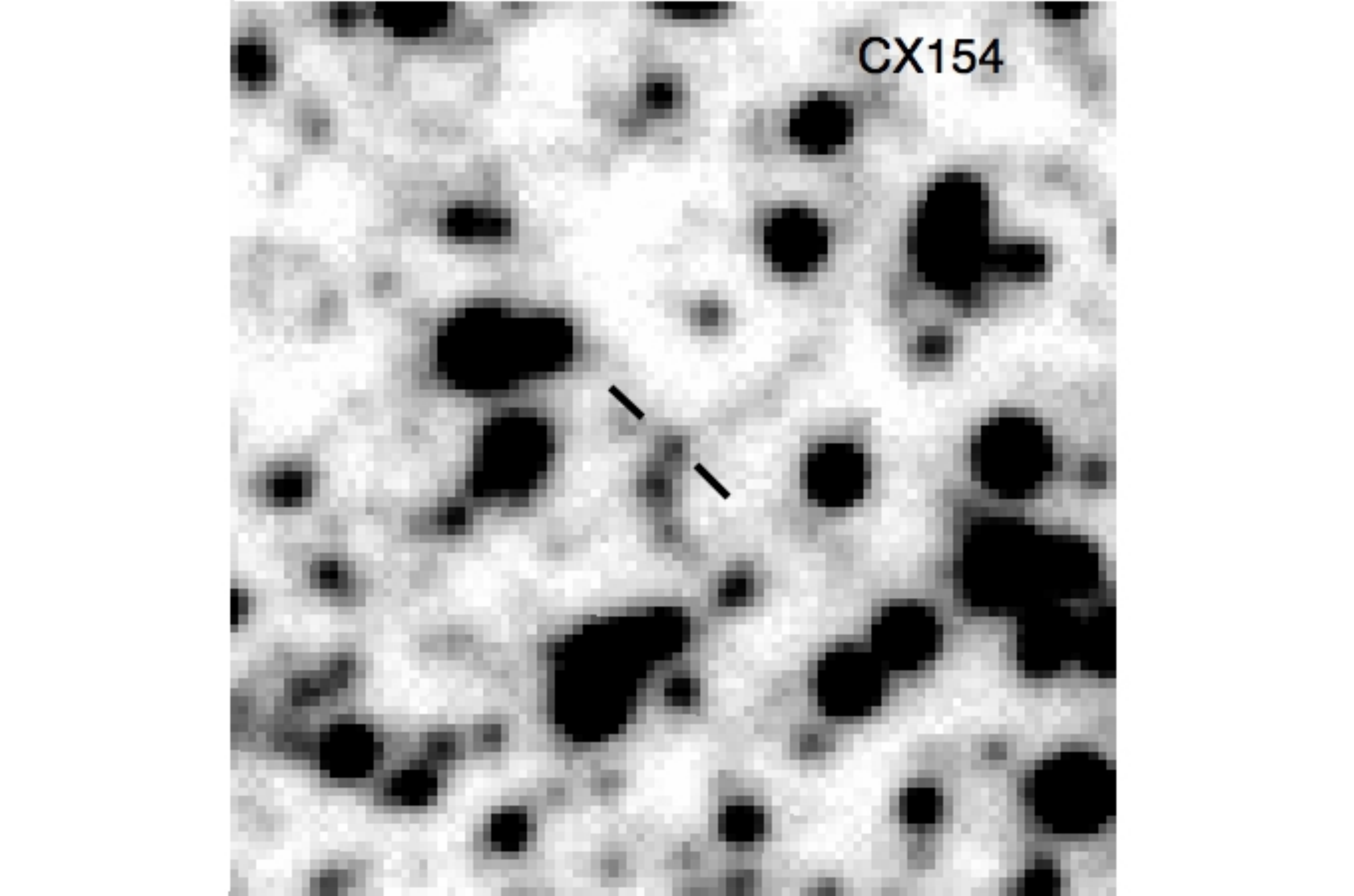} \\

\end{array}$
\end{center}
\caption{  Finding charts.}
\label{licus}
\end{figure*}

\begin{figure*}
\begin{center}$
\begin{array}{ccc}
\vspace{1cm}
\includegraphics[width=2.5in, angle=0.0]{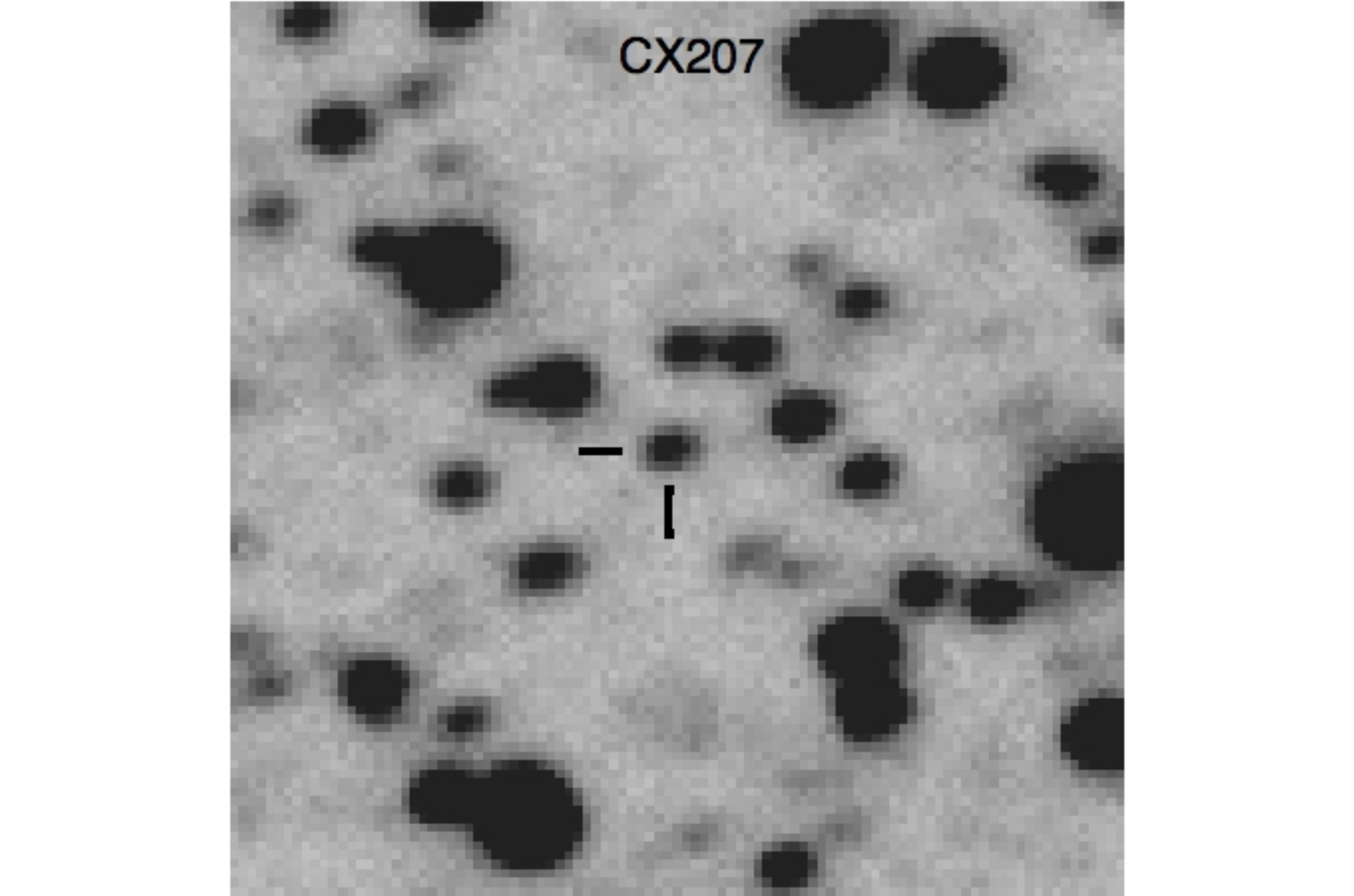} &
\includegraphics[width=2.5in, angle=0.0]{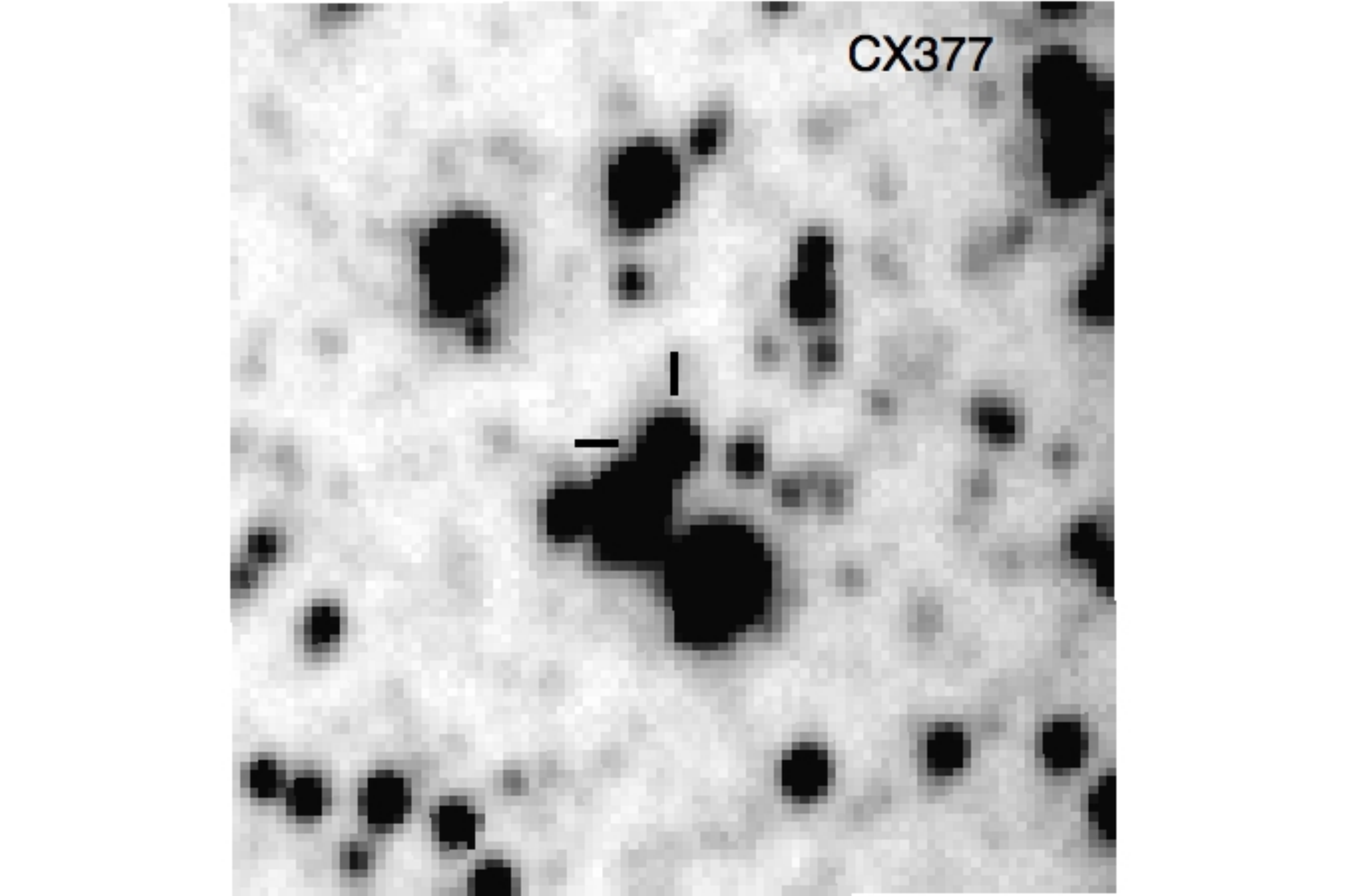}  &
\includegraphics[width=2.5in, angle=0.0]{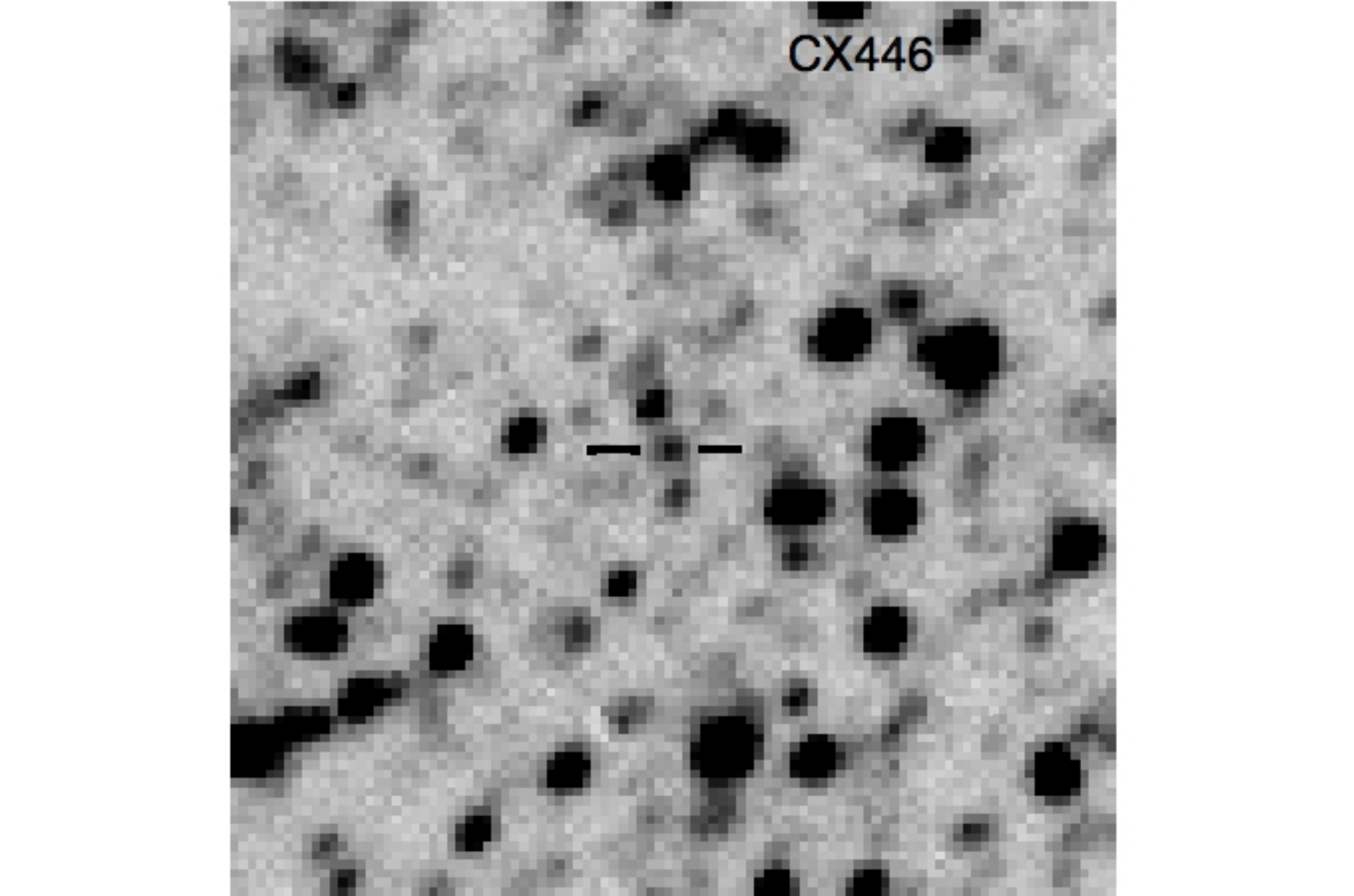} \\
\vspace{1cm}
\includegraphics[width=2.5in, angle=0.0]{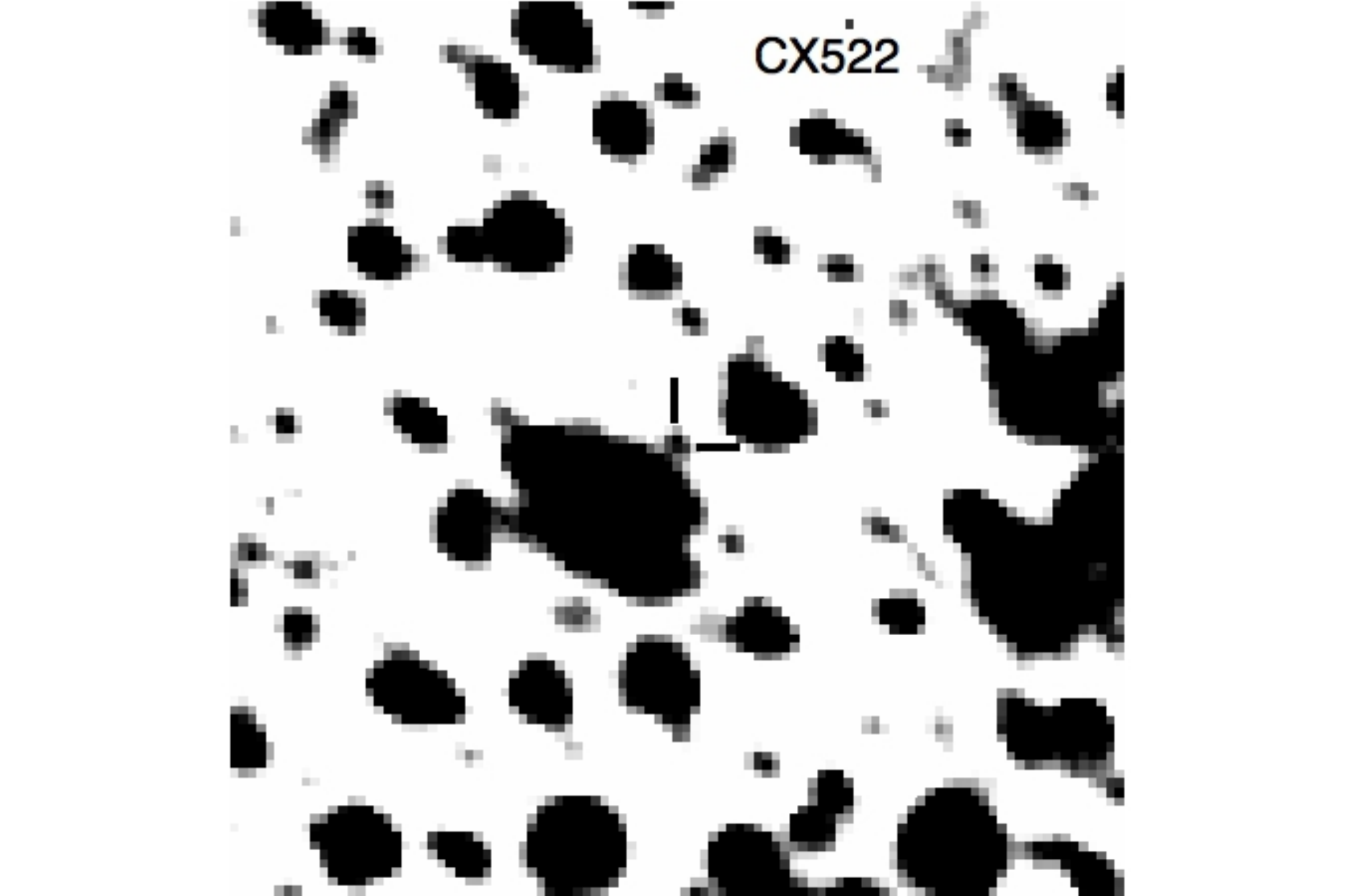} &
\includegraphics[width=2.5in, angle=0.0]{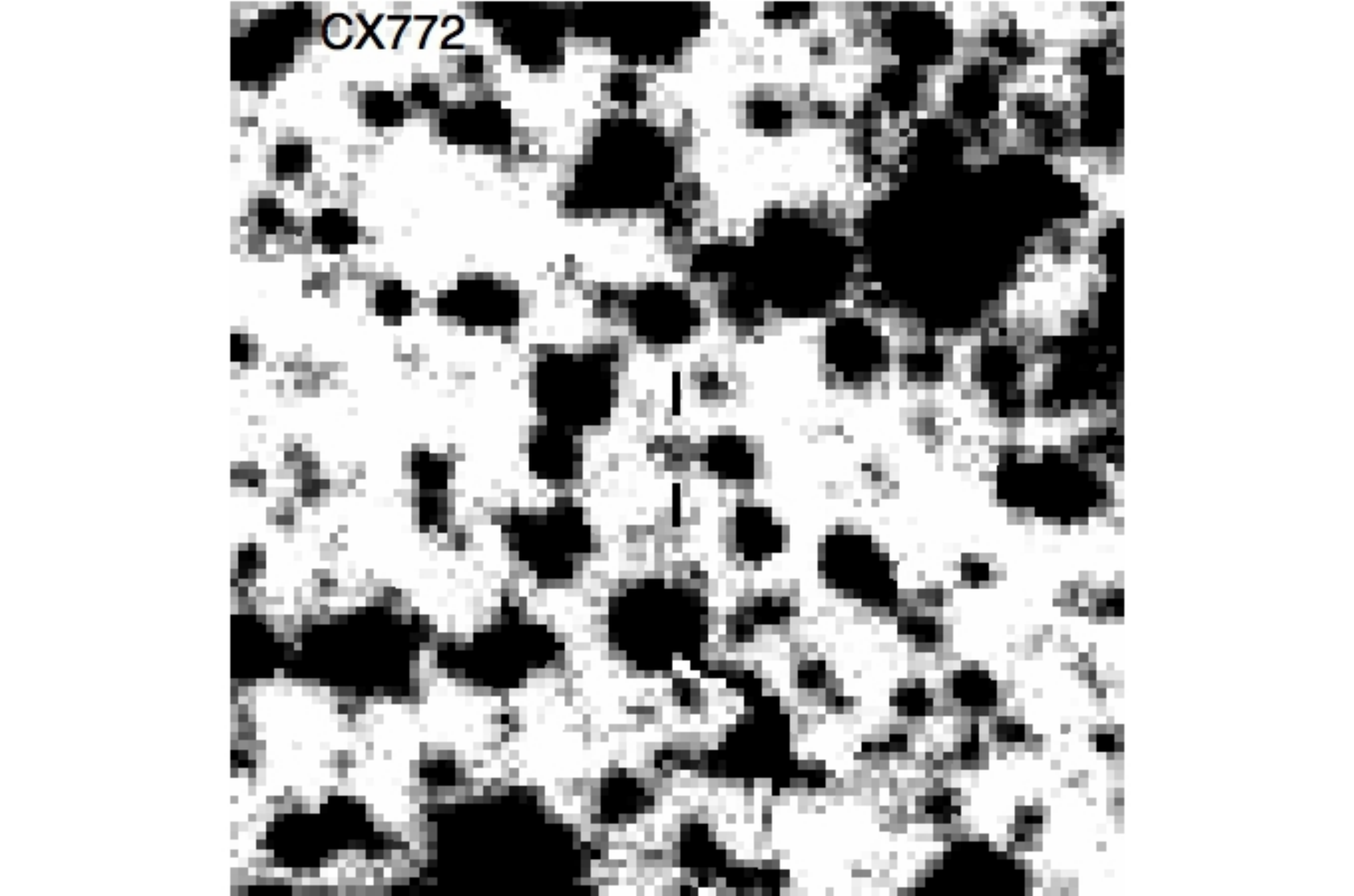} &
\includegraphics[width=2.5in, angle=0.0]{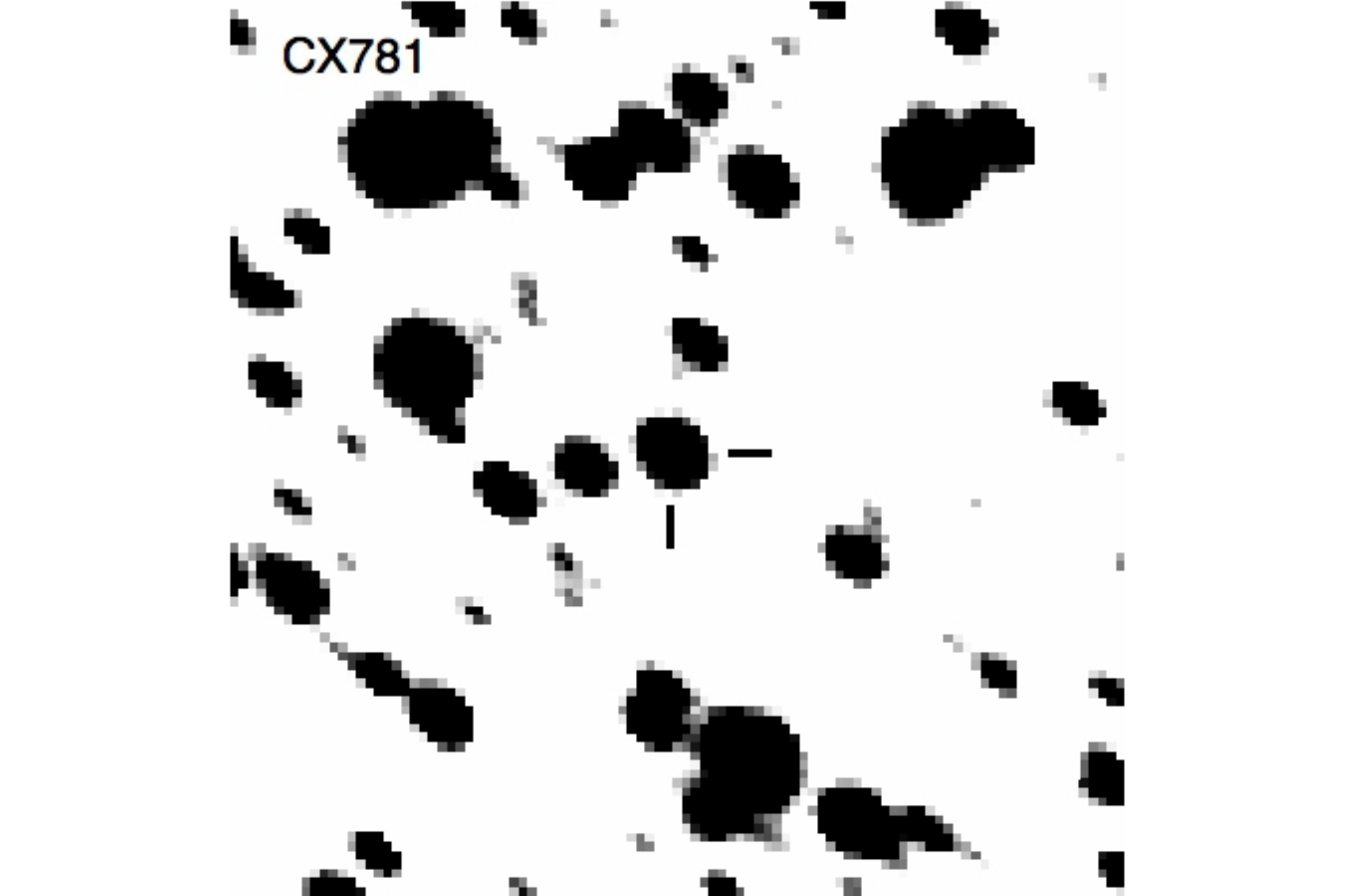} \\
\vspace{1cm}
\includegraphics[width=2.5in, angle=0.0]{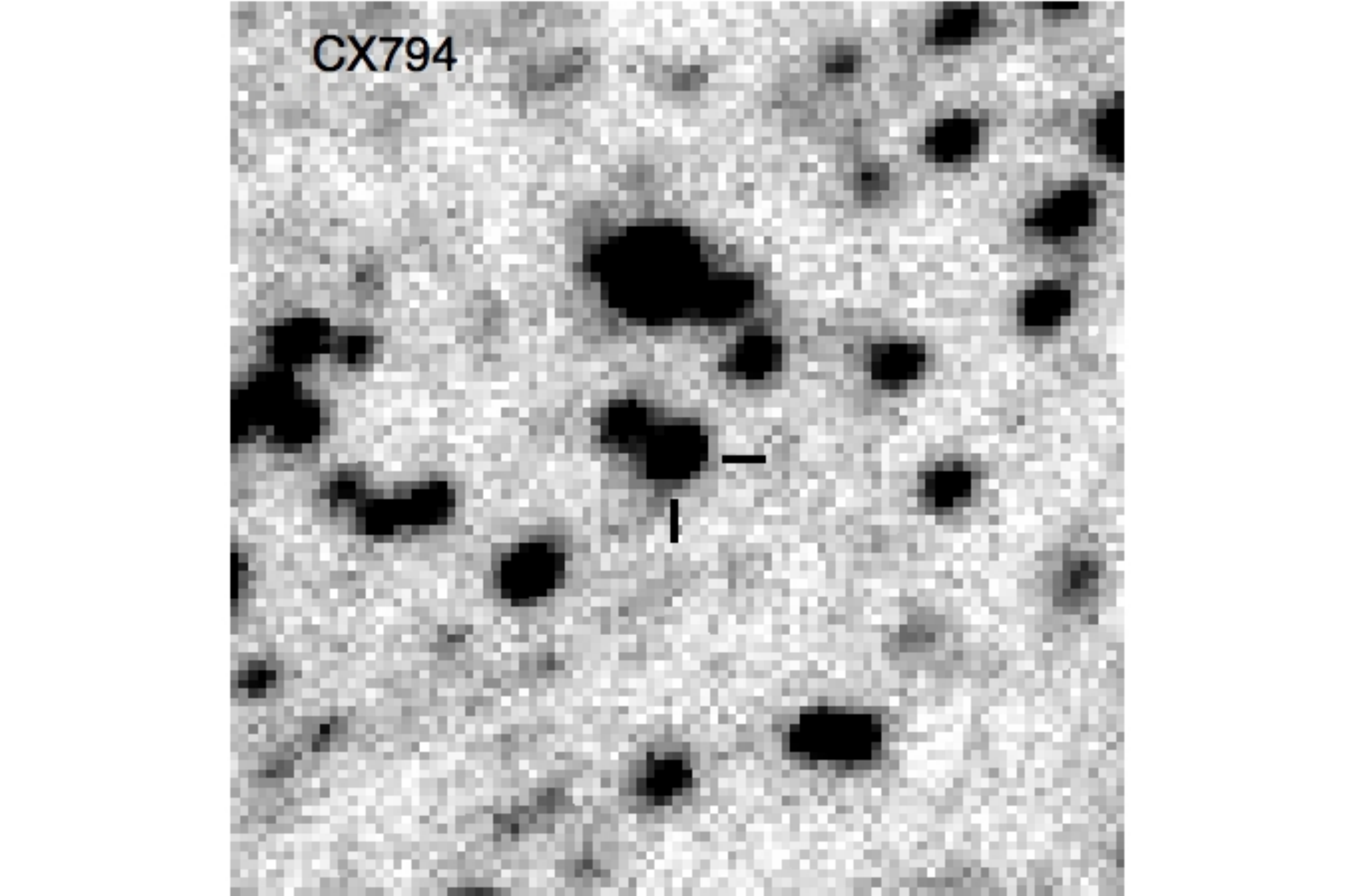} &
\includegraphics[width=2.5in, angle=0.0]{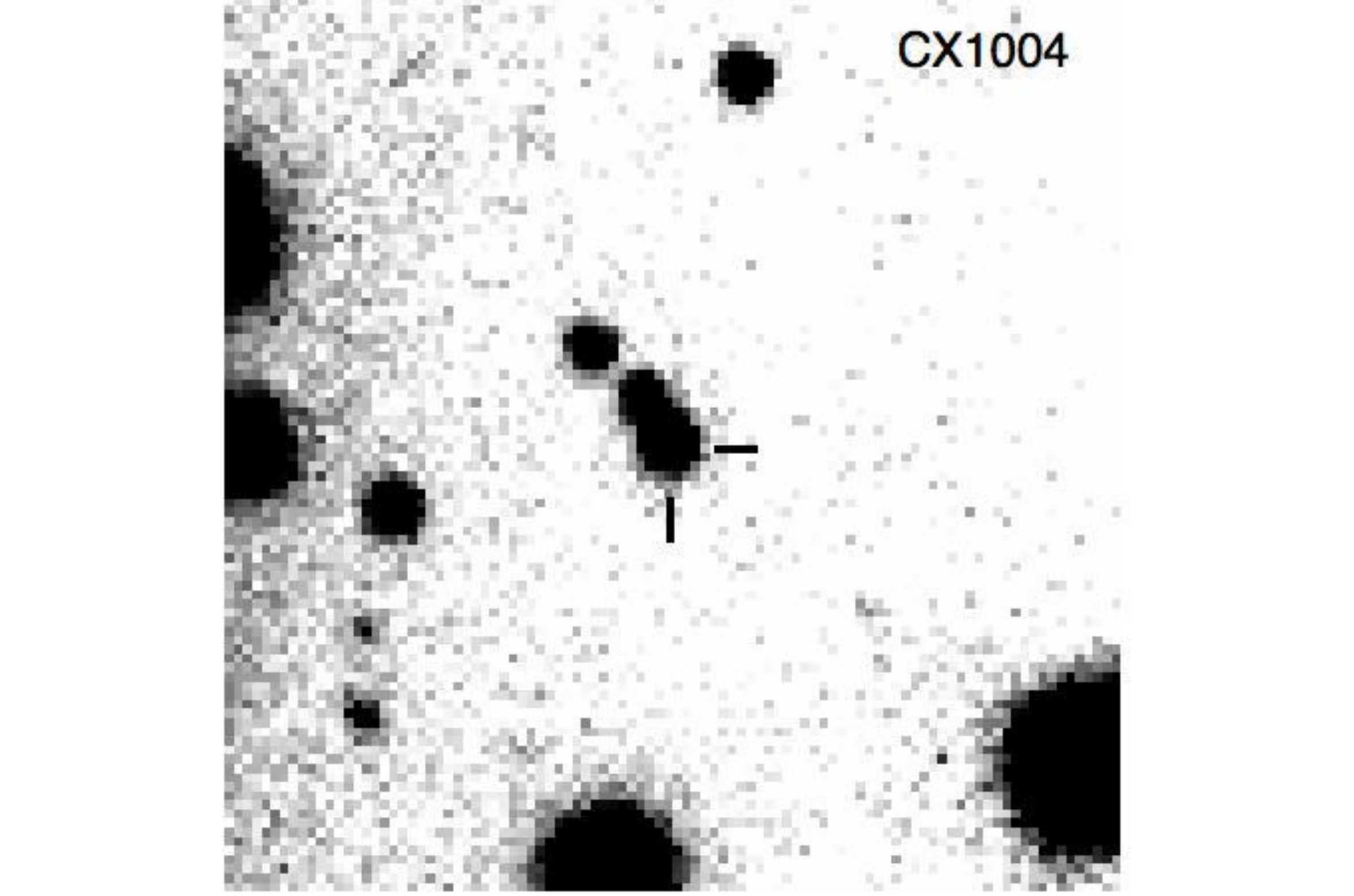} &
\includegraphics[width=2.5in, angle=0.0]{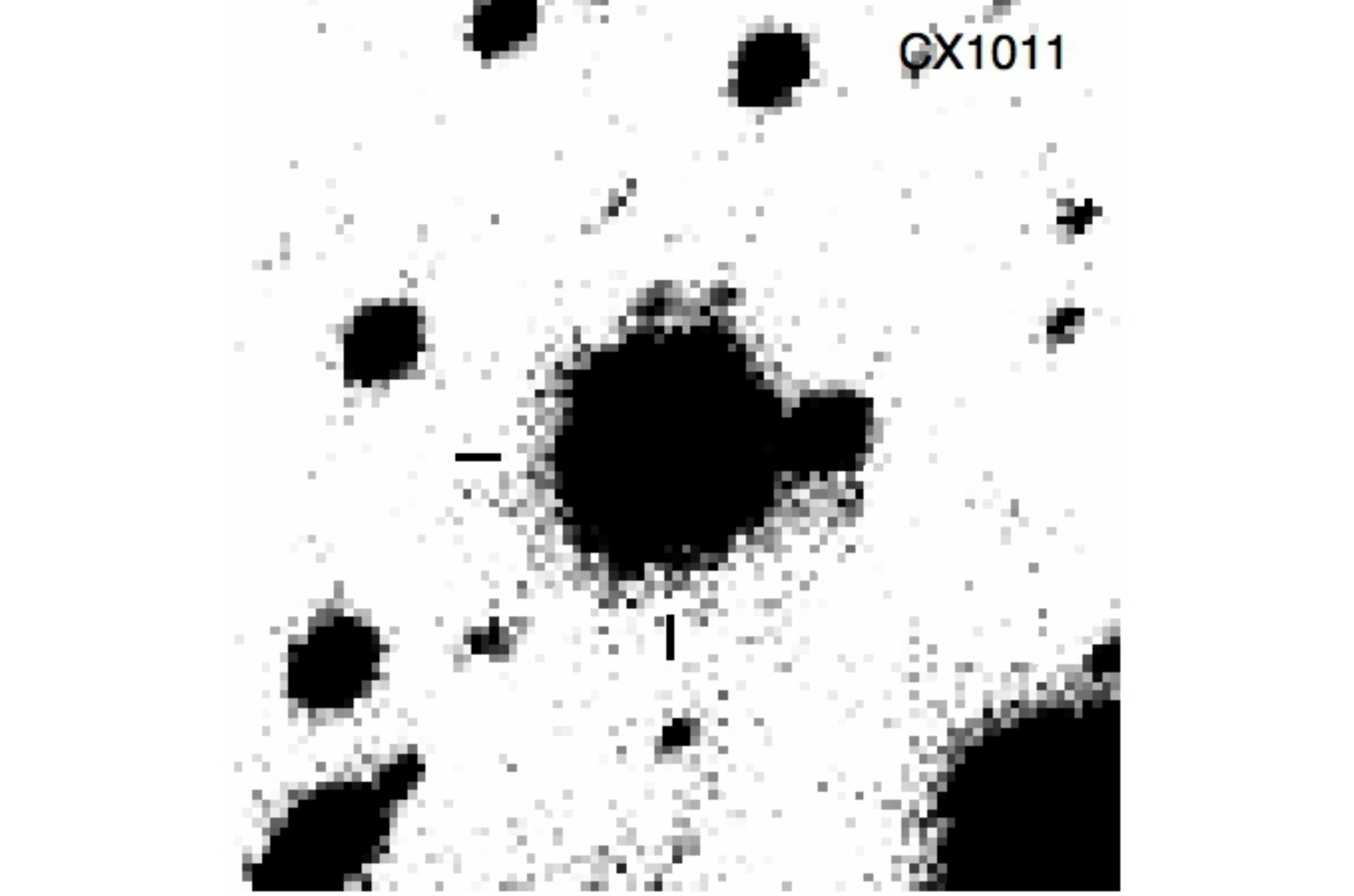} \\

\end{array}$
\end{center}
\caption{  Finding charts (continued).}
\label{licus}
\end{figure*}

%\bibliographystyle{mn} 
%\bibliography{gbs.bib}

\end{document}